%% file: Acta_paper.tex
\documentclass[twoside,fleqn]{ActaStyle}
\usepackage[pdftex]{graphicx}      
\usepackage{times}
\usepackage{amsfonts}
\usepackage{amssymb}
\usepackage{amsmath}
\usepackage[square,numbers,sort&compress]{natbib}
\newcommand{\bR}{{\bf R}}

\def\authorheading{M. Bajdich and L. Mitas}
\def\shorttitle{Quantum Monte Carlo}
\DeclareGraphicsExtensions{.pdf}

\usepackage[pdftex,                
bookmarks=true,
bookmarksnumbered=true,
hypertexnames=false,
breaklinks=true,%
citebordercolor={0 0 1},                     
linkbordercolor={0 0 1}]{hyperref} 

\hypersetup{
pdfauthor   = {Michal Bajdich, Lubos Mitas},
pdftitle    = {Electronic Structure Quantum Monte Carlo},
pdfsubject  = {Review},
pdfkeywords = {QMC, Nodes, Pfaffians, Backflow, FeO, QWalk},
}

\begin{document}
\pagerange{1}{87}   

\title{Electronic Structure Quantum Monte Carlo}

\author{Michal Bajdich, Lubos Mitas\email{lmitas@ncsu.edu}}
              {Department of Physics, North Carolina State University, Raleigh, 27695}

\abstract{Quantum Monte Carlo (QMC) is an advanced simulation methodology
 for studies of many-body quantum systems. 
The QMC approaches combine analytical insights with stochastic 
computational techniques for efficient solution of several classes 
of important many-body problems such as the stationary
Schr\"odinger equation.  QMC methods of various flavors
have been applied to a great variety 
of systems spanning continuous and lattice quantum models, 
molecular and condensed systems, BEC-BCS ultracold condensates,
nuclei, etc. In this review,  we focus on the 
electronic structure QMC, i.e., methods relevant for
systems described by the electron-ion Hamiltonians. 
Some of the key QMC achievements include direct treatment of electron correlation,
accuracy in predicting energy differences and favorable scaling 
in the system size.
Calculations of atoms, molecules, clusters and solids  
have demonstrated QMC applicability to real systems with
hundreds of electrons while providing  
90-95\% of the correlation energy and energy differences typically within a few 
percent of experiments.  Advances in accuracy beyond these limits 
are hampered by the so-called fixed-node approximation which is used to  
circumvent the notorious fermion sign problem. Many-body nodes 
of fermion states and their properties have therefore become 
one of the important topics for further progress in predictive power
and efficiency of QMC calculations. Some of our recent  
results on the wave function nodes
and related nodal domain topologies will be briefly reviewed. This
includes analysis of
  few-electron systems and descriptions of exact and approximate nodes
using transformations and projections of the highly-dimensional 
nodal hypersurfaces into the
3D space. Studies of fermion nodes offer new insights into topological 
properties of eigenstates such as explicit demonstrations that
 generic fermionic ground states exhibit the minimal 
number of two nodal domains.  
Recently proposed trial wave functions based on 
pfaffians with pairing orbitals are presented
and their nodal properties are tested in calculations of first row atoms and 
molecules. 
Finally, backflow ``dressed'' coordinates are introduced 
as another possibility for capturing correlation effects
and for decreasing the fixed-node bias.}

\newpage
\tableofcontents

\include{Introduction/Introduction}

\include{methods_of_qmc/methods_of_qmc}
 \include{nodeschapter/nodeschapter}
 \include{pfaffianchapter/pfaffianchapter}

 \include{backflow/backflow}

 \include{qwalk/qwalk}
 \include{Conclusions/Conclusions}

\bibliographystyle{apsrev}
\bibsep=.1em
\bibliography{Acta_paper}
\end{document}

%% file: Introduction/Introduction.tex
\section{Introduction}\label{ch:intro}

Properties and behavior of quantum many-body systems
are determined by the laws of quantum physics which have been known since the 1930s.  
The time-dependent and stationary Schr\"odinger equations describe molecules,
condensed matter systems, surfaces and interfaces, but also  
artificial structures such as quantum dots, nanowires, ultracold condensates
and so forth. However, the task of solving the 
 Schr\"odinger equation for systems of electrons and ions and predicting the key quantities of interest 
such as cohesion and 
binding energies, optical gaps, chemical reaction barriers, types of magnetic order or 
presence of quantum condensates seems to be beyond formidable.
Indeed, Paul Dirac recognized this already in 1929: ``The general theory of quantum
mechanics is now almost complete. The underlying physical laws necessary for the mathematical
theory of a large part of physics and the whole chemistry are thus completely known, and
the difficulty is only that the exact application of these laws leads to equations much
too complicated to be soluble.'' Let us therefore try to understand the key obstacles. The first 
complication comes from the fact that
a typical system of interest involves large number of particles. Even moderate size molecules
or periodic supercells of solids contain from tens to thousands or more quantum particles.
Second, the particles interact and for interesting problems the interactions are strong
and influence the systems in profound ways.
Third, the solutions have to conform to the required
quantum symmetries such as fermionic antisymmetry and others. This is a fundamental departure from 
classical systems and makes the problem particularly challenging. 
Fourth, the required accuracy for sensible predictions, which are comparable with experiments,
is very high.

Past studies of quantum systems were often 
based on ideas of reducing the interactions into various types of effective mean-fields. 
These approaches gradually developed into a high level of sophistication and
despite limitations, they are still very useful and important on their own or as starting points 
for building new theories. 
At the same time,
 advent of computer technology has offered us a new window of opportunity for studies
of quantum (and many other) problems. It spawned a ``third way'' of doing science which is 
based on simulations, in contrast to analytical approaches and experiments.
In a broad sense, by simulations we mean computational models of reality based on fundamental 
physical laws.
 Such models have value when they enable to make predictions or to provide new 
information which is otherwise impossible or too costly to obtain otherwise. 
In this respect, 
QMC methods represent an illustration and an example of what is the potential of such methodologies. 

Some of the ideas used in QMC methods go back to the times 
before the invention of electronic computers. Already in 1930s
Fermi noticed similarities between
the imaginary time Schr\"odinger equation and stochastic processes in 
statistical mechanics. Apparently, scientists at the Los Alamos National Laboratory
attempted to calculate the hydrogen molecule by a simple version of 
QMC in the early 1950s, around the same time when a seminal work on the first  
Monte Carlo study of classical systems was published by Metropolis and coworkers. 
In the late 1950s Kalos initiated
QMC simulations for a few-particle systems and laid down the statistical and mathematical
background of the Green's function Monte Carlo method.  Pioneering
QMC calculations of the correlation energy of  homogeneous electron gas by 
Ceperley and Alder in 1980 started a new era of stochastic methods applied to electronic
structure problems. QMC methods were then further developed for quantum lattice models,
quantum liquids, nuclear and other systems with contributions from many scientists.   

Currently, the term ``quantum Monte Carlo'' covers several related stochastic methodologies
adapted to determine ground or excited states 
or finite-temperature equilibrium properties for a variety of quantum systems.
Note that the word ``quantum'' is important since QMC approaches differ
in many ways from the Monte Carlo methods for classical systems.
In this overview we present QMC methods which enable to solve the stationary
Schr\"odinger equation for interacting Hamiltonians in continuous space. 
QMC reaches the high accuracy by a direct treatment of many-body effects and 
by employing stochastic computational techniques which 
enable to capture many-body effects well beyond the scope of mean-field or analytical methods.
Conceptually very straightforward is the variational Monte Carlo (VMC) which involves construction
of {\em trial (variational) wave functions} and evaluations of corresponding 
expectation values by stochastic integration.
More advanced approaches are based on projection techniques such as  
the diffusion Monte Carlo (DMC) which 
projects out the ground state from a given trial wave function.
The method is formally exact, however, for fermions it suffers
from a well-known fermion sign problem. 
The reason is that in using statistical methodology we tacitly assume
that the sampled distributions are nonnegative. Unfortunately, for fermions this not true and
in order to circumvent this obstacle
many practical implementations of DMC employ the so-called fixed-node approximation.
This introduces the 
{\em fixed-node error} which is one of the focal points of this overview. 

Elimination or, at least, alleviation of
 the fixed-node errors is therefore of interest from both fundamental and practical point 
of views---since this is essentially the {\em only} approximation which is not deeply understood 
and also because its full control requires inordinate computational resources. 
At first, the nodal problem looks intractable since many-body wave functions nodes are complicated
high-dimensional hypersurfaces determined by the full solution of the Schr\"odinger equation.
However, study of nodal structures has revealed a number of important properties of fermionic
states such as nodal domain topologies
and their relationships to ideas in geometry and spectral theory of operators.
Therefore, some of the latest research developments on fermionic wave function nodes will be presented and explained. 

Along the same line of the research, we outline current developments in construction
of efficient and accurate wave functions which explore ideas
beyond the traditional determinants of single-particle orbitals.
Recently proposed {\em Pfaffian} wave functions built from pairing
 orbitals, which provide higher accuracy and also correct nodal topologies, 
will be explained
 ~\cite{pfaffianprl,pfaffianprb}. In the subsequent part,
 a  fermion coordinate transformation of the
{\em backflow type} which was
recently tested for inhomogeneous systems is introduced.
We generalize its application to Pfaffian wave functions and illustrate the results on
some of the first benchmarks of these ideas.

Fixed-node QMC simulations are computationally more demanding by two to three orders 
of magnitude when compared to the mainstream electronic structure
methods which rely on mean-fields treatments of electron-electron interactions.
However, traditional mean-field theories often lack the needed accuracy and also require frequent
re-tuning, sometimes just to qualitatively reconcile the results with experiments.
Although at present, building of accurate many-body wave functions is far from being 
automatic and demands very significant effort as well, the results often provide high accuracy 
and become benchmarks for other methods. 
For cohesive and binding energies, band gaps, and other energy differences 
the agreement with experiments is typically within 1-3\%~\cite{jeff_benchmark,allelectronFNDMCbenchmark,qmcrev}.
The computational cost of QMC increases basically as a third power of the system size, making
calculations with hundreds of electrons tractable and
large clusters~\cite{williamson01} and solids~\cite{williamson_fse}
up to 1000 electrons have been already studied. 
Producing accurate ``numbers'' has its own merit and the predictive power is 
particularly valuable for experiments and for practical purposes. However, even more important  
are new insights into quantum phenomena and detailed analysis of 
theoretical ideas. Indeed, QMC is very much in the line of ``it from bit'' paradigm,
alongside, for example, of substantional computational efforts in quantum chromodynamics 
which not only predict hadron masses but,  
at the same time, contribute to the validation of the fundamental theory. 
Similar simulations efforts exist in other areas of physics as well.
Just a few decades ago it was almost unthinkable that one would be able to solve
Schr\"odinger equation for hundreds of electrons in an explicit, many-body wave function framework.
Today, such calculations are feasible using available computational resources. At the same time, much 
more remains to be done, of course, to make the methods more insightful, more efficient and
their application less laborious. We hope this overview will contribute
to the growing interest in this rapidly developing field of research.

\newpage
\noindent{The review is organized as follows:}
\begin{itemize}
\item The remainder of this Section provides brief introduction into several traditional methods of solving 
the stationary Schr\"odinger problem for electron-ion systems.
\item The next section follows with the description of the methodology 
behind the VMC and DMC methods. 
\item Results of studies on exact and 
approximate nodes of fermionic wave functions are summarized in the Sec.~\ref{ch:nodes}.
\item Section~\ref{ch:pfaffians} discusses 
generalized pairing wave functions and their properties.
\item Overview of preliminary results on the 
backflow wave functions is given in the section~\ref{ch:bf}. 
\item Section~\ref{ch:feo} shows the example of large scale QMC application: FeO solid.
\item A QMC package Qwalk is introduced in Sec.~\ref{ch:qwalk}.
\item The last section concludes with comments and ideas on current developments and outlooks.
\end{itemize}

\subsection{Many-Body Schr\"odinger Equation}
Let us consider a system of quantum particles such as electrons and ions 
with Coulomb interactions. Since the masses of nuclei and electrons 
differ by three orders of magnitude or more,
let us simplify  the problem by
the {\em Born-Oppenheimer approximation} which allows to separate the electrons
 from the slowly moving system of ions.
The electronic part of non-relativistic Born-Oppenheimer Hamiltonian 
  is given by
\begin{equation}\label{eq:hamiltonian}
{\mathcal H}=T+V=T+V_{ext}+V_{ee}= - \frac 1 2 \sum_i \nabla^2_i+\sum_{i,I}
        \frac{-Z_I}{|r_i-R_I|} +  \sum_{j >
        i}\frac{1}{|r_i-r_j|} ,
\end{equation}
where $i$ and $j$ label the electrons and while $I$
sums over the ions with charges $Z_I$.
(Throughout the review we employ the atomic units, a.u., with
$\hbar=m_e=e=1$.)
 The basic properties of such systems can be found 
from eigenstates $\Psi_n$ of the stationary
 Schr\"odinger equation 
\begin{equation}\label{eq:sch}
 {\mathcal H} \Psi_n= E_n\Psi_n.
\end{equation}
Colloquially, we call such solutions (exact or approximate) and derived properties  
as electronic structure.

The history of electronic structure  {\it ab initio} methods (see, for example, Ref.~\cite{martin})  
started soon after the discovery of quantum mechanics by pioneering works of Heitler
and London~\cite{london}, Hartree~\cite{hartree}, and
Hylleraas~\cite{hylleraas,node1s2s_1}. An important step forward 
was made by Fock~\cite{fock} who established the simplest antisymmetric
wave function and formulated the Hartree--Fock
(HF) theory which takes into account
the Pauli {\em exclusion
principle}~\cite{pauli}. The theory was further developed by 
Slater~\cite{slater} and others and it has become a starting point of many sophisticated
approaches not only in condensed matter physics or quantum chemistry but also in
other areas such as nuclear physics. 
The HF theory replaces the hard problem of many
interacting electrons with a system of independent electrons in an effective,
self-consistent field (SCF).

For periodic systems 
the effective free electron theory 
and the {\em band theory} of Bloch~\cite{bloch}
were the first critical steps towards understanding real crystals. In
1930s, the foundation of the basic classification of solids into
metals, semiconductors and insulators was laid down and later  Wigner
and Seitz~\cite{wigner1,wigner2} performed the first quantitative
calculation of electronic states in sodium metal. 
Building upon homogeneous electron gas model,
the density functional theory (DFT) was invented by Hohenberg and
Kohn~\cite{kohn_dft} and further developed by Kohn and Sham~\cite{kohn_sham}. 
Subsequent developments proved to be very successful and  
DFT has become the mainstream method for many applications which 
cover not only condensed systems but also molecules, surfaces, even nuclear
and other systems. DFT together with HF and
post-HF methods are relevant for our discussion of quantum
Monte Carlo, which uses the results from these approaches as a reference and also
for building wave functions. What follows is their brief overview.

\subsection{Hartree--Fock Theory}\label{sec:intro:HF}
Due to the Pauli exclusion principle, any solution of the
Schr\"odinger equation with Hamiltonian~(\ref{eq:hamiltonian}) has to
be antisymmetric under coordinate exchange of any two electrons as
\begin{align}
\Psi(\ldots,i,\ldots,j,\ldots)=-\Psi(\ldots,j,\ldots,i,\ldots),
\end{align} 
where $i,j$ denote both spatial and spin degrees of freedom.
The Hartree--Fock theory ~\cite{fock,slater} is based on the simplest
antisymmetric wave function, sometimes called the Slater determinant, 
\begin{align}\label{slaterdet}
\Psi(1,2,\ldots,N)&= \frac{1}{\sqrt{N!}} \sum_P (-1)^P \tilde
\varphi_{i_1}(1)\tilde \varphi_{i_2}(2)\ldots\tilde \varphi_{i_N}(N)\\
\nonumber
&=\frac{1}{\sqrt{N!}}\begin{vmatrix} \tilde\varphi_1(1) &
\tilde\varphi_1(2) & \ldots & \tilde\varphi_1(N)\\ \tilde\varphi_2(1)
& \tilde\varphi_2(2) & \ldots & \tilde\varphi_2(N)\\ \vdots & \vdots &
\vdots & \vdots \\ \tilde\varphi_N(1) & \tilde\varphi_N(2) & \ldots &
\tilde\varphi_N(N)\\
\end{vmatrix}
\equiv {\rm det}[\tilde\varphi_1(1),\ldots, \tilde\varphi_N(N)],
\end{align}
to approximate the state of $N$-electron system.  $\{ \tilde
\varphi_i(j)\}$ are one-particle spin-orbitals, each being a product 
of the spatial $\varphi_i^\sigma(j)$ and spin $\alpha_i(\sigma_j)$ orbitals.
(Note that $\varphi_i^\sigma(j)$ is independent of spin $\sigma$ 
in closed-shell cases. In open-shell systems, this assumption
corresponds to {\em the spin-restricted Hartree--Fock approximation}).
The Hartree--Fock total energy can be found
from the following variational condition
\begin{equation}\label{eq:hfvariation}
\langle \Psi |{\mathcal H} |\Psi \rangle = {\rm min}; \quad \langle \varphi_j|\varphi_j\rangle =1, \; j=1,...,N.
\end{equation}
If the Hamiltonian  is spin-independent or it is diagonal 
in the spin basis $\sigma=\{|\uparrow\rangle,|\downarrow\rangle\}$, the
expectation value of  Hamiltonian~(\ref{eq:hamiltonian}) with the wave function~(\ref{slaterdet}) is given by
\begin{align}
E_{HF}=&\sum_{i,\sigma}\!\int \varphi^{\sigma*}_i({\bf r})\left[
-\frac{1}{2} \nabla^2 + V_{ext} \right] \varphi^{\sigma}_i({\bf r})\,{\rm
d}{\bf r} \nonumber \\ &+\frac{1}{2}\sum_{i,j,\sigma_i,\sigma_j}\!\int
\!\!\!\int \frac{\varphi_i^{\sigma_i*}({\bf r}) \varphi_j^{\sigma_j*}({\bf r'})
\varphi_i^{\sigma_i}({\bf r}) \varphi_j^{\sigma_j}({\bf r'})}{|{\bf r}-{\bf r'}|} \,{\rm
d}{\bf r} \,{\rm d}{\bf r'}\nonumber \\
&-\frac{1}{2}\sum_{i,j,\sigma}\!\int \!\!\!\int
\frac{\varphi_i^{\sigma*}({\bf r}) \varphi_j^{\sigma*}({\bf r'}) \varphi_j^{\sigma}({\bf
r}) \varphi_i^{\sigma}({\bf r'})}{|{\bf r}-{\bf r'}|} \,{\rm d}{\bf r} \,{\rm
d}{\bf r'}.
\end{align}
The 
expression for the HF energy contains three different terms. The first one
represents one-particle contributions,
the second is the 
Coulomb interaction also called the Hartree term and the last contribution
is the {\em exchange} or Fock term. Note that for
 $i=j$ (self-interaction) the last two terms
explicitly cancel each other.
Variation of the HF total energy leads to one-particle
HF equations with the resulting spectrum
$\{\varphi_i^{\sigma},\epsilon_i\}$ of occupied and virtual orbitals
obtained by solving 
\begin{equation}
\left[
-\frac{1}{2} \nabla^2 + V_{ext} + V_{HF}( \{ \varphi_j^{\sigma} \} ) \right] \varphi^{\sigma}_i({\bf r})=\epsilon_i
\varphi_i^{\sigma}({\bf r})
\end{equation}
The effective field operator $V_{HF}$ contains both Coulomb and exchange terms and due
to the self-consistency depends also on the set of occupied orbitals.  The resulting operator is both nonlinear
and nonlocal as manifested by the fact that $V_{HF}$, in general, differs for each state $i$.
This complicated dependence makes the solution of HF equations rather involved. The 
solutions are usually expanded in an appropriate basis sets and iterated to self-consistency.
Sophisticated and powerful tools such as well-developed
and extensive quantum chemical or solid state packages are necessary for such solutions.
For solids the HF equations were out of reach for a long time and only in late
eighties the first package had emerged which was able to find reasonably accurate and 
converged solutions.

\subsection{Post Hartree--Fock Methods}\label{sec:intro:postHF}
The HF energy
is variational, i.e, above the exact eigenvalue,
and whatever is left out is referred to as {\em electronic
correlation}. For a given state, the correlation energy is therefore defined as a difference
of the HF energy and the exact eigenvalue, 
\begin{equation}
E_{corr}=E_{HF}-E.  
\end{equation}
In electron-ion systems $E_{corr}$ is only a small fraction of the total energy,
of the order of $\approx$ 1-3 \%, however, correlations impact almost all quantities in a major way.
For example, correlation accounts for a large portion of
cohesions, excitations, barrier heights and other important energy differences. Many other effects
such as magnetism or superconductivity are driven by correlations so that its accurate 
description becomes crucial. Revealing the nature of correlation effects and how they influence
the behavior of quantum systems
is therefore one of the central challenges of modern electronic structure research.

The missing correlation in the HF theory can be accounted for by expanding the wave
function into linear combination of Slater determinants with one-, two-, ..., $N$-particle excitations.
Providing the  one-particle set of states is complete and all possible
excitations are covered one can formally reach the exact solution. 
Unfortunately, the convergence of such
methods is slow and the computational demands grow exponentially with the system size.
 Nevertheless, for small systems such methods are effective and provide
valuable checks for other approaches. 
 The most frequently used post-HF approaches are
configuration interaction (CI), multi-configurational self-consistent
field (MCSCF), and coupled cluster (CC) methods.  The idea behind CI
is to diagonalize the $N$-electron Hamiltonian on the space of
orthogonal Slater determinants.  These determinants are typically
constructed as few-particle excitations from the reference
determinant (usually the HF solution) using the virtual HF orbitals.  
The CI method thus provides the most optimal expansion on such 
determinantal space.
The most commonly used is CI
with all the single and double excitations (CISD). Besides the slow convergence  
in  one-particle orbitals the method becomes problematic with increasing 
$N$ since the recovered correlation is proportional  to 
  $\sqrt{N}$ (instead of $N$), thus leading to the well-known 
{\em size-consistency problem} (i.e., a wrong scaling of total energy with $N$).

The MCSCF is  a generalization of the expansion in excited determinants 
and differs from CI by optimization of both of the orbitals and the
determinantal coefficients.  Such optimizations are quite difficult and the demands 
often limit the size of the problem which can be studied. The special case
of the MCSCF is the complete active space SCF (CASSCF), which is size consistent, at
the price of exponential scaling in the number of particles.  

The size-consistency problem can be overcome by going beyond linear expansions as it 
is done in the  coupled cluster method~\cite{cizek}. The wave function ansatz is based
on exponentiated excitation operators and the corresponding optimized parameters are found by
solving a set of coupled nonlinear equations.  
The computational cost for CCSD (with singles and doubles) scales as 
$N^6$ and therefore constitutes a formidable challenge for larger systems. The commonly used 
CCSD(T) (CC with single, double and perturbative triple excitations) is often used as a standard
in accuracy for small systems. 
For an overview and detailed introduction into the above mentioned
quantum chemical methods we refer the reader, for example,
 to the book by A. Szabo and N. S. Ostlund~\cite{szabo}.


\subsection{Density Functional Theory}\label{sec:intro:DFT}
Previous methods were examples of the wave function based theories. 
For the density functional theory (DFT) the primary object is the {\em one-particle electron density}, i.e.,
the diagonal part of one-particle density matrix. 
DFT is formally an exact method based on Hohenberg and Kohn theorem~\cite{kohn_dft} 
that states that the ground state properties of a many-electron system can be obtained by 
minimizing the total energy functional 
\begin{equation}
E[\rho({\bf r})]=\int V_{ext}({\bf r}) \rho({\bf r})\,{\rm d}{\bf r} + F[\rho({\bf r})],
\end{equation}
where the $F[\rho({\bf r})]$ is some unknown, universal functional of the electronic density $\rho({\bf r})$. 
Assuming such functional would be known, the total energy $E$ would exhibit a minimum for the exact 
ground state 
density  $\rho({\bf r})$. 
 
Kohn and Sham~\cite{kohn_sham} introduced an {\it ansatz} for the density in
terms of one-electron orbitals of an auxiliary non-interacting system as
\begin{equation}
\rho({\bf r})=\sum_i^N |\varphi_i({\bf r})|^2.
\end{equation}
 The total energy of electronic system can be then expressed as  
\begin{equation}
E[\rho({\bf r})]=-\frac{1}{2} \sum_i \int \varphi_i^{*}({\bf r}) \nabla^2\varphi_i({\bf r})\,{\rm d}{\bf r}
+\int \rho({\bf r})V_{ext}({\bf r})\,{\rm d}{\bf r}
+ \frac{1}{2}\int\!\!\!\int \frac{\rho({\bf r})\rho({\bf r'})}{|{\bf r}-{\bf r'}|}\,{\rm d}{\bf r}{\rm d}{\bf r'}
+E_{xc}[\rho({\bf r})],
\end{equation}
where the first two terms represent the familiar energy of a non-interacting system in the external potential, 
the third term being the Hartree term while the rest is an unknown universal {\em exchange-correlation 
functional}. 
By minimization of the total energy as a functional of normalized occupied orbitals one can derive
the Kohn-Sham equations
\begin{equation}
\left[
-\frac{1}{2} \nabla^2 + V_{ext} + V_{DFT}(\rho({\bf r}))\right]\varphi_i^{\sigma}({\bf
r})=\epsilon_i
\varphi_i^{\sigma}({\bf r}).
\end{equation}
The structure of the equations appears similar to the HF equations, the key difference is that
the effective exchange-correlation operator is nonlinear yet local, since it depends only 
on the density but not on individual states. This feature of the DFT theory simplifies the calculations 
considerably and it has enabled to expand the DFT applicability to very large systems. 
If the functional  $E_{xc}[\rho({\bf r})]$ would be precisely known, 
we would have one to one mapping between difficult many-electron system and an effective 
one-particle problem. The fundamental difficulty is that the exact
exchange-correlation functional is unknown
and therefore we have to rely on approximations.

The simplest is the local density approximation (LDA),
\begin{equation}
E_{xc}[\rho({\bf r})]=\int \epsilon_{xc}^{heg}[\rho({\bf r})]\rho({\bf r})\,{\rm d}{\bf r},
\end{equation}
where $\epsilon_{xc}^{heg}$ is the exchange-correlation energy of the 
homogeneous electron gas of the same density. 
It is interesting that basically in all practical implementations of LDA, the correlation portion 
of this energy functional comes from the pioneering QMC calculations
of the homogeneous electron gas~\cite{ceperley80}. 

It is intuitively suggestive to generalize the exchange-correlation functional to include also the density
derivatives, i.e., to 
express  $\epsilon_{xc}[\rho({\bf r}), \nabla \rho({\bf r})]$ as a function of the local density and its gradient. 
This was the essential idea behind the generalized gradient approximation (GGA)~\cite{becke_3parm,becke_exchange,LYP,PBE}, 
which provide improved results for some quantities and has enabled wide-spread use both in condensed matter 
physics and quantum chemistry. 

There are many flavors and generalizations of the density functional theory, 
e.g., hybrid-functionals~\cite{PBE,becke_3parm,LYP}, time-dependent DFT (TD-DFT)~\cite{tddft1,tddft2} for calculations 
of excited states, 
extensions to non-local density functionals (averaged density approximation (ADA)~\cite{ADA_WDA} and weighted density approximation (WDA)~\cite{ADA_WDA}),
or orbital-dependent functionals (e.g. self-interaction corrections (SIC)~\cite{self_interaction} and LDA+U~\cite{lda_plus_u}).   
For some systems the DFT methods are very effective and provide reasonable accuracy at very low cost. Unfortunately,
there are many cases where DFT requires further elaboration: introduction of ad hoc parameters
such as in LDA+U approach or perturbative corrections such as in the so-called GW method. Even then for many systems
such as  
transition metal compounds, weakly interacting van der Waals systems
and other cases the DFT has profound difficulties in reaching the required accuracy. 

%% file: methods_of_qmc/methods_of_qmc.tex
\section{Quantum Monte Carlo Methods}\label{ch:qmc}

\subsection{Introduction}
Quantum Monte Carlo (QMC) is a suite of approaches  based on two complementary strategies for 
solving quantum many-body problems. The first key component is an analysis 
of many-body effects at the wave function level: wave function analytical structure and impact of 
potential singularities, 
efficient functional forms, handling of symmetries and boundary conditions, 
explicit multi-particle correlations and construction of intermediate effective Hamiltonians 
which enable to build-up more accurate many-body treatment. The second key
ingredient is based on employing 
stochastic techniques whenever it is impractical or inefficient to use analytical and 
explicit constructions. 
This is done on several fronts: evaluations of high-dimensional
integrals, solving of quantum equations by mapping onto equivalent stochastic processes,
development of algorithms and numerical approaches for
high accuracy and efficiency. 

Over the last three decades, the impact of QMC has been steadily growing as scientists
explored new algorithms and produced results which were otherwise impossible
to obtain by traditional methods.
Currently,  QMC covers  a great variety of specialized approaches tuned to particular 
Hamiltonians and applications. 
In this review we will briefly describe the variational and diffusion 
quantum Monte Carlo methods. 

In the variational Monte Carlo (VMC) the expectation values 
are evaluated via stochastic integration over
$3N$-dimensional space using trial (variational) wave functions constructed to capture
the desired many-body phenomena.
The variational theorem ensures that the expectation value of the Hamiltonian 
is, within the statistical error bar, a true upper bound to the exact ground-state energy.
The results are determined 
by the choice and quality of the trial wave function construction
and therefore are subject to a {\em variational bias}.
The second described method, the diffusion Monte Carlo, removes most of the variational bias
by projecting out the ground state component from any trial function with nonzero overlap
within the given 
symmetry class.
The fixed-node approximation enables to avoid the fermion sign problem and makes the method
scalable to large sizes.


\subsubsection{Metropolis Sampling}
Basic Monte Carlo multi-variate integration approaches
 are based on the following simple idea. 
Assume 
that we are interested in finding an 
expectation value of $A({\bf R})$ for  
a known non-negative distribution ${\mathcal P}({\bf R})$ 
\begin{equation}
\langle A \rangle =\frac{\int A({\bf R}) {\mathcal P}({\bf R}) {\rm d}{\bf R}}
{\int {\mathcal P}({\bf R}) {\rm d}{\bf R}},
\end{equation}
where  ${\bf R}$ is a position in $d-$dimensional space. This is a common expression for evaluation
of observables encountered
in  classical or quantum statistical mechanics. If the dimension 
of space is large and the integral cannot be reduced to sums and/or products of 
low-dimensional integrals, its straightforward evaluation scales exponentially
in the dimension $d$. This is easy to see by considering a simple rectangular integration rule 
with $N_{g}$ grid points in each dimension. The total number of evaluations will be $(N_{g})^d$
and therefore not feasible even for moderate values of $N_g$ and $d$. The key advantage of
  stochastic methods is that they are immune to the dimensionality problem under reasonable
 conditions which are easy to verify
 (see below). Let us first consider how to generate {\em samples} of the distribution  ${\mathcal P}({\bf R})$.
This is most conveniently done by the Metropolis algorithm~\cite{metropolis} which generates a Markov
chain of  samples 
by a stochastic walk in the configuration ${\bf R}$-space. The stochastic samples are sometimes 
colloquially referred to as random {\em walkers}.
The algorithm enables us to sample any desired multi-dimensional 
probability distribution ${\mathcal P}({\bf R})$ without a prior knowledge of its normalization. 
Given some transition rule $P({\bf R}\to{\bf R}')$ from ${\bf R}$ to ${\bf R}'$, 
which satisfies {\em ergodicity} and {\em detailed balance},
\begin{align}
P({\bf R}\to{\bf R}'){\mathcal P}({\bf R})=P({\bf R}'\to{\bf R}){\mathcal P}({\bf R}'),
\end{align}
the desired probability density ${\mathcal P}({\bf R})$ will converge to an equilibrium state
given by
\begin{align}
\int P({\bf R}\to{\bf R}'){\mathcal P}({\bf R})\,{\rm d}{\bf R}={\mathcal P}({\bf R}'),
\end{align}
see, for example, \cite{kaloswhitlock}.
The transition rule $P({\bf R}\to{\bf R}')$ can be further written as a product of the sampling distribution $T({\bf R}\to{\bf R}')$ 
(typically Gaussian distribution centered around {\bf R}) and the probability of an acceptance $A({\bf R}\to{\bf R}')$ of the proposed step
\begin{align}
P({\bf R}\to{\bf R}')=T({\bf R}\to{\bf R}')A({\bf R}\to{\bf R}').
\end{align}
The conditions of detailed balance are satisfied by choosing the $A({\bf R}\to{\bf R}')$ to be
\begin{equation}
A({\bf R}\to{\bf R}')=\min\left[1,\frac{T({\bf R}'\to{\bf R}){\mathcal P}({\bf R}')}{T({\bf R}\to{\bf R}'){\mathcal P}({\bf R})}\right].
\end{equation}
In this manner, the Markov chain initialized at some random state evolves towards an equilibrium. 
After some initial convergence and equilibration period 
we can start to collect the statistics. Assuming that we can apply the 
{\em central limit theorem}, the usual estimators for the mean, 
variance and error of the mean 
are given by  
\begin{align}
\langle A \rangle &=\frac{1}{M}\sum_m^M A({\bf R}_m) + {\mathcal O}(1/\sqrt{M}),\\ \nonumber
\langle \sigma_A^2\rangle& \approx \frac{1}{M-1}\sum_m^M (A({\bf R}_m)-\langle A \rangle)^2, \\ \nonumber
\epsilon_A&\approx \frac{\sigma_A}{\sqrt{M}},
\end{align}
where $M$ represents the number of statistically independent sampling points.  

The computational demands 
depend on the desired error $\epsilon_A$ and on the variance $ \sigma_A^2$.
Seemingly, the estimators and the error bar 
do not depend on the dimensionality $d$ of the problem. In general, however, one needs
to take into account the impact of
increasing $d$ for the following: i) amount of calculation needed
to obtain statistically independent sample of ${\mathcal P}({\bf R})$ and
for evaluation of $A({\bf R})$; and  
ii) growth of variance $\sigma_A^2$. 
Providing all of these grow as some polynomials of $d$,
the method guarantees the
 overall polynomial scaling of the algorithm. In its basic form
the expression for computational demands is then
\begin{equation}\label{eq:timedem}
T_{comp}= \frac{\sigma^2_A}{\epsilon_A^2} \; T_{sample},
\end{equation}
where $T_{sample}$ lumps the time needed to obtain statistically independent sample and 
to evaluate required quantities.

We note that it is often possible to decrease the variance 
by the importance sampling methods and thus increase
the efficiency as explained, for example, in Ref. \cite{kaloswhitlock}.

\subsection{Variational Monte Carlo}
Let us consider a {\em trial variational wave function} $\Psi_T({\bf R})$ which 
approximates the lowest eigenstate of ${\mathcal H}$ for a given symmetry.
We will elaborate
on constructions and properties of
 $\Psi_T$ in Sec.~\ref{sec:twf}. In
 variational Monte Carlo (VMC) the expectation values 
for the given $\Psi_T({\bf R})$ are evaluated by stochastic sampling. 
For example, the total energy is given by
\begin{equation}
E_{VMC}=\frac{\int \Psi_T^*({\bf R}){\mathcal H}\Psi_T({\bf R})\,{\rm d}{\bf R}}{\int \Psi_T^*({\bf R})\Psi_T({\bf R})\,{\rm d}{\bf R}}
=\frac{\int |\Psi_T({\bf R})|^2 E_L({\bf R})\,{\rm d}{\bf R}}{\int  |\Psi_T({\bf R})|^2\,{\rm d}{\bf R}}\geq E_0,
\end{equation}
where we have introduced the so-called local energy 
\begin{align}\label{eq:localenergy}
E_L({\bf R})=\frac{{\mathcal H}\Psi_T({\bf R})}{\Psi_T({\bf R})}.
\end{align}
The VMC estimator $E_{VMC}$ is then given as 
\begin{equation}
E_{VMC}=\frac{1}{M}\sum_m^M E_L({\bf R}_m) + {\mathcal O}(1/\sqrt{M}),
\end{equation}
assuming the set of $M$ configurations is distributed according to
 ${\mathcal P}({\bf R})=|\Psi_T({\bf R})|^2$.


\subsubsection{Correlated Sampling}
Correlated sampling exploits the fact that the 
variance of difference of correlated random variables $X$ and $Y$
decreases as their correlation increases. 
For example, consider variational energy energy difference for two close wave functions
$\Psi_T^{(1)}$ and $\Psi_T^{(2)}$, e.g., for
different ionic positions or with small external field when evaluating responses.
The energy difference can be written as
\begin{align}
E_1-E_2&=\frac{\int |\Psi_T^{(1)}({\bf R})|^2 E_L^{(1)}({\bf R})\,{\rm d}{\bf R}}{\int |\Psi_T^{(1)}({\bf R})|^2\,{\rm d}{\bf R}} 
-\frac{\int |\Psi_T^{(2)}({\bf R})|^2 E_L^{(2)}({\bf R})\,{\rm d}{\bf R}}{\int |\Psi_T^{(2)} ({\bf R})|^2\,{\rm d}{\bf R}} \nonumber \\
&=\int |\Psi_T^{(1)}({\bf R})|^2 \left[\frac{E_L^{(1)} ({\bf R}) }{\int |\Psi_T^{(1)}({\bf R})|^2\,{\rm d}{\bf R}}
      -\frac{w({\bf R})E_L^{(2)}({\bf R})}{\int w({\bf R})|\Psi_T^{(1)}({\bf R})|^2\,{\rm d}{\bf R}}\right]\,{\rm d}{\bf R},
\end{align} 
where  
the re-weighting factor is given by $w({\bf R})=|\Psi_T^{(2)}({\bf R})|^2/|\Psi_T^{(1)}({\bf R})|^2$. 
Therefore, we can obtain the above energy difference as a sum over weighted local energy differences. 
The estimations can be carried out using the set of samples drawn, for example,
from the distribution $(\Psi_T^{(1)})^2$. There are other possibilities such as sampling $|\Psi_T^{(1)}\Psi_T^{(2)}|$
and modifying the expressions accordingly.  By taking the difference, 
the noise mostly cancels out, since for 
very similar wave functions the random fluctuations will be very similar
and  the weights will be close to unity.
Consequently, the variance of the difference 
can be much smaller than the variance of each individual quantity alone. The actual decrease in variance 
depends on particular application but gains can be
two orders of magnitude or more.  The algorithm can be further
generalized for more complicated estimators and has been developed also for the diffusion Monte Carlo 
method presented below
~\cite{filippi_force}. 


\subsection{Diffusion Monte Carlo}
\subsubsection{Imaginary Time Schr\" odinger Equation}
Diffusion Monte Carlo method belongs to a class of projection and Green's function QMC approaches.
Let us consider the imaginary time Schr\" odinger equation 
\begin{equation}\label{eq:timesch}
-\frac{\partial\Psi({\bf R},\tau)}{\partial\tau}=({\mathcal H}-E_T)\Psi({\bf R},\tau),
\end{equation}
where $\tau$ is a  real variable 
(imaginary time) and $E_T$ is an energy offset.
The effect of imaginary time is an exponentially fast projection of the lowest state of 
a given symmetry from
any trial function with non-zero overlap.
This can be 
easily seen by expanding $\Psi({\bf R},\tau)$ 
in eigenstates $\{\Phi_n\}$ of ${\mathcal H}$ so that for time $\tau \to \infty$ one finds
\begin{align}
\lim_{\tau \to \infty}\Psi({\bf R},\tau)&=\lim_{\tau \to \infty}\sum_n  \langle \Phi_n| \Psi(0) \rangle \,\,{\rm e}^{-\tau(E_n-E_T)} \,\Phi_n({\bf R}) \\ \nonumber
&=\lim_{\tau \to \infty}\langle \Phi_0| \Psi(0) \rangle \,\,{\rm e}^{-\tau(E_0-E_T)} \Phi_0({\bf R}).
\end{align}
Although the absolute ground states of the Hamiltonians we consider are bosonic, 
by keeping the wave function in the antisymmetric sector, 
one can find the lowest antisymmetric state. 
The parameter $E_T$ can be adjusted to $E_0$  so that
 the exponential weight of $\Psi$ becomes asymptotically constant. 
It is convenient to rewrite the equation~(\ref{eq:timesch})
 into into an integral form using the Green's function
\begin{align}
G({\bf R}\leftarrow {\bf R}',\tau)&=\langle {\bf R}| \,{\rm e}^{-\tau({\mathcal H}-E_T)}|{\bf R}'\rangle.
\end{align}
Then after propagation by a time interval $\Delta\tau$ the wave function is given by
\begin{equation}
\Psi({\bf R},\tau+\Delta\tau)=\int G({\bf R}\leftarrow {\bf R}',\Delta\tau)\Psi({\bf R}',\tau)\,{\rm d}{\bf R}'.
\label{eq:timeevol2}
\end{equation}
and the propagation can be continued by iterations. 


Let us first consider the Schr\"odinger equation with non-interacting particles
\begin{equation}
\frac{\partial\Psi({\bf R},\tau)}{\partial\tau}= \frac{1}{2}\nabla^2 \Psi({\bf R},\tau).
\end{equation}
The equation has a familiar form of diffusion 
 in ${\bf R}$-dimensional space. Assuming that we have 
$N$ particles in a 3D space the corresponding Green's function is a $3N-$dimensional Gaussian 
with variance $\sigma^2=\tau$. The equation describes a well-known process of diffusing particles 
in a random noise medium  with the diffusion constant equal to $1/2$. The key departure from 
traditional methods of solving such differential equations is in the solution representation. The quantum
amplitude $\Psi({\bf R},\tau)$ is described by sampling, i.e., as a density of random walkers (delta functions in ${\bf R}-$space)
\begin{equation}
\Psi({\bf R},\tau)={\rm dens}[\sum_{m=1}^M \delta({\bf R}-{\bf R}_m(\tau))] +{\mathcal O}(1/\sqrt{M}),
\end{equation}
where by ${\rm dens}$ we denote the density estimator, which can be carried out
 by appropriate binning of the walker ensemble. The function
is therefore represented only statistically so that the values are known up to the statistical uncertainty term
as indicated 
 above. If we substitute the ensemble of delta functions
 into the integral Eq.(\ref{eq:timeevol2}), we obtain
 Gaussians of width $\tau$ centered at the locations $\{{\bf R}_m\}$. In order to restore the 
walker representation back we then {\em sample} the Gaussians, i.e., each walker evolves to a new position as 
given by
\begin{equation}
{\bf R}_m(\tau+\Delta\tau) = {\bf R}_m(\tau) +\sqrt{\Delta\tau}\chi, 
\end{equation}
   where $\chi$ is a $3N$-dimensional vector of independent
 random numbers with Gaussian distribution and unit variance. The process is iterated until the long-time limit
$\tau\to \infty$ is reached.

Considering now more interesting Hamiltonian with external potential and interaction terms
represented by $V$, one can find the following 
short-time approximation~\cite{reynolds82} of
the  Green's function using the Trotter-Suzuki formula as
\begin{align}\label{eq:GFshortsimple}
G({\bf R}\leftarrow {\bf R}',\tau)&=
\langle {\bf R}| \,{\rm e}^{-\tau(T+V-E_T)}|{\bf R}' \rangle  \\ \nonumber
&\approx\,{\rm e}^{-\tau(V({\bf R})-E_T)/2}\langle {\bf R}| \,{\rm e}^{-\tau T}|{\bf R}'\rangle
 \,{\rm e}^{-\tau(V({\bf R}')-E_T)/2}\\ \nonumber
&\approx\, (2\pi\tau)^{-3N/2} \exp\left[-\frac{({\bf R}-{\bf R}')^2}{2\tau}\right]
\exp\left[-\tau\left(\frac{V({\bf R})+V({\bf R}')}{2}-E_T \right)\right].
\end{align}
The approximation is of the order of ${\mathcal O}(\tau^3)$. For sufficiently small time step
we can propagate the walkers even for Hamiltonians with complicated local interactions.
The last term shows that the propagation involves two processes: Gaussian diffusion caused by
the kinetic energy operator, as in the case
of free particles, and reweighting, which originates from the presence of the potential. Note that
the weights will vary very strongly whenever the potential varies. This is true, in particular, close to 
any potential singularities and makes such process very noisy and inefficient. 
In order to make the sampling more efficient we need to smooth out the terms entering 
the weights and to modify the the algorithm.

\subsubsection{DMC Importance Sampling}
Let us introduce {\em importance sampling}~\cite{Grimm,qmchistory1,reynolds82} 
Instead of sampling the wave function $\Psi({\bf R},\tau)$ we will sample a mixed distribution 
$f({\bf R},\tau)=\Psi_T({\bf R})\Psi({\bf R},\tau)$, where $\Psi_T({\bf R})$ is some trial function.
Multiplying the Schr\"odinger equation with $\Psi_T$ and rearranging the terms leads to
the following equation for the mixed distribution $f$ 
\begin{align}\label{eq:dmcevol}
-\frac{\partial f({\bf R},\tau)}{\partial\tau}=& -\frac{1}{2}\nabla^2 f({\bf R},\tau)
+\nabla \cdot [{\bf v}_D({\bf R})f({\bf R},\tau)] \\ \nonumber
&+(E_L({\bf R})-E_T)f({\bf R},\tau),
\end{align}
where ${\bf v}_D({\bf R})={\bf \nabla} \ln|\Psi_T({\bf R})|$ 
represents a drift velocity term and $E_L({\bf R})$ is the local 
energy given by Eq.~(\ref{eq:localenergy}). 
The corresponding short-time approximation of the Green's function is given by
\begin{align}\label{eq:GFshort}
\tilde G({\bf R}\leftarrow {\bf R}',\Delta\tau)
\approx &(2\pi\Delta\tau)^{-3N/2} \exp\left[-\frac{({\bf R}-{\bf R}'-\Delta\tau{\bf v}_D({\bf R}'))^2}
{2\Delta\tau}\right]\\ \nonumber
&\times \exp\left[-\Delta\tau\left(\frac{E_L({\bf R})+E_L({\bf R}')}{2}-E_T\right)\right]\\ \nonumber
&=\tilde G_D({\bf R}\leftarrow {\bf R}',\Delta\tau)\times 
\tilde G_B({\bf R}\leftarrow {\bf R}',\Delta\tau).
\end{align}
Here, we can again interpret the action of the short-time Green's function as a 
diffusion-drift process evolved according to $\tilde G_D$
and the re-weighting process $\tilde G_B$ controlled by the  
average of local energies at ${\bf R}$, ${\bf R}'$ and $E_T$. 
This transformation has several implications. 
First, due to the drift, the density of walkers is enhanced in the regions 
where $\Psi_T({\bf R})$ is large.  
Second, the re-weighting term contains the average of local energies, which for $\Psi_T$ approximating
$\Psi_0$, is close to a constant and therefore much better behaved than possibly unbounded potential $V({\bf R})$. 
By a proper choice and construction of $\Psi_T$ one can eliminate large amount of statistical noise,
potential singularities, etc. For any large-scale calculation accurate $\Psi_T$ is important and improves
the efficiency by two orders of magnitude or more. 

In actual simulations, the walkers are initially distributed according to $f({\bf R},0)=|\Psi_T({\bf R})|^2$
with the weights of all walkers set to unity. The propagation then proceeds by iterative application of the Green's 
function.  
First, the position of each walker evolves according to 
the diffusion-drift $\tilde G_D$ from Eq.~(\ref{eq:GFshort}). However, accuracy of $\tilde G_D$ is not uniform
and decreases in regions
with strongly varying drift. In order to correct for this finite time step bias
 the walker move is accepted with the Metropolis probability
\begin{equation}
A({\bf R}\leftarrow{\bf R}')=\min\left(1,\frac{\tilde G_D({\bf R}'
\leftarrow{\bf R})|\Psi_T({\bf R}')|^2}{\tilde G_D({\bf R}\leftarrow {\bf R}')|\Psi_T({\bf R})|^2}\right).
\end{equation} 
The time step $\Delta\tau$ is chosen to be small so that
the acceptance in the DMC method is high, typically 99\% or so. The Metropolis step improves the Green's
function approximation appreciably and decreases the time step bias due to the fact that it enforces the detailed 
balance condition which is fulfilled by the exact Green's function.

After evolving the walker position we need to update the walker weight
according to 
\begin{equation}\label{eq:reweiht}
w_m(\tau+\Delta\tau)=\tilde G_B({\bf R}_m\leftarrow {\bf R}'_m,\Delta\tau)\,w_m(\tau).
\end{equation}
Note that over the course of the simulation some weights  will grow very rapidly due to the exponential nature
of the weight product. As a result, in the long-time limit, the weight of a single walker will exponentially
dominate the rest.  It is therefore necessary 
to introduce importance sampling also for the weights and adjust the walker population to improve the efficiency
of the reweighting.
One possible way how to accomplish this is stochastic {\em branching} and {\em elimination} of walkers.
Consider a  walker evolving from $ {\bf R}'_m$ to $ {\bf R}_m$ according to
$G_D({\bf R}_m\leftarrow {\bf R}'_m)$. The weight for the step becomes  
$w_m=G_B({\bf R}_m\leftarrow {\bf R}'_m)$.
 We can now calculate           
 $n_m={\rm int}[\eta +w]$, where $\eta$ is a uniform random number from the interval $[0,1]$, and 
${\rm int}$ is the integer part. The resulting integer $n_m$ is the {\em the number of walker copies} placed at the
position ${\bf R}_m$ and subsequently evolved as independent walker(s). In case $n_m=0$, the walker
is eliminated from the ensemble. In this manner, the
branching/elimination step enables to keep all the weights $w_m=1, m=1,...,M$ throughout the
simulation.  More elaborated approaches for sampling the weights
can be found elsewhere~\cite{unr}.
%

After an equilibration period, we can start to collect the statistics needed for the 
calculation of the projected ground state energy
\begin{align}
E_{DMC}&=\lim_{\tau \to \infty}\frac{\int \Psi^*({\bf R},\tau){\mathcal H}\Psi_T({\bf R})\,{\rm d}{\bf R}}{\int \Psi^*({\bf R},\tau)\Psi_T({\bf R})\,{\rm d}{\bf R}} 
\\ \nonumber
&=\lim_{\tau \to \infty} \frac{\int f({\bf R},\tau) E_L({\bf R})\,{\rm d}{\bf R}}{\int  f({\bf R},\tau)\,{\rm d}{\bf R}}\\ \nonumber
&=\ \frac{1}{M} \sum_m  E_L({\bf R}_m) +{\mathcal O}(1/\sqrt{M})
\end{align} 
or other quantities.

\subsubsection{Fixed-Node Approximation}\label{subsec:qmc:fixednode}
For simplicity, we assume that the Hamiltonian has the time reversal symmetry (i.e., no magnetic field,
induced currents, etc) so that the states can be chosen real.
For bosonic systems, the ground state wave function 
has no nodes and is positive everywhere. However, 
the fermionic wave function is by definition antisymmetric, 
and therefore can no longer be represented 
by the walker distribution, which has to be non-negative everywhere. 
Clearly, for fermions, the wave function will have both positive and negative amplitude domains
defined by the nodes---subsets of the configuration space 
where the wave function vanishes, i.e., $\Psi({\bf R})=0$.  (The nodal properties of wave functions will
be studied in detail in the subsequent section.)

\begin{figure}[t]
\begin{center}
\includegraphics[width=0.80\columnwidth]{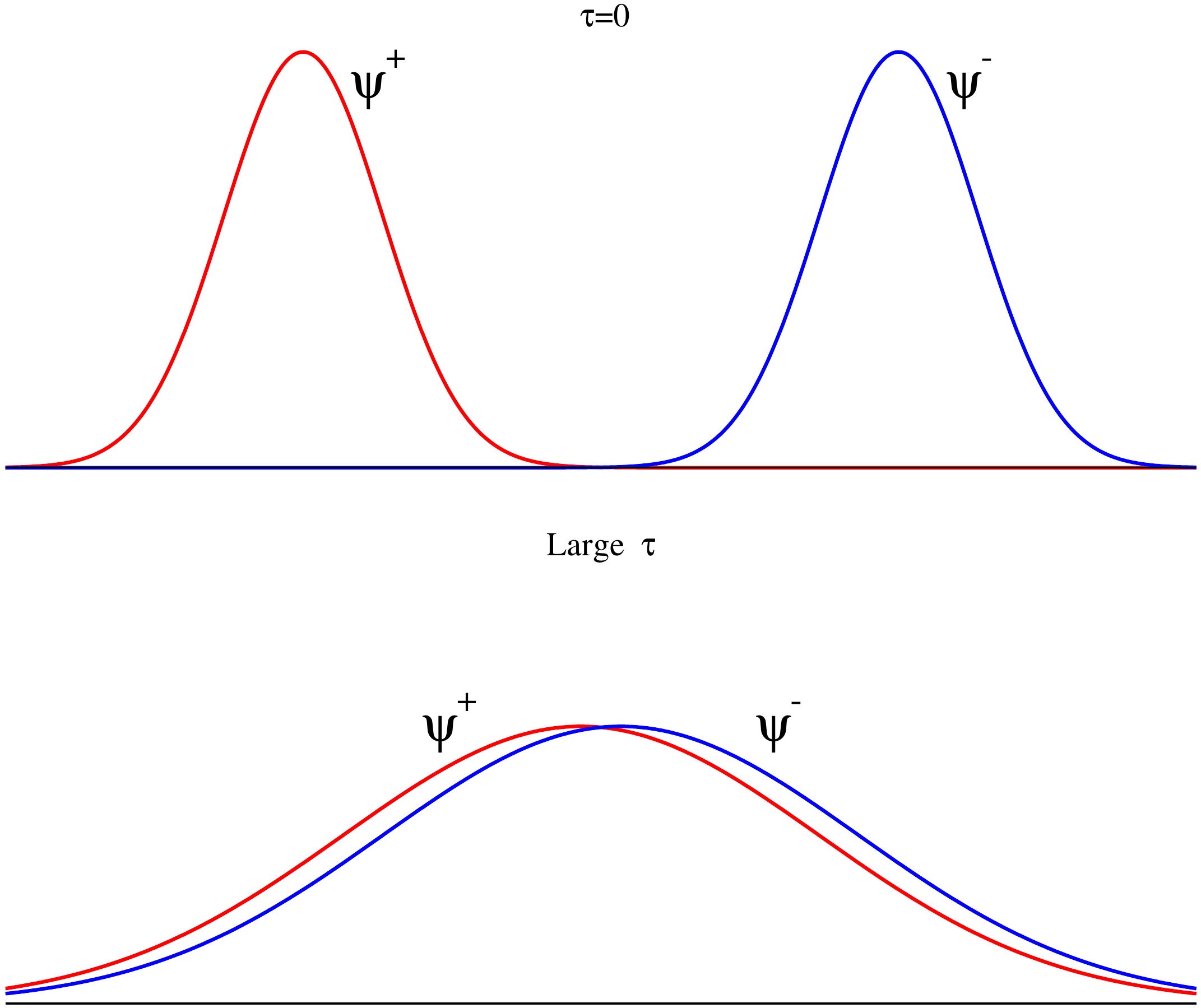}
\caption{Imaginary time behavior of the walker distributions $\Psi^+$ (red) and $\Psi^-$ (blue). Upper figure: at the beginning 
of the propagation ($\tau=0$).
Lower figure: after some imaginary time evolution. The fermionic signal 
decays as $(\Psi^+ - \Psi^-) \propto \exp[-\tau(E_0^F-E_0^B)]$, where $E_0^F-E_0^B$ is the
difference  of fermionic and bosonic ground state energies.}
\label{fig:sign_problem}
\end{center}
\end{figure}

We could try to overcome this problem by decomposing $\Psi({\bf R},\tau)$ into 
positive functions $\Psi^+({\bf R},\tau)$ and $\Psi^-({\bf R},\tau)$ and 
assign each walker a positive or a negative weight accordingly
\begin{align}
\Psi({\bf R},\tau)=\Psi^+({\bf R},\tau)-\Psi^-({\bf R},\tau).
\end{align}
However, since the Schr\"odinger equation is linear, both $\Psi^+({\bf R},\tau)$ and $\Psi^-({\bf R},\tau)$ 
independently converge to the same bosonic distribution, hence making the fermionic component to decay
in time (see Fig.~\ref{fig:sign_problem}).

\begin{figure}[!h]
\begin{minipage}{\columnwidth}
\centering
\includegraphics[width=\columnwidth]{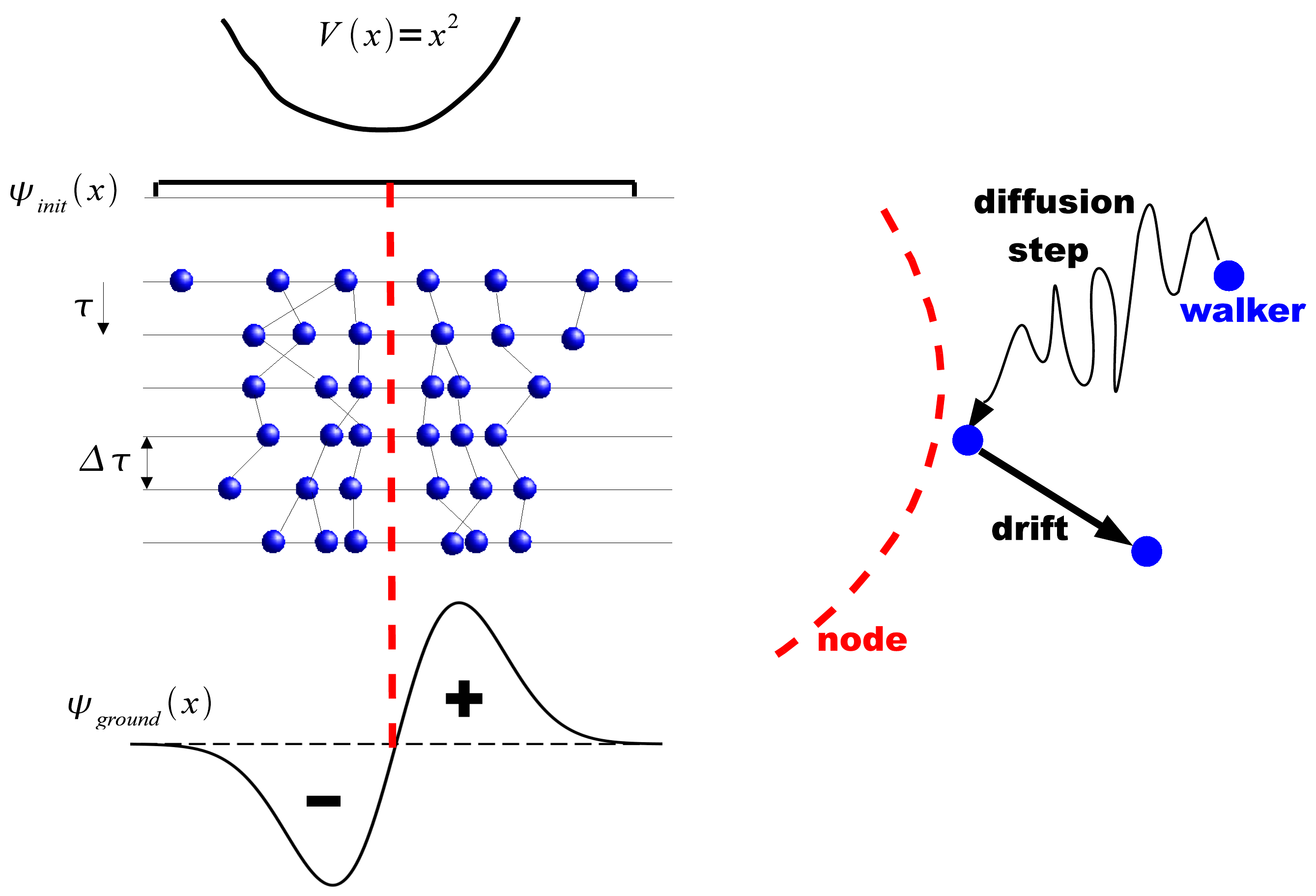}
\end{minipage}
\caption{
On the left is a
 sketch of the fixed-node diffusion Monte Carlo method. The first excited state of
1D harmonic oscillator  serves as a fermionic ground state toy model.
Walkers are generated from a constant
initial distribution and are divided into ``plus'' and ``minus'' populations according
to the trial function (in 1D the node is a point). 
Enforcing the fixed node condition implies that the two populations never mix.
The plot on the right shows walker 
diffusion and drift steps close to the node which is depicted by the dashed line.
}
\label{fig:diffusiondrift}
\end{figure}

This is a well-known {\em fermion sign problem} where the fermionic ``signal'' disappears exponentially quickly
in the projection time. The method is possible to use providing one can estimate quantities 
of interest before the efficiency deteriorates, nevertheless, for large systems the method becomes 
very inefficient.
The inefficiency can be eliminated at the price of introducing  
the so-called  {\em fixed-node approximation}~\cite{jbanderson75,jbanderson76,moskowitz82,reynolds82}, 
which restricts the evolution of positive and negative 
walkers within the separate regions defined by the sign domains of the trial wave function $\Psi_T({\bf R})$.
This is achieved by enforcing the mixed distribution $f({\bf R},\tau)=\Psi_T({\bf R})\Psi({\bf R},\tau)$ 
to be non-negative everywhere
\begin{align}\label{eq:fixednode}
f({\bf R},\tau)\ge 0.
\end{align} 
In other words, $\Psi({\bf R},\tau)$  shares the same zeros and sign structure as 
$\Psi_T({\bf R})$ for all $\tau$, so that Eq.~(\ref{eq:fixednode}) holds.  
This is an important conceptual change in the sampling process:
the fixed-node approximation replaces
the (nonlocal) antisymmetric condition by the local boundary 
condition in Eq.~(\ref{eq:fixednode}). 
Note that the drift term diverges at the node $\Psi_T({\bf R})=0$ and therefore 
repels the walkers away from the nodal regions. 
The fixed-node approximation has several appealing properties. 
The change of symmetries into boundaries effectively {\em bosonizes} the fermionic problem and enables
us to apply the statistical machinery to solve the evolution equation.
The sketch of the fixed-node DMC in Fig.~\ref{fig:diffusiondrift} reminds us Feynman path intergrals
and  the walker paths really are closely realated to quantum paths except that they happen
in imaginary instead of real time.

The fixed-node solution provides the lowest possible energy within the given
boundary conditions and as such is a rigorous upper bound to the true ground
state energy~\cite{reynolds82,moskowitz82}.
This result also implies that resulting fixed-node energy bias is proportional to the square of the nodal displacement error
and in actual applications the method has proved to be very successful as further elaborated later.

The basic form of the fixed-node DMC method is conceptually rather straightforward. Let us therefore sketch a simple version 
of the algorithm. We assume that $\Psi_T({\bf R})$ approximates
the desired ground state and conforms to the required symmetries. The algorithm is as follows.

\begin{itemize}

\item[(1)]
 Generate a set of walkers $\{ {\bf R}_m \}_{m=1}^M $ distributed according to $\Psi_T^2$,  for example, by
using the VMC method.

\item[(2)] Evaluate the drift ${\bf v}({\bf R})=\nabla_{\bf R} \ln|\Psi_T({\bf R})|$  for each walker.

\item[(3)] Evolve each walker from its position ${\bf R}'$ to a new position  ${\bf R}$  according to 
\begin{align}
{\bf R} \leftarrow {\bf R}' + \Delta\tau {\bf v}({\bf R}) + \chi,
\end{align}
where $\chi$ is a 3N-dimensional vector with components drawn from the normal distribution with zero mean 
and variance $\Delta\tau$.

\item[(4)] Accept the move with the probability
\begin{align}
P_{acc}({\bf R}\leftarrow {\bf R}')= {\rm min} \left[ 1,\; 
\frac{G_D({\bf R}\leftarrow{\bf R}') \Psi_T^2({\bf R})}
{G_D({\bf R}'\leftarrow{\bf R})\Psi_T^2({\bf R}')}\delta_{ss'}
\right],
\end{align}
where $s={\rm sign} [\Psi_T({\bf R})]$ and  $s'={\rm sign} [\Psi_T({\bf R}')]$.

\item[(5)] For each walker calculate the number of copies positioned at ${\bf R}$
as 
\begin{align}
n({\bf R})=  {\rm int}\{\eta+\exp[-\tau(E_L({\bf R}')+E_L({\bf R})-2E_T)/2)]\}, 
\end{align}
where $\eta$ is drawn from a uniform random distribution on the interval 
$[0,1]$.

\item[(6)] Evaluate averages of interest over the ensemble of walkers and update
$E_T$ as 
\begin{align}
E_T(\tau+\Delta\tau) = E_T(\tau) -C_{adj} \ln(M_{act}/M_{ave}),
\end{align}
where $M_{act}$ is the actual walker population,  $M_{ave}$ is the desired
walker population and $C_{adj}$ is a positive constant, which tunes how
quickly the population would reach the desired value. The constant is typically
chosen to rebalance the population to the desired number in a few steps, say,
for 10 steps its value is $\approx 10/\tau$. 

\item[(7)] Repeat the steps (2) to (6) until the error bars of the calculated quantities
reach the desired values. 
\end{itemize}

Note that in the step 4) we correct the short-time Green's function approximation
 in two ways. Besides the acceptance step explained previously,
the fixed node approximation is enforced for the case when a walker violates the 
nodal boundary.
This occasionally happens, again, due to the Green's function inaccuracy.
Since the evolution is discretized,
the inherent fluctuations in the diffusion step can cause the walker to cross
the nodal boundary---although for the exact
Green's function the probability of such an event vanishes.
As mentioned above, the time step $\Delta\tau$ is usually 
chosen to be small enough so that the acceptance is 99\% or higher. 
Any remaining time step bias can be eliminated by extrapolating  $\Delta\tau\to 0$
from repeated simulations with decreasing $\Delta\tau$.

In order to clarify the fixed-node DMC scaling of the computational time in the system size  
we invoke the previously given formula Eq.(\ref{eq:timedem}). We assume a requirement
of the fixed error bar $\epsilon$ for the total energy. The growth of variance 
$\sigma^2$ is basically proportional to the number of electrons $N$, since for 
large systems fluctuations from different electrons will be approximately independent,
so that $\sigma^2 \propto N$. The time $T_{sample}$ then includes evaluations of  the wave function
and operators---such as the kinetic energy, etc---and we can write for $T_{sample}=c_2 N^2+c_3 N^3$.
The quadratic component comes from the pair terms in the Jastrow factor and also from
filling the Slater matrices, while the cubic part is from evaluating the determinant.
For systems with up to a few thousand electrons $c_2>>c_3$ so that the quadratic term dominates.
There is also some cost associated with an increase in statistical correlation of samples which we 
estimate as being of the order of $N$. The overall scaling then looks like $N^4$. However,
this appears to be a rather conservative estimate. 
For larger systems one can recast the one-particle
orbitals into the Wannier orbitals which are localized and  also impose
sufficiently loose cut-off on the range 
of the Jastrow correlations. That makes the Slater matrix sparse
\cite{williamson01} and basically eliminates one power of $N$. If we are interested
in intensive quantities such as cohesion (energy per particle) that eliminates another 
factor of $N$. For some energy differences correlated sampling shows similar improvements
and so forth. The algorithms and implementation has a significant impact here and in practice
the fixed-node DMC calculations scale approximately as $N^p$ where $2\le p\le 3$.

\subsubsection{Beyond The Fixed-Node Approximation}
The fixed-node approximation was has been an important step on the way to overcome the 
fundamental difficulties with fermion signs. 
Besides improving the nodes with more accurate
trial wave functions (see Sec.~\ref{ch:pfaffians} and~\ref{ch:bf}) 
or attempts to understand them (see Sec.~\ref{ch:nodes}), several ideas how to reach beyond the limits
of  fixed-node approximation were proposed and tested.

Ceperley and Alder~\cite{ceperley_node_release} developed the {\em released-node method}.
As the name suggests, the walkers are allowed to sample the regions of opposite sign
by using the guiding distribution, which is positive everywhere including the nodal regions.
The method suffers from the fermion sign problem but enables to somewhat better control
the loss of the fermion signal by appropriate choice of the guiding distribution and 
other tune-ups.  The algorithm produces transient estimators 
so that the decay of fermionic excitations 
 has to be faster then the decay of the fermionic ground state component.  
Let us explain the basic idea of the release node approaches. 
The fixed-node condition is eliminated and instead of $\Psi_T$ the 
importance sampling is carried out with another
 importance distribution $\Psi_G$ which is sometimes
called the {\em guiding} wave function. Unlike $\Psi_T$, which is antisymmetric, $\Psi_G$ is
{\em symmetric} and positive in the whole configurations space, 
allowing thus the walkers to reach any point.
The DMC simulation then runs according to the importance sampled     
 Eq.(\ref{eq:dmcevol}), where $\Psi_T$ is replaced by $\Psi_G$.  The resulting release
node $\Psi_{RN}$ solution is found from $f(R,\tau)=\Psi_{RN}\Psi_G$. 
The release node solution is transient and can be       
qualitatively expressed as          
\begin{equation}\label{eq:relnodes}
\Psi_{RN} \approx c(\tau)\Psi_{F}(\tau) + (1-c(\tau))\Psi_B(\tau),
\end{equation}
where $c(\tau) \approx \exp[-(E_F-E_B)\tau]$ and indices $F$ and $B$ correspond to 
fermions and bosons,
respectively. The energy of the fermionic component can be evaluated from
\begin{align}\label{eq:relen}
E_{RN}(\tau)&=
\frac{\int \Psi_G\Psi_{RN}
\frac{\Psi_T}{\Psi_G}
\frac{{\mathcal H}\Psi_T}{\Psi_T}
{\rm d}{\bf R}}
{\int \Psi_G\Psi_{RN}\frac{\Psi_T}{\Psi_G}{\rm d}{\bf R}}
\\ \nonumber
&= \frac{\int f({\bf R},\tau) \frac{\Psi_T}{\Psi_G}  E_L({\bf R})\,{\rm d}{\bf R}}
{\int  f({\bf R},\tau)\frac{\Psi_T}{\Psi_G}\,{\rm d}{\bf R}}\\ \nonumber
&=\ \frac{\sum_m s_m 
\frac{\Psi_T({\bf R}_m)}{\Psi_G({\bf R}_m)} 
E_L({\bf R}_m)}
{\sum_m s_m\frac{\Psi_T({\bf R}_m)}{\Psi_G({\bf R}_m)}} 
+ {\mathcal O}(e^{\alpha\tau}/\sqrt{M}),
\end{align}
where $\alpha \propto (E_F-E_B)$.
Each weight $s_m$ carries a {\em sign} which is assigned
initially as $w_m(\tau=0) ={\rm sign} [\Psi_T({\bf R}_m)]$ and kept constant during the simulation. 
A simple guiding function can be written as $\Psi_G=\sqrt{\Psi_T^2 +\alpha}$, where the
parameter $\alpha$ controls the average rate of walkers passing into the
opposite sign region. More sophisticated forms are necessary for inhomogeneous systems.
The transient sampling efficiency can be measured by the ratio
\begin{equation}\label{eq:relnodes}
\kappa = 
\frac{1}{M}
\sum_m s_m
\frac{\Psi_T({\bf R}_m)}{{\Psi_G}({\bf R}_m)}
\end{equation}
which, ideally, should be close to unity (as is the case for the fixed-node algorithm
when $\Psi_G=|\Psi_T|$). 
As the simulation proceeds, the ratio decreases
and the fermionic signal decays while the statistical noise grows.
Despite this, the method can be useful providing one can reach the fermionic ground state estimator    
before the error bars become too large. 
 In fact, the release node method was used
for the famous calculations of the homogeneous electron gas correlation energies \cite{ceperley80}
 and for small atoms and molecules later \cite{ceperley_node_release}.
Note that the simplicity of the homogeneous electron gas enabled to apply
this method to hundreds of electrons already in 1980. Unfortunately, applications to 
inhomogeneous systems are much more difficult and have progressed at much slower pace. 

Another idea advanced by Kalos and collaborators   
is based on the {\em cancellation of paired walkers} with opposite signs~\cite{kalos_pairing1}.   
This method has been subsequently improved to work with fully interacting ensemble of walkers~\cite{kalos_pairing2}.
The method delays the fermionic decay and enhances the fermion
solution component, however, the current understanding is that the scaling appears to be exponential.
An important improvement is, however, that the exponent is smaller than the difference between
the bosonic and fermionic energies and therefore the technique has a potential for efficiency gains
in some applications.
\cite{assaraf}.

Anderson and Traynor~\cite{anderson91} combined some of the ideas in the fixed-node, released-node and 
cancellation methods in an algorithm which employed improvements such as relocation after the node crossing,
 self and multiple cancellations, and maximum use of symmetry.
Among the systems of interest, the authors calculated 
the excited state of H$_2$ $^3\Sigma_u^+$ and the barrier height for the simplest possible 
chemical reaction reaction ${\rm H}+{\rm H}_2\to{\rm H}_2+{\rm H}$~\cite{anderson92}. 
The accuracy of this essentially exact calculation was exceptional
with resulting statistical uncertainty of only
0.01 kcal/mol.

Among other alternative attempts, let us also mention the 
method of {\em adaptive nodes} (see, e.g., Ref.~\cite{adaptive_nodes}).
The key idea is based on representing the approximate nodal function directly by the walker ensemble  
(e.g., by positioning a Gaussian on each walker). This results in the adaptive description of the nodes
that does not depend upon {\it a priori} knowledge of the wave function. The main difficulty 
of the method, however, comes from scaling to larger systems
 since one basically has to populate the whole configurations space with walkers.   

In simulations of quantum problems
the fermion sign problem is pervasive. It appears in various incarnations and determines 
computational complexity of the underlying problem.  
It has been recently shown~\cite{troyer_nphard} explicitly, that QMC simulations of
random coupling Ising model with fermion sign problem belongs to the class
of {\em NP-hard} problems, i.e., with non-polynomial computational complexity.              
There are exceptions which enable to eliminate the fermion
sign problem exactly such as separable Hamiltonians or problems with certain symmetries
(the half-filled Hubbard model, for example). However, all such known cases are 
``special'' having some underlying symmetries which enable to eliminate the fermion signs.
It is therefore possible that any {\em exact} QMC approach with broad applicability 
belongs to the class of NP-hard problems and therefore finding such method might be an 
elusive and exceedingly challenging,
if not impossible goal. The next best goal should be therefore based
on a more pragmatic view on what does it mean to ``solve'' a given many-body quantum problem.   
Perhaps the right question to ask can be formulated as follows:
 What is the most optimal scaling of a method which can calculate
expectations within a prescribed error?
 (say, energy differences with errors below 0.1 eV, 0.01 eV and so forth). 
Here by the error we mean the total error
which contains both {\em  systematic biases} from approximations {\em and statistical uncertainties}.
The answer to this question is a lot easier, at least in a qualitative sense, since 
already at present the electronic structure QMC methods can achieve
accuracies of 0.1-0.2 eV for energy differences with low-order polynomial scaling in the system size. 
Similar progress has been achieved also in simulations in other areas of physics.

%
   
\subsection{Trial Variational Wave Functions}\label{sec:twf}
The great advantage of QMC methods is the capability 
to use explicitly correlated trial wave functions.  
Their quality is the most important factor which influences both the achieved accuracy and 
efficiency. On the other hand,  repeated evaluations of $\Psi_T$ (and $\nabla \Psi_T$, $\nabla^2 \Psi_T$, etc) 
is  the most costly part of QMC calculation and therefore restrict possible alternatives for $\Psi_T$ to forms 
which can be evaluated rapidly and scale with the system size.

\subsubsection{Basic Properties}

$\Psi_T$ has to be antisymmetric with respect
to exchange of both spin and spatial coordinates
\begin{align}
\Psi_T(P{\bf R},P{\boldsymbol \Sigma})=(-1)^P \Psi_T({\bf R},{\boldsymbol \Sigma}),
\end{align}
where ${\boldsymbol \Sigma}=(\sigma_1,\ldots,\sigma_N)$ are discrete spin variables with values $\pm \frac{1}{2}$
and $P$ is an arbitrary permutation with the sign equal to $(-1)^P$. Assuming that the 
Hamiltonian (or other quantities of interest) does not involve spin-dependent terms, 
any expectation value requires only the spatial antisymmetry since the summation
over spin variables becomes just a permutation of indices inside the same spatial
integral. That enables to assign the spins to particular electrons and requires
imposing antisymmetry only with respect to spatial interchanges between the
electrons of the same spin. 
The spins of all electrons are therefore fixed to the total spin projection
equal to $\frac{1}{2}(N_\uparrow-N_\downarrow)$.
The first $N_{\uparrow}$ particles are labeled as spin-up and the rest  $N-N_\downarrow$
as spin-down. Throughout this paper we assume that
the spin variables are factored out by this labeling scheme.
Using an appropriate expansion in determinants the wave function can be made
an eigenstate of the total spin $S^2$ and possibly other integrals of motion such as 
the total angular momentum $L^2$ and $L_z$ in atoms.

We demand that $\Psi_T$ approximates a given bound state so that  
$\int \Psi_T^* \Psi_T $, $\int \Psi_T^* {\mathcal H} \Psi_T$
and $\int \Psi_T^* {\mathcal H}^2 \Psi_T$ do exist. 
The last condition is necessary for having a finite variance. 
Further, $\Psi_T$ and $\nabla \Psi_T$ have to be continuous, wherever the potential is finite. 
In addition, the continuity of the potential implies also the continuity of $\nabla^2 \Psi_T$.
Important corollary is that as $\Psi_T$ approaches 
an exact solution, the local energy becomes constant everywhere. We can express this also  
as $ \sigma ^2 =\langle ({\mathcal H} -\langle {\mathcal H} \rangle)^2\rangle \to 0$ for $\Psi_T \to \Phi_0$, 
and it is often called as {\em zero variance property}. Note that both VMC and DMC estimators have the zero variance property.
(This is not automatically true for many QMC approaches, in particular, this 
does not apply whenever there are non-localities in the Hamiltonian.
For example, many quantum lattice models such as the Hubbard model are nonlocal 
due to hopping terms.) 

We wish that the local energy is close to the eigenvalue
and also that the variance as small as possible. Therefore we demand
that any singularity from 
the potential is to be canceled by the opposite sign singularity in the kinetic energy, what 
gives rise to Kato {\em cusp conditions}~\cite{Kato57,Pack66}.
As the distance of electron to nucleus vanishes, $r_{Ii} \to 0$, the potential energy divergence  
cancels out when
\begin{align}\label{eq:necusp}
{\frac{1}{\Psi_T}{\frac{\partial \Psi_T}{\partial r_{Ii}} }\bigg\arrowvert}_{r_{Ii}=0}=-Z_I.
\end{align}
This electron-nucleus cusp condition [Eq.~(\ref{eq:necusp})] can be satisfied by a proper 
choice of orbitals, removed by pseudopotentials (see Sec.~\ref{sec:pseudo}) or enforced by Jastrow factor. 
Of course, the hydrogenic solutions 
such as, for example, 1s and 2p orbitals given as
$\phi_{1s}(r)=\exp(-Z_Ir)$ and  $\phi_{2p}(r)=\exp(-Z_Ir/2)Y_{1m}$,   
have the correct cusp. In addition, it is clear that any linear combination of orbitals
 with the correct cusp has the same property. In particular, if all orbitals in a Slater determinant 
have correct cusps then the same is true for the determinant what is easy to verify by
an expansion in cofactors.

Similarly, as the distance $r_{ij}$ of any two electrons vanishes, there is a divergence in the 
electron-electron Coulomb potential. 
Let us first assume that electrons $i$ and $j$ have unlike spins. In general, the wave function 
does not vanish at such a collision point. Close to the singularity we can write it as an 
 s-wave in  $r_{ij}$ times a smooth function which depends on the rest of the 
variables $\tilde{\bf R}$, i.e., 
$lim_{r_{ij}\to 0} \Psi({\bf R})=exp(cr_{ij})F(\tilde{\bf R})$. Since there are 
two contributing Laplacians $\Delta_i+\Delta_j$ we get 
\begin{align}\label{eq:cupsupup}
{\frac{1}{\Psi({\bf R})}{\frac{\partial \Psi({\bf R})}{\partial r_{ij}} }
\bigg\arrowvert}_{r_{ij}=0}=c=\frac{1}{2}.
\end{align}

For two like spins electrons with $r_{ij}\to 0$ the wave function vanishes.
The behavior around the collision point 
has a character of a p-wave. Let us transform the coordinates ${\bf r}_i$ and ${{\bf r}_j}$
to ${\bf r}_{ij}={\bf r}_i-{\bf r}_j$ and ${\bf r}_{ij}^+={\bf r}_i+{\bf r}_j$.  
The wave function can be then written as
$lim_{ r_{ij} \to 0} \Psi({\bf R})= f({\bf r}_{ij}) \exp(cr_{ij}) F(\tilde{\bf R})$. 
By an appropriate coordinate rotation the linear behavior 
around $r_{ij}\to 0$ can be expressed as
 $f({\bf r}_{ij})=(x_i-x_j) + {\mathcal O}((x_i-x_j)^2)$. 
The like spins cusp condition therefore modifies to
\begin{align}\label{eq:cupsupdown}
{\frac{1}{\Psi}{\frac{\partial \Psi}{\partial r_{ij}} }\bigg\arrowvert}_{r_{ij}=0}=c=\frac{1}{4}.
\end{align}

\subsubsection{Trial Wave Functions Forms}\label{subsec:twf}
It is convenient to express $\Psi_T({\bf R})$ as a
product of an antisymmetric part $\Psi_A$ and an exponential of a symmetric
correlation function $U_{corr}({\bf R})$, often called the Jastrow factor, as given by 
\begin{align}
\Psi_T({\bf R})=\Psi_A \times \exp[U_{corr}({\bf R})].
\end{align}

Let us describe the forms for $\Psi_A({\bf R})$, $U_{corr}({\bf R})$ and 
their generalizations as commonly used in calculations (and also implemented 
in our QMC code QWalk~\cite{qwalk}). The 
Jastrow part~\cite{jastrow,reynolds82,moskowitz,cyrus1} explicitly depends on electron-nucleus and 
electron-electron distances 
\begin{align}
U_{corr}(\{r_{ij}\},\{r_{iI}\},\{r_{jI}\})=\sum_{iI} \chi(r_{iI})+ \sum_{i\neq j} u(r_{ij}) 
+ \sum_{i\neq j,I} w(r_{ij},r_{iI},r_{jI}) 
\end{align}
and is written as the sum of one-body (electron-nucleus), two-body (electron-electron) and three-body (electron-electron-nucleus) terms.
These correlation functions describe several effects. First, the form captures
 the electron-electron cusp conditions [Eqs.~(\ref{eq:cupsupup}) and~(\ref{eq:cupsupdown})]. 
As derived above, these are satisfied providing 
\begin{align}
{{\frac{\partial u_{\sigma_i \sigma_j}}{\partial r_{ij}} }\bigg\arrowvert}_{r_{ij}=0}=
\left\{
\begin{array}{cc}
1/4 & \sigma_i=\sigma_j,\\
1/2 & \sigma_i=-\sigma_j,
\end{array} \right.
\end{align}
and while all the remaining functions have zero derivatives for $r_{ij}=0$. Second, the 
terms which depend 
on electron-ion distances enable to vary the electron density which needs to 
be adjusted once
the electron-electron terms are included. Third, the triple electron-electron-ion correlations 
enable to decrease the correlation effects and decrease the local energy fluctuations.
Expanded in the basis of one dimensional functions the components of the Jastrow factor take form
\begin{align}\label{eq:jas2}
\chi(r)&=\sum_k c_k^{en} a_k(r),  \\
u(r)&=f_{cusp}(r) + \sum_k c_k^{ee} b_k(r), \\
w(r_{ij},r_{iI},r_{jI})&=\sum_{klm} c_{klm}^{een} [a_k(r_{iI})a_l(r_{jI})+a_k(r_{jI})a_l(r_{iI})]b_m(r_{ij}).
\end{align}
The above Jastrow factor has proved to be very efficient 
in describing significant portion of correlation effects with small number of variational 
parameters. More details about the 
implemented basis functions can be found below or 
in the QWalk documentation~\cite{qwalk}.

The simplest antisymmetric functional form is the determinant so that one can
expand the wave function in spin-factorized 
Slater determinants as 
\begin{align}
\Psi_A^{Slater}({\bf R})=\sum_k d_k \, {\rm det}[ \ldots, \varphi_{k}^{\uparrow}({\bf r}_i), \ldots ]
{\rm det}[\ldots, \varphi_{l}^{\downarrow}({\bf r}_j),  \ldots ]
\end{align}
where $i=1,\ldots, N_\uparrow$ and $j= N_\uparrow+1, \ldots, N)$.
Each spin-up and spin-down Slater determinant
is a function of one-particle orbitals typically found by (post-)HF or DFT methods. 
In the introductory part we have touched upon some of the 
properties of $\Psi_A^{Slater}$.  Although formally
the Slater determinants provide a complete set of antisymmetric functions, 
 in practice, we are limited to use only finite expansions, very often to a single determinant.  
It is remarkable that the {\em single determinant
Slater-Jastrow} form 
 proved to be very effective for many applications, perhaps beyond expectations. 
It provides surprisingly accurate
 energy differences such as cohesion and binding, excitations, etc, due to partial cancellations of 
the fixed-node errors.

On the other hand, multi-determinant expansions which are used to improve the fixed-node 
energies require rather involved and technically difficult processing such as 
the following tasks:
\begin{itemize}
\item[(a)] Large determinantal expansions often need computationally expensive 
 reoptimizations of the determinantal coefficients. At present, with such wave functions
we are thus bound to small or intermediate sizes of chemical systems. 

\item[(b)] For very high accuracy calculations one-particle orbitals need to be 
reoptimized. Typically, orbitals come from methods which do not incorporate the explicit correlations 
and therefore are not optimal with Jastrow correlation factor included.
However, full optimization of orbitals in many-particle systems is 
still a formidable challenge. What has been successfully tried is to parameterize
the effective one-particle Hamiltonian such as hybrid DFT functionals 
and find optimal orbitals in such restricted parameter space.
\end{itemize}

The above limitations motivate to think about  possibilities
beyond $\Psi^{S-J}_T$. Since the determinant is the simplest antisymmetric form constructed 
from one-particle orbitals, it is interesting to ask 
what is the simplest
antisymmetric object built from two, three or $n-$particle orbitals ?
For two particle orbitals such antisymmetric form is called {\em Pfaffian} and the resulting 
 {\em Pfaffian pairing wave function} is written as
\begin{align}
{\rm pf}_i(1,\ldots, N)&={\mathcal A}[\tilde{\phi}(1,2), \tilde{\phi}(3,4), \ldots,
                            \tilde{\phi}(N-1,N)]\\
&={\rm pf}[\tilde{\phi}(1,2), \tilde{\phi}(3,4), \ldots,
                            \tilde{\phi}(N-1,N)],
\end{align}
where ${\mathcal A}$ is the antisymmetrizing operator and the arguments of the pairing spin-orbital $\tilde{\phi}(i,j)$
denote both spatial and spin variables. Note that in the most basic form the Pfaffian contains only
{\em one} pair orbital unlike the Slater determinant which requires as many independent 
orbitals as is the number of particles.


Similarly to multi-determinantal expansions, we can expand the  wave function  
in a linear combination of Pfaffians
\begin{align}\label{eq:mpf1}
\Psi_A^{PF}({\bf R})=\sum_m w_m \, {\rm pf}_m(\ldots, \tilde{\phi}_m(i,j), \ldots).
\end{align} 
Note that each Pfaffian in the expansion~(\ref{eq:mpf1})
contains different pairing orbital.
The form will be later simplified to spatial antisymmetries only
such as in the case of determinants by employing spin-singlet and spin-triplet pairing orbitals. 
The description of Pfaffians together with applications  
to several systems and possible extensions is discussed in more detail in Sec.~\ref{ch:pfaffians}.

Another way how to boost the variational freedom of trial functions
is based on the so-called backflow coordinates. The idea goes back to Feynman and has 
been employed 
for description of  homogeneous systems~\cite{feynman,schmidt_bf,panoff,moskowitz,kwon1,kwon2,kwon3}.
Feynman suggested that correlations in
quantum liquids can be approximately captured 
by replacing the actual coordinates ${\bf R}$ in $\Psi_A(${\bf R}$)$ by quasi-particle 
coordinates ${\bf X}$ which take into account the influence from the rest of the particles.
As a consequence, the 
nodes and other properties of $\Psi_A({\bf X})$ will be, in general, different from those of $\Psi_A({\bf R})$.
Recently, some progress was also reported in applying this idea to 
 chemical (inhomogeneous) systems~\cite{drummond_bf,rios_bf}. 
In this review we report some of the results we obtained with these new developments.
The quasi-particle coordinate of the $i$-th electron at position ${\bf r}_i$ is given as 
\begin{align}
{\bf x}_i&={\bf r}_i+{\boldsymbol \xi}_i({\bf R}) \nonumber \\
&={\bf r}_i+{\boldsymbol \xi}_i^{en}({\bf R})+{\boldsymbol \xi}_i^{ee}({\bf R})+{\boldsymbol \xi}_i^{een}({\bf R}),
\end{align}
where we have again divided the contributions to one-body (electron-nucleus), two-body (electron-electron) 
and three-body (electron-electron-nucleus) backflow terms.
They can be further expressed as 
\begin{align}
{\boldsymbol \xi}_i^{en}({\bf R})&=\sum_I \chi(r_{iI}) {\bf r}_{iI} \\
{\boldsymbol \xi}_i^{ee}({\bf R})&=\sum_{j\ne i} u(r_{ij}) {\bf r}_{ij} \\
{\boldsymbol \xi}_i^{een}({\bf R})&=\sum_I \sum_{j\ne i} [w_1(r_{ij},r_{iI},r_{jI}) {\bf r}_{ij} + w_2(r_{ij},r_{iI},r_{jI}) {\bf r}_{iI}],
\end{align}
where ${\bf r}_{ij}={\bf r}_i-{\bf r}_j$.  $\chi$, $u$ and $w_1$ with $w_2$  are similar to one, two and three-body Jastrow terms. The calculational 
and implementation details together with some results are further discussed in Sec.~\ref{ch:bf}.

\subsubsection{Basis Functions for the Jastrow Correlation Factor}\label{appendix:functions}

The one-particle orbitals and also the correlations has to be represented in an appropriate
{\em basis} functions. There is a vast amount of literature on basis sets for expressing the one-particle orbitals.
Typically, the localized basis sets are based on Gaussians, while for
periodic systems plane waves are frequently used. Other options exist such as numerical orbitals, 
mixed basis, etc. We have little to add to this technically
involved subject except that QWalk is able to import several types of these. We refer the reader
to the literature or manuals for the actual packages. 

\begin{figure}[!t]
\begin{center}
\includegraphics[width=0.8\columnwidth]{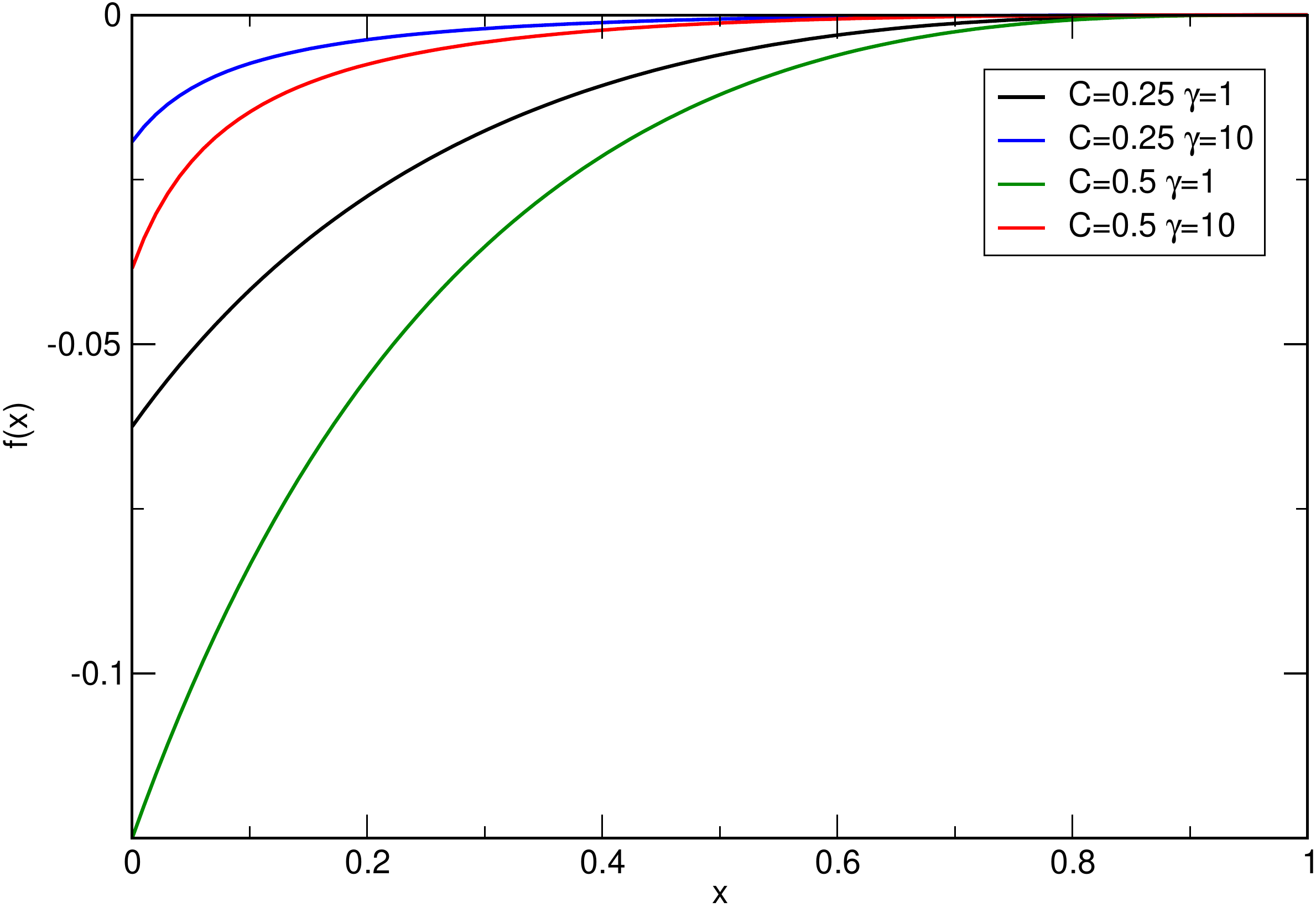}
\end{center}
\caption{Cusp functions with smooth cut-off for two different values of the 
 curvature parameter ($\gamma=1$ and $\gamma=10$) for like ($c=\frac{1}{4}$) and
unlike ($c=\frac{1}{2}$) spins.}
\label{fig:cutofffunc}
\end{figure}

However, we briefly outline the basis which is used for correlations in the Jastrow factor.
The exact electron-electron cusp conditions are satisfied by the choice of following cusp function [see Fig.~(\ref{fig:cutofffunc})]
with variable $x=r/r_{cut}$,  where $r_{cut}$ is some cutoff radius and $\gamma$ is the curvature as
\begin{align}\label{eq:appendix:functions:1}
f_{cusp}(r)= f_{cusp}(x,\gamma)=c\left(\frac{x-x^2+x^3/3}{1+\gamma(x-x^2+x^3/3)}-\frac{1}{\gamma+3}\right)
\end{align}
The cusp constant is $c=\frac{1}{4}$ for electrons with like and $c=\frac{1}{2}$ for electrons with unlike spins.

\begin{figure}[!b]
\begin{center}
\includegraphics[width=0.8\columnwidth]{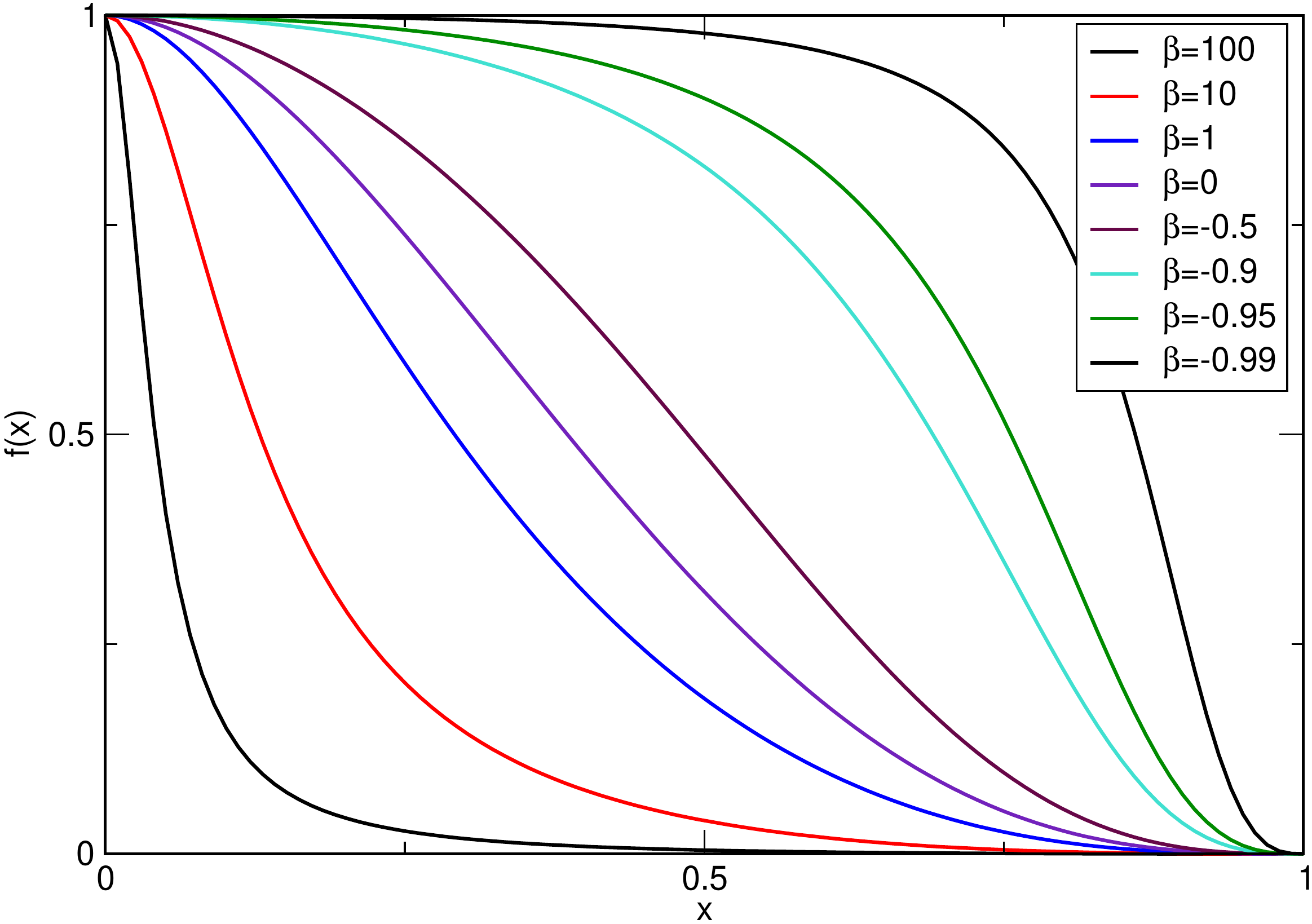}
\end{center}
\caption{Polynomial Pad\'e functions with curvatures $\beta$ ranging from -0.99 to 100.}
\label{fig:bessel}
\end{figure}

Polynomial Pad\'e functions for the same variable $x=r/r_{cut}$ and curvature $\beta$ had proved to
be a convenient 
choice for describing the electron-electron and electron-nucleus correlation. In calculations, we use
the form
\begin{align}
a_k(r), b_k(r) = f_{poly-Pade}(x,\beta)=\frac{1-x^2(6-8x+3x^2)}{1+\beta x^2(6-8x+3x^2)}
\end{align}
The $f_{poly-Pade}(0)=1$ with derivative $f'_{poly-Pade}(0)=0$ 
and also goes smoothly to zero as $r \to r_{cut}$ [see Fig.~(\ref{fig:bessel})].
These conditions are necessary for preserving the 
cusp conditions already fixed by the cusp functions and 
by the choice of orbitals or pseudopotentials.
Polynomial  Pad\'e functions are then used as basis functions in expansions for correlation 
part in the Jastrow factor, for example, for the electron-nucleus part we have 
\begin{align}\label{eq:jas2}
\chi(r)&=\sum_k c_k^{en} a_k(r), 
\end{align}
where $c_k$ is an expansion coefficient and basis functions $ a_k(r)$ with different $k$ differ in
curvatures $\beta_k$. 
Similarly for other components as given by Eq.~(\ref{eq:jas2}).

\subsection{Pseudopotentials}\label{sec:pseudo}
To a high degree of accuracy,
most properties of real systems are
determined by the states in valence electronic sub-shells only.  Atomic cores are also expensive 
in QMC simulations since the computational cost
increases with the atomic number $Z$ as $\approx Z^6$
(see Ref.~\cite{ceperley86,hammond87}) and makes the all-electron calculations
unnecessarily difficult for heavy atoms.  
The presence of core electrons complicates calculations for several reasons.
The short length scales associated with
atomic cores states require decrease in the QMC time step. 
Even more difficult problem is the increase in fluctuations of the local energy as $Z$ gets higher
due strongly varying kinetic and potential terms close
to the nucleus. This implies also basic efficiency issue 
since almost all computational time would go to sampling energy fluctuations in the core
instead of the important valence effects.

In analogy with theories such as Hartree--Fock and DFT, 
in QMC the atomic core states can be eliminated by introducing
an {\em pseudopotential (or effective core potential)} operator which represents
the impact of core on valence states.
The action 
of pseudopotential is typically different for states with different angular momenta. 
It is customary to divide the pseudopotential operator into local (common to all angular momenta) and
nonlocal (different for each angular momentum) parts. For a given ion we can therefore eliminate 
the core states and replace the ionic term as follows 
\begin{equation}
\frac{-Z}{r}\;\; \to \;\; U^{ps}  =  V^{ps}(r) +W^{nloc},
\end{equation}
where $V^{ps}(r)$ is the local part while 
\begin{align}
W^{nloc}=\sum_l^{l_{max}} V_{l}^{ps}(r)\sum_m |lm\rangle \langle lm|
\end{align}
is the nonlocal pseudopotential with projectors over $lm$ states of angular momentum.
The nonlocal portion of the local energy $E_L({\bf R})$ from
the electron $i$ in the field of the ion $I$  is given by  
\begin{align}\label{eq:ppint}
\frac{W^{nloc}\Psi_T({\bf R})}{\Psi_T({\bf R})}=&\sum_l V^{ps}_l( r_{iI}) \frac{2l+1}{4\pi} \int_{4\pi} P_l[\cos(\theta'_{iI})]
\times \frac{\Psi_T({\bf r}_1,\ldots,{\bf r}_i',\ldots,{\bf r}_N)}{\Psi_T({\bf r}_1,\ldots,{\bf r}_i,\ldots,{\bf r}_N)} {\rm d}\Omega'_{iI},
\end{align}
where $P_l$ denotes the Legendre polynomial of degree $l$.
The angular integral above is over
the surface of the sphere passing through the $i$-th electron and centered on the ion.
The integral is evaluated numerically using 
Gaussian spherical quadrature grids. 
When the orientation of axes of a given quadrature are chosen randomly,  
the result is an unbiased Monte Carlo estimation of the given integral (\ref{eq:ppint}). 
For more details, see original Ref.~\cite{lubos_ps}.

In the DMC method, the application of the nonlocal operator $W^{nloc}$ causes the matrix element 
$\langle {\bf R}| \,\exp[-\Delta\tau W^{nloc}]|{\bf R}'\rangle$, and therefore also the 
propagator, to have a fluctuating sign. This would create its own fermion sign problem
even for a single electron. 
Consequently, the imaginary time Schr\" odinger equation  for the 
 function $f({\bf R},\tau)=\Psi_T({\bf R})\Psi({\bf R},\tau)$  is rearranged as given by
\begin{align}\label{eq:locapprox}
-\frac{\partial f({\bf R},\tau)}{\partial\tau }=& -\frac{1}{2}\nabla^2 f({\bf R},\tau)
+\nabla \cdot [{\bf v}_D({\bf R})f({\bf R},\tau)] +[E_L({\bf R})-E_T]f({\bf R},\tau) \nonumber \\
&+\Bigg\{ \frac{V_{nloc} \Psi_T({\bf R})}{\Psi_T({\bf R})} 
-\frac{V_{nloc} \Psi({\bf R},\tau)}{\Psi({\bf R},\tau)} \Bigg\}f({\bf R},\tau),
\end{align}
where $V_{nloc}$ denotes the sum of all nonlocal pseudopotential components  from all ions. 
Note that, unfortunately,
we do not know how to calculate the last term since it contains the solution which
is a priori unknown.
We therefore introduce the {\em localization approximation} by neglecting the last term in Eq.~(\ref{eq:locapprox}). 
It has been shown~\cite{lubos_ps} that the error in energy from the
localization approximation $\propto (\Psi_T -\Phi_0)^2$. However, since the Hamiltonian is
modified due to the neglected term, the obtained energy is not necessarily an upper bound
to the true eigenenergy of the original Hamiltonian. Fortunately, accurate Jastrow-Slater trial wave 
functions enable to carry out calculations without appreciable impact from the localization error.
Calculations have shown that the 
 localization approximation bias is typically smaller then the fixed-node bias~\cite{lubosFe,flad&dolg}. 
Therefore this approximation proved to be less of a problem in 
practice since the localization error is overshadowed by the fixed-node error. 
The issue of the upper bound has been resolved some years ago once
ten Haaf, van Bemmel and co-workers \cite{bemmelprl,bemmelprb} suggested modification of Hamiltonians with nonlocal terms which does not violate 
the variational principle.
 Let us write the matrix element of  the nonlocal part
as 
\begin{align}\label{eq:nonloc3}
\langle {\bf R}|V_{nloc}|{\bf R}'\rangle =
\langle {\bf R}|V_{nloc}^{+}|{\bf R}'\rangle + \langle {\bf R}|V_{nloc}^{-}|{\bf R}'\rangle
\end{align}
where $\langle {\bf R}|V_{nloc}^{+}|{\bf R}'\rangle$ are matrix elements for which
$\langle {\bf R}|V_{nloc}^{+}|{\bf R}'\rangle\Psi_T({\bf R})\Psi_T({\bf R}')>0$ and 
{\it vice versa} for $\langle {\bf R}|V_{nloc}^{-}|{\bf R}'\rangle$. 
For small times we can write 
\begin{align}\label{eq:nonloc4}
\langle {\bf R}|-\Delta\tau V_{nloc}|{\bf R}'\rangle = 
\delta({\bf R}-{\bf R}')- \langle {\bf R}|V_{nloc}|{\bf R}'\rangle\Delta\tau+
{\mathcal O}[(\Delta\tau)^2]
\end{align}
so that it is clear that the 
{\em  negative}  matrix elements
do not flip the walker sign and can be sampled without introducing any sign problems. The 
sign flipping (positive) part is evaluated by the projection onto the trial function very much like 
in the localization approximation, i.e.,

\begin{align}\label{eq:nonloc}
V_{eff}^{+}=    \Psi_T({\bf R})^{-1}\int \langle {\bf R}|V_{nloc}^+|{\bf R}'\rangle \Psi_T({\bf R}'){\rm d}{\bf R}' 
\end{align}
so that the new modified Hamiltonian has a form
\begin{align}\label{eq:nonloc1}
 \tilde {\mathcal H} =  {\mathcal H}_{loc} + V_{eff}^{+} +\int \langle {\bf R}|V_{nloc}^-|{\bf R}'\rangle {\rm d}{\bf R}' 
+V_{node}^{\infty}(\Psi_T({\bf R})=0) 
\end{align}
where $  {\mathcal H}_{loc}$ contains all the local parts such as kinetic energies and local interactions including 
the local components 
of pseudopotentials. The 
 the node term enforces the fixed-node approximation explicitly
(otherwise $V_{eff}^{+}$ would be unbounded at the nodes). Interestingly, one can explicitly show that
\begin{align}\label{eq:nonloc2}
 \langle \Psi_T| \tilde {\mathcal H} |  \Psi_T\rangle \ge  \langle \Psi_T|  ({\mathcal H}_{loc}+V_{nloc}) |  \Psi_T\rangle  
\end{align} 
for any $\Psi_T$ while the equality takes place when $\Psi_T=\Psi_0$, i.e., the energy
of the modified Hamiltonian is always an upper bound to the original ${\mathcal H}={\mathcal H}_{loc}+V_{nloc}$.
New algorithmic developments 
by Casula and coworkers~\cite{casula_lrdmc} enabled to implement this treatment of nonlocal operators into the 
DMC methods as the so-called $T-$moves. 

Pseudopotentials are also very useful for heavy elements which require 
inclusion of relativistic effects.             
The scalar relativistic corrections can be built-in into the pseudopotentials
and for very heavy elements this is basically the only choice if one wants to avoid the full
relativistic treatment.

Although this section seem somewhat specialized and technical, the results have wider implications. Note that the 
last development enables us to work with {\em any type of nonlocal operators} using the decomposition 
 from Eq.~(\ref{eq:nonloc3}) and the effective upper-bound Hamiltonian from  Eq.~(\ref{eq:nonloc1}).
Clearly, this opens possibilities for simulations with more complicated  Hamiltonians,  beyond just electrons and ions.

\subsection{Optimization of Variational Wave Functions}\label{sec:qmc:opt}
The quality of trial wave functions has a significant impact on the efficiency of VMC and DMC methods 
as well as on the accuracy of the final results. The presented functional forms typically depend 
on a number of linear and non-linear parameters which has to be optimized. 
The optimization of functions is a classic 
numerical problem and many approaches can be found in the literature. However,
in combination with random fluctuations and objective functions defined by the finite ensemble 
of walkers, the problem becomes more complicated. We will therefore outline a few methods used for 
this purpose.

\subsubsection{Variance Minimization}
Let us denote the set of variational parameters $\{c\}$ of some
real valued trial wave function $\Psi_{\{c\}}$. 
The mean value of the local energy  with respect to $\Psi_{\{c\}}$ 
is given by
\begin{align}\label{eq:energymin}
E_{VMC}^{\{c\}}=\frac{\int \Psi_{\{c\}}^2 E_L^{\{c\}}\,{\rm d}{\bf R}}{\int \Psi^2_{\{c\}}\,{\rm d}{\bf R}}
=\langle E_L^{\{c\}} \rangle\equiv \bar{E}^{\{c\}}.
\end{align}
The variance of the local energy can be written as
\begin{align}\label{eq:varmin}
\sigma^2_{\{c\}}=\frac{\int \Psi^2_{\{c\}} (E_L^{\{c\}}-\bar{E}^{\{c\}})^2\,{\rm d}{\bf R}}{\int  \Psi^2_{\{c\}}\,{\rm d}{\bf R}}
=\langle(E_L^{\{c\}}-\bar{E}^{\{c\}})^2\rangle.
\end{align}

Since the variance directly determines the 
efficiency one possible approach for wave function optimizations
is
 minimize the variance~\cite{cyrus1} in Eq.~(\ref{eq:varmin}) using a finite
set of fixed configurations, where the walkers are distributed according to $\Psi^2_{\{c_0\}}$.
The set of starting variational parameters is denoted as $\{c_0\}$. The variance optimization is a rather robust
strategy since it is always positive and bounded from below. In practice, 
the  mean value of the local energy is not {\it a priori} known,
so we replace $\bar{E}^{\{c\}}$ in Eq.~(\ref{eq:varmin}) by a reference energy $E_{ref}$. 

The method can be further improved.
One possibility is to use weights which account for the wave function change after the parameter updates. 
The {\em re-weightened variance} with the new set 
of parameters $\{c_{new}\}$ is then given as 
\begin{align}\label{eq:varmin2}
\sigma^2_{\{c_{new}\}}=\frac{\int \Psi^2_{\{c_0\}}W_{\{c_{new}\}}({\bf R})(E_L^{\{c_{new}\}}-E_{ref})^2\,{\rm d}{\bf R}}
{\int  \Psi^2_{\{c_0\}}W_{\{c_{new}\}}({\bf R})\,{\rm d}{\bf R}},
\end{align}
where we have introduced the weighting function
\begin{align}\label{eq:weights}
W_{\{c_{new}\}}({\bf R})=\frac{\Psi^2_{\{c_{new}\}}({\bf R})}{\Psi^2_{\{c_0\}}({\bf R})}.
\end{align}
The advantage of the reweighting scheme is a more accurate value of variance at each optimization step.
However, once the wave function departs significantly from its original distribution one has to update 
the walker positions as well to reflect the updated wave function otherwise the method becomes unreliable.  

For minimization algorithms we 
use two different methods. The first one is 
a modified version of a {\em quasi-Newton method}~\cite{va10a}. 
It uses only the values and numerical gradients of the objective function (in this case the variance)
and builds up the inverse Hessian from updates along the optimization trajectory.
Second algorithm is based on the general Levenberg-Marquardt (LM) method.
It can be smoothly tuned between {\em Newton} and {\em steepest descent} approaches with a built-in stabilization parameter. 
However, it requires both gradient and Hessian and is therefore more costly to calculate. 
The gradient of the variance with respect to $i$-th parameter is readily obtained by differentiating Eq.~(\ref{eq:varmin}) as 
\begin{align}
\sigma^2_i=2\bigg[ \langle E_{L,i}(E_L-\bar{E})\rangle +\Big \langle \frac{\Psi_i}{\Psi}E_L^2\Big\rangle 
-\Big \langle \frac{\Psi_i}{\Psi} \Big\rangle \Big \langle  E_L^2 \Big\rangle 
-2\bar{E}-\Big \langle \frac{\Psi_i}{\Psi} (E_L-\bar{E}) \Big\rangle
\bigg], 
\end{align}
where the subscript $i$ denotes $\frac{\partial}{\partial c_i}$.
Since the variance minimization method can be viewed
as a fit of the local energy for a fixed set 
of configurations~\cite{cyrus1}, 
by ignoring the change of the wave function
an alternative expression for the same gradient follows as
\begin{align}\label{eq:vargrad}
\sigma^2_i=2\langle E_{L,i}(E_L-\bar{E})\rangle. 
\end{align}
The Hessian derived from the gradient (\ref{eq:vargrad}) is then given as 
\begin{align}\label{eq:varhess}
\sigma^2_{ij}=2\langle (E_{L,i}-\bar{E})(E_{L,j}-\bar{E})\rangle. 
\end{align}
An important property of this approximate Hessian (\ref{eq:varhess}) is that it is 
symmetric in $i$ and $j$ and positive definite even for a finite set of configurations what is
not automatically true when using the formally exact expression.
    
\subsubsection{Energy Minimization}
In general, 
a straightforward  minimization of the energy in Eq.~(\ref{eq:energymin}) can be 
quite unstable. This is somewhat counterintuitive and has its origin in the finite
sample. The reason is that for a sufficiently flexible variational wave function 
it is possible to seemingly lower the energy for the fixed set of configurations,  
while, in fact, raising the true expectation value. As we have mentioned earlier, the 
variance is bounded from below and therefore this problem is not present even for the finite sample~\cite{cyrus1}.
Although one can frequently re-sample the wave function so that the method is always close
to the actual minimum one can do better with some modifications. 

We can improve the method convergence by employing the LM method
and information about gradient and Hessian of the local energy mean. 
The corresponding gradient can be readily obtained from Eq.~(\ref{eq:energymin}) as
\begin{align}\label{eq:engrad}
\bar{E}_i&=\Big \langle  \frac{\Psi_i}{\Psi}E_L +  \frac{H\Psi_i}{\Psi} -2\bar{E}\frac{\Psi_i}{\Psi}\Big\rangle \nonumber \\
&=2\Big \langle \frac{\Psi_i}{\Psi}(E_L-\bar{E})\Big\rangle 
\end{align}
where we used the fact that $H$ is Hermitian.
Note, that the expression in the last step of Eq.~(\ref{eq:engrad}) has a favorable property of 
zero fluctuations as $\Psi$ approaches the true eigenstate. By
differentiating the  Eq.~(\ref{eq:engrad}) we get the Hessian as 
\begin{align}\label{eq:hess1}
\bar{E}_{ij}=&2\bigg[ \Big \langle \Big( \frac{\Psi_{ij}}{\Psi}+\frac{\Psi_i \Psi_j}{\Psi^2}\Big) (E_L-\bar{E}) \Big\rangle \nonumber \\
&-\Big \langle \frac{\Psi_i}{\Psi}\Big\rangle\bar{E}_j 
-\Big \langle \frac{\Psi_j}{\Psi}\Big\rangle\bar{E}_i + \Big\langle \frac{\Psi_i}{\Psi}E_{L,j}\Big\rangle \bigg ].
\end{align}
It is clear that the above Hessian is not symmetric in $i$ and $j$ when approximated by a finite sample
and the expression leads to quite large noise.
Umrigar and Filippi~\cite{cyrus2} have recently demonstrated that the fully symmetric Hessian 
written entirely in the terms of covariances ($\langle ab \rangle - \langle a \rangle \langle b \rangle$) 
has much smaller 
fluctuations then the Hessian from Eq.~(\ref{eq:hess1}). 
The modified Hessian from Ref.~\cite{cyrus2} is given by
\begin{align}\label{eq:hess2}
\bar{E}_{ij}=&2\bigg[ \Big \langle \Big( \frac{\Psi_{ij}}{\Psi}+\frac{\Psi_i \Psi_j}{\Psi^2}\Big) (E_L-\bar{E}) \Big\rangle
-\Big \langle \frac{\Psi_i}{\Psi}\Big\rangle\bar{E}_j 
-\Big \langle \frac{\Psi_j}{\Psi}\Big\rangle\bar{E}_i 
\bigg ]\nonumber \\
&+\Big[ \Big \langle \frac{\Psi_i}{\Psi}E_{L,j}\Big\rangle
 -\Big \langle \frac{\Psi_i}{\Psi}\Big\rangle \langle  E_{L,j} \rangle \Big]
 +\Big[ \Big \langle \frac{\Psi_j}{\Psi}E_{L,i}\Big\rangle 
 -\Big \langle \frac{\Psi_j}{\Psi}\Big\rangle \langle  E_{L,i} \rangle \Big],
\end{align}
where we have added three additional terms 
with vanishing expectation value (for proof, see e.g. Ref.~\cite{lin}). 
The terms make the Hessian symmetric while at the same time decreasing the variance.
 Hence, addition of terms with zero expectations helps to cancel 
most of the finite sample fluctuations and makes the minimization method vastly more efficient.

Another useful rearrangement of the Hessian~(\ref{eq:hess2}) is
\begin{align}\label{eq:hess3}
\bar{E}_{ij}=&2\bigg[ \Big \langle \Big( \frac{\Psi_{ij}}{\Psi}-\frac{\Psi_i \Psi_j}{\Psi^2}\Big) (E_L-\bar{E}) \Big\rangle 
 \nonumber \\
&+2 \Big\langle \Big( \frac{\Psi_i}{\Psi} -\Big\langle \frac{\Psi_i}{\Psi} \Big\rangle \Big) 
                \Big( \frac{\Psi_j}{\Psi} -\Big\langle \frac{\Psi_j}{\Psi} \Big\rangle \Big)
(E_L-\bar{E}) \Big\rangle \bigg ]\nonumber \\
&+\Big[ \Big \langle \frac{\Psi_i}{\Psi}E_{L,j}\Big\rangle
 -\Big \langle \frac{\Psi_i}{\Psi}\Big\rangle \langle  E_{L,j} \Big\rangle \Big] 
 +\Big[ \Big \langle \frac{\Psi_j}{\Psi}E_{L,i}\Big\rangle 
 -\Big \langle \frac{\Psi_j}{\Psi}\Big\rangle \langle  E_{L,i} \Big\rangle \Big].
\end{align}
If the minimization procedure involves only the 
linear parameters in the exponential (i.e., of the form $\exp[c(i)f]$) such
as linear coefficients in the Jastrow factor,
the two terms on the first line of Eq.~(\ref{eq:hess3}) cancel out,
 eliminating thus the need for evaluation of the $\Psi_{ij}$ term. 
For details on Levenberg-Marquardt method and other minimization see, for example, Ref.~\cite{madsen}.



%% file: nodeschapter/nodeschapter.tex
\section{Nodal Properties of Fermionic Wave Functions}\label{ch:nodes}
\subsection{Introduction}\label{nodessec:level0}
For a long time,
nodes of fermionic wave functions and also related nodes of 
density matrices were considered uninteresting, since they are 
very complicated and their usefulness or connection to 
 physical effects were unclear. For a given real state $\Psi({\bf R})$, the node 
is defined as the subset of ${\bf R}$-space for which $\Psi({\bf R})=0$. 
The node divides the configuration space into {\em nodal domains} with constant wave function
sign, ``+'' or ``-''. 
Interestingly, the first ideas on nodal properties of eigenstates go back 
to Hilbert and later to Courant. In particular, Courant proved the famous 
theorem that the $n-$th eigenstate of the 2D Laplacian on simply connected
convex 2D regions with Dirichlet or  Neumann
boundary conditions have $n$ or less nodal domains. Later this has been further
expanded into higher dimensions and more complicated Hamiltonians.
In the context of physics problems, Breit analyzed the  
nodes of two-electron states in 1930 and he actually found the first exact node for
the  atomic 
$^3P(2p^2)$ state \cite{Breit30}.
More recently, the behavior of nodes
for highly excited (quasi-classical) states were studied and various bounds 
on nodal (Hausdorff) volumes for $n\to \infty$ were established. 
As we have seen, the nodes play an important role in eliminating the fermion
sign problem and
knowledge of exact nodes would enable us  
to obtain fermionic ground states  essentially exactly.

In relation to QMC methods, the nodes were perhaps for the first time encountered
and identified as a ``problem'' in the paper by Anderson in 1976 ~\cite{jbanderson76}.
The fermion nodes of small systems, mostly atoms, were investigated in several
other  papers~\cite{node1s2s_1,node1s2s_2,jbanderson75,andersp,lesternode}.
The fermion nodes for degenerate and excited states were further studied by Foulkes and co-workers~\cite{foulkes}. 
Bressanini, Reynolds and Ceperley demonstrated differences in the nodal surface
topology between Hartree--Fock and correlated wave functions for the Be atom and explained thus 
the large impact of the $2s,2p$ near-degeneracy on the fixed-node DMC energy~\cite{dariobe}.
More recently, improvements in fixed-node DMC energies of small systems were 
studied by using CI expansions~\cite{Bressanininew,umrigarC2},
and pairing wave functions~\cite{sorellabcs1,sorellabcs2,pfaffianprl,pfaffianprb}.

Some of the general properties of fermion nodes were analyzed in an 
extensive study by Ceperley~\cite{davidnode},
which included proof of the tiling and connectivity properties as well as generalizations of the 
 notion of nodes for density matrices. From few-electron systems as well as from
numerical studies of free particle electron gas in 2D and 3D ~\cite{davidnode} 
came the conjecture that fermionic ground states have only two 
 nodal cells (one ``+'', one ``-''). The paper laid down some of the ideas, 
however, explicit demonstrations were missing
even for non-interacting systems. Explicit proofs 
of this conjecture
for several important cases was 
found later \cite{lubos_nodeprl,lubos_nodeprb}, when
one of us used the property of connectivity to show analytically that a number 
of spin-polarized non-interacting and mean-field systems 
(non-interacting homogeneous electron gas, HF atomic states, 
harmonic fermions and fermions on a sphere surface)
have ground state wave functions with 
{\em the minimal number of two nodal cells}~\cite{lubos_nodeprl,lubos_nodeprb}. 
Furthermore,
using the Bardeen--Cooper--Schrieffer wave function, it was shown 
that for in spin-unpolarized systems (i.e., with both spin channels occupied) an 
arbitrarily weak interaction 
reduces the four non-interacting nodal cells again to the minimal two. 
Finally, in the same paper was also demonstrated that the minimal number of nodal cells 
property extends to the temperature density matrices.

From the point of view of QMC methods, the two key issues are the {\em nodal topologies} and 
the {\em nodal shapes} and their efficient description. The topologies are important since
they determine how the configuration space domains are connected, property crucial for the
correct sampling process. In addition, the topologies are also related to physical effects such as 
multiple particle exchanges and pairing. The shapes are important for the accuracy and so far
our ability to describe these with sufficient flexibility is quite limited.  
Some recent advances on this aspect will be presented later in the review.  

This section is organized as follows. In Sec.~\ref{nodessec:level1},
we outline some of the general properties of fermion nodes. In Sec.~\ref{nodessec:level2},
{\em exact} fermion nodes are described for few-electron spin-polarized atomic states, including
the only known three-electron case discovered very recently.
In Sec.~\ref{nodessec:level3}, we categorize the nodal surfaces for the several half-filled 
sub-shells relevant for atomic and molecular states. 
In Sec.~\ref{nodessec:level5},
we show how opposite spin correlations eliminate the redundant nodal structure of HF wave functions 
for cases of spin-unpolarized states.

\subsection{General Properties}\label{nodessec:level1}
Let us consider a system of $N$ spin-polarized fermions in $d$ dimensions  
described by a real antisymmetric wave function $\Psi({\bf R})$. We assume that $\Psi({\bf R})$
is an eigenstate of a Hamiltonian which, for simplicity, contains only local interactions.
Consequently, there exists a subset of electron configurations  
called a fermion node,
for which the wave function  vanishes so that we can specify the nodal subset by an implicit equation 
\begin{align}\label{eq:node1}
\Psi({\bf R})=0.
\end{align}
Assuming that $\Psi({\bf R})$ describes the ground state, from
the nodal subset we omit regions where the wave function might vanish because of other
reasons (e.g., external potential, additional symmetries, etc). 
Note that when we talk about the nodes,
we have in mind the nodes of many-body wave functions {\em in the space of all particles}.  
These nodes have nothing to do with 
familiar nodes of one-particle orbitals such as the angular nodes of spherical harmonics $Y_{lm}$ or 
radial nodes of radial one-particle orbitals.
In general, the many-particle state node is 
an $(Nd-1)$-dimensional hypersurface defined  by the implicit Eq.~(\ref{eq:node1}).
Whenever positions of any two electrons 
coincide (i.e., ${\bf r}_i={\bf r}_j$), 
the antisymmetry ensures that the $\Psi({\bf R})=0$. For $d>1$, these coincidences
do not fully specify the nodes, 
but only define $(Nd-d)$-dimensional subspaces of {\em coincidence planes}. As we will see,
the coincidence planes will play an important role for the nodes in 1D systems. 

Another definition which
we will need is the {\em nodal domain}. The node divides
the configuration space into subsets in which the wave function sign is constant and which
are called nodal domains (or cells). The nodal domain definition can be reformulated also
in terms of {\em paths in the configuration space}. A nodal domain
$\Omega({\bf R}_t)$ is a subset of configurations which can be reached
from the point ${\bf R}_t$ by a continuous path without crossing the node.

Let us now introduce the basic properties of wave function nodes.
\begin{itemize}
\item[(a)] {\em Manifold almost everywhere}--- Uhlenbeck \cite{uhlenbeck} proved that the node is 
generically a manifold, i.e., it can be locally mapped onto a $(dN-1)$-dimensional plane except, possibly,  
for a subset of zero measure. This assumes reasonable Hamiltonians and a few other general conditions.
Note that this {\em does not preclude} two or more hypersurfaces to cross 
since crossing is automatically lower dimensional, i.e., of zero measure.  
\item[(b)] {\em Nodal crossing angles}---If two nodal surfaces cross each other, they are orthogonal
at the crossing. If $n$ nodal surfaces cross each other, the crossing 
angles are all equal to $\pi/n$. 
\item[(c)] {\em Symmetry of the node}--- The symmetry of the node cannot be lower than
the symmetry of the state. Typically, these symmetries are the same, however,
there are exceptions. There are known exact cases with the node symmetry being higher than the state
symmetry.
\item[(d)] {\em Tiling property for non-degenerate ground states}---
The tiling property
says that by applying all possible particle permutations to an arbitrary
nodal cell of a ground state wave function one covers the complete configuration
space $\mathfrak{R}$ (i.e., $\sum_P\Omega(P{\bf R}_t)+\Omega({\bf R}_t)=\mathfrak{R}$).
Note that this does not specify how many nodal cells are there: the number of nodal domains
can be any even number between two and $N!$. We will use this property later. 
\end{itemize}

Let us analyze the nodes for several models. 
It is instructive to start with 1D systems
for which the exact ground state 
nodes can be found explicitly and for arbitrary interaction. 
The reason is that in 1D the coincidence and the nodal hypersurfaces
are identical due to geometric constrains of 1D space.
 Let us first consider $N$ spin-polarized fermions
in a $1D$ harmonic oscillator well.
The wave function is given by the Slater determinant
\begin{equation}
\Psi_{1D}(1, ...,N)= {\rm det}[\phi_k({\bf r}_i)]=
A\prod_i e^{-x_i^2/2}{\rm det}[1,2x, ...,H_{N-1}(x)],
\label{eq:oned1}
\end{equation}
where $H_n(x)$ is a Hermite polynomial of degree $n$ and $A$
is the normalization.
We omit the prefactors and transform
the Slater matrix
to monomials so that
the wave function is given by the Vandermonde determinant
\begin{equation}
\Psi_{1D}(1, ...,N)={\rm det} [1,x,x^2, ..., x^{N-1}]
= \prod_{i<j}(x_j-x_i).
\label{eq:oned}
\end{equation}
The node is encountered whenever two fermions pass through
each other and the wave function has
 $N!$ nodal cells, since any permutation requires at least
one node-crossing. In general, the derived node is exact for other
$1D$ models including systems with arbitrary {\em local} interactions 
(which are not too singular). For periodic boundary conditions
one can find a similar expression. Let us consider
$N=(2k_F+1)$ particles in a 1D periodic box $(-\pi,\pi)$, so that the coordinates are dimensionless.
In this setting
 the Fermi momentum becomes an integer,
$k_F=1,2 ...$.  The one-particle occupied states are written
as $\phi_n(x)=e^{inx},$ $n=0,\pm 1,...,\pm k_F$
and the spin-polarized ground state is given by
\begin{equation}
\Psi_{1D}(1,...,N)={\rm det}\left[ \phi_n(x_j)\right],
\end{equation}
 where
$x_j$ is the $j$-th particle coordinate and $j=1,...,N$.
The term 
$\exp(-ik_F\sum_jx_j)$ can be factorized out of the determinant so that the
Slater matrix elements become powers of $z_j=e^{ix_j}$.
This leads to the Vandermonde determinant 
and after some rearrangements we find
\begin{equation}
\Psi_{1D}(1,...,N)=
e^{-ik_F\sum_jx_j}\prod_{j>k}(z_j-z_k)=
A\prod_{j>k}
\sin(x_{jk}/2)
\label{eq:van1d}
\end{equation}
where
$x_{jk}=x_j-x_k$ and $A$ is 
an unimportant constant prefactor. The periodic boundary conditions 
result in the same 1D nodal hypersurface which is also manifestly periodic, 
as expected.  Again, for  1D periodic ground states these nodes are exact
irregardless of interactions.

Properties of nodes for $d>1$ are very different and also
far more difficult to analyze than in one dimension. The key difference is that 
the node dimensionality is higher than the dimensionality of the coincidence hyperplanes. 
Clearly, for $d>1$ particles have enough space to ``avoid each other'' and that has 
implications on the nodal topologies and the number of nodal cells. In addition,
this spatial or dimensional flexibility also means
that the nodal hypersurfaces can be and are
affected by interactions.

We first consider spin-polarized non-interacting fermions in a $2D$ harmonic well.
The one-particle states are
simply $\phi_{nm}=C_{nm}H_{n}(x)H_{m}(y),$ $n,m=0,1, ... $
where $C_{nm}$ includes the Gaussian and normalization which are
absorbed into a common prefactor and omitted.
The Slater matrix elements can be rearranged
to monomials as follows
\begin{equation}
\Psi_{2D}(1,...,N)={\rm det}[1,x,y,..., x^ny^m, ...].
\label{eq:2dmono}
\end{equation}
The closed-shell states and the system size are conveniently labeled by
$M=1,2, ...$
where $n+m\le M$, with the corresponding number of fermions given by $N=(M+1)(M+2)/2$.

Let us  
 illustrate 
the qualitative difference in nodal topologies between higher dimensions
and 1D systems on the following example of three harmonic particles.
For $M=1$, the three-particle wave function is given by
$\Psi_{2D}(1,2,3)={\rm det}[1,x,y]$.
In order to understand the nodal domain topology, it is convenient to
extend the $2D$ coordinates
by a ``dummy'' third dimension as ${\bf r}_i=(x_i,y_i,0)$.
Then $\Psi_{2D}(1,2,3)={\bf z}_0\cdot ({\bf r}_{21}\times
{\bf r}_{31})$ where ${\bf z}_0$ is the unit vector in
the third dimension
and ${\bf r}_{ij}={\bf r}_i-{\bf r}_j$. Note that while in 1D three particle
state exhibits $3!=6$ nodal cells, the result we just derived implies
something very different.  There are only {\em two  nodal
domains} since the set of vectors ${\bf z}_0, {\bf r}_{21}, {\bf r}_{31}$ is
either left- or right-handed, and the node is encountered whenever
the three particles become
collinear, i.e., ${\bf r}_{21}\times{\bf r}_{31}=0$.
Another important property which will be explored in detail below,
is the fact that it is possible to
carry out a triple exchange
 $1 \to 2$, $2 \to 3$, $3 \to 1$ without node-crossing 
(e.g., rotate an equilateral triangle).

\begin{figure}[!t]
\begin{center}
\includegraphics[width=0.8\columnwidth]{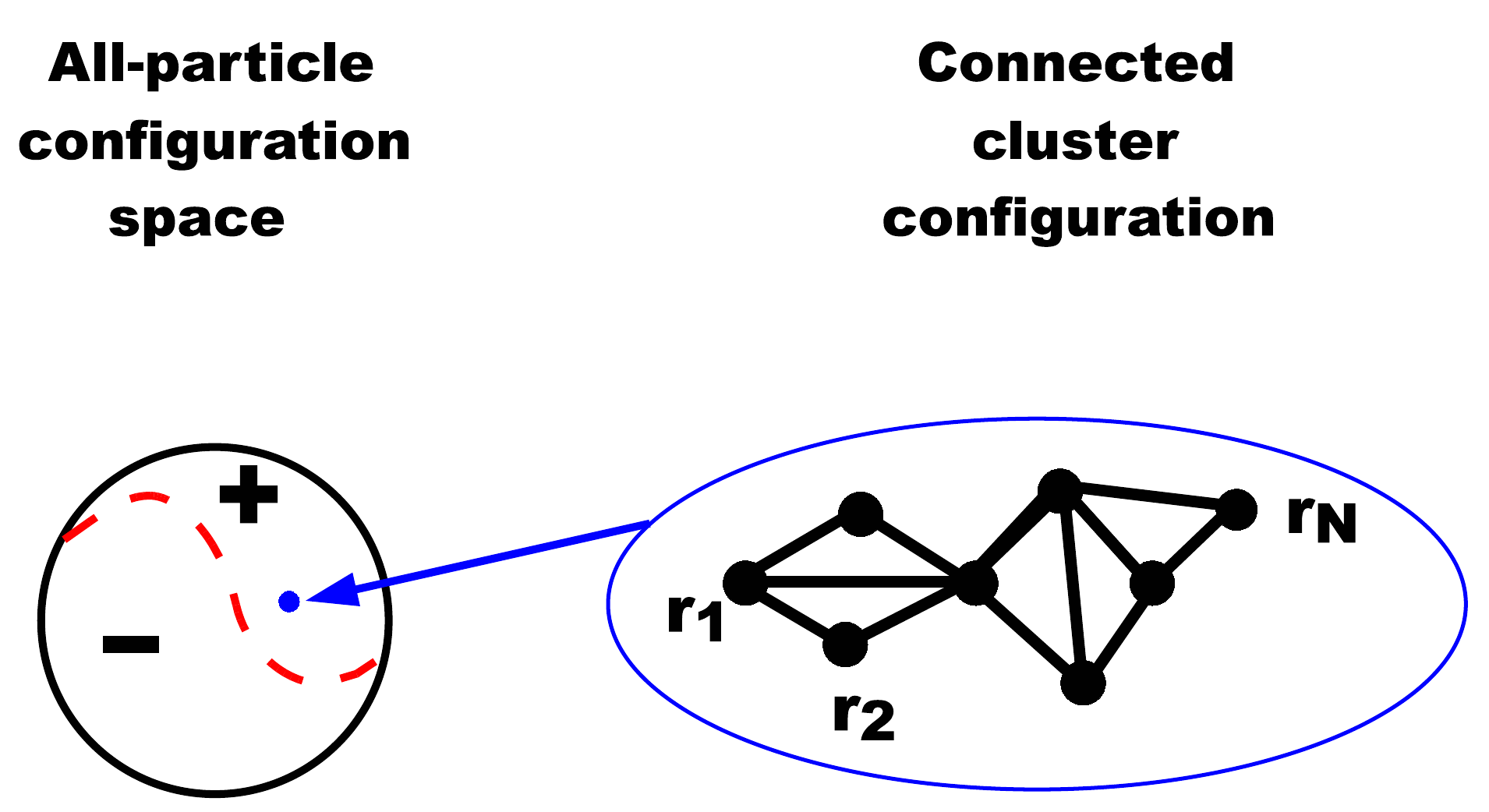}
\end{center}
\caption{
Sketch of a connected cluster of triple exchanges. Each triangle
means that there is a triple exchange path which does not cross the node.
}
\label{fig:connectivity}
\end{figure}

The triple exchanges are important for proving the two-nodal cell property in
general cases for arbitrary number of particles. 
In order to explain this aspect, we introduce a 
notion of {\em particles connected by triple exchanges}. 
The three particles $i$, $j$, $k$ are called {\em connected} if there exist a cyclic
exchange path $i \to j$, $j \to k$, $k \to i$, which does not cross the node.
Let us now suppose that there exist a point ${\bf R}_t$ inside a nodal domain such 
that  all particles are
connected into a single cluster of triple exchanges 
(see Fig.~\ref{fig:connectivity}).  Then $\Psi({\bf R})$ has only
two nodal cells. In other words, the whole configuration space is covered
by only one positive and one negative nodal cell (i.e., the nodal cells are {\em maximal}).
Let us try to understand why this is the case.
First, any triple exchange is equal to two pair exchanges.
The existence of the point ${\bf R}_t$ with all particles connected then means
that all positive (even) permutations $P_+{\bf R}_t$ can be carried
out within the same nodal cell. Second, the tiling property implies that once
all particles are connected for ${\bf R}_t$ the same is true for entire
cell $\Omega({\bf R}_t)$. Therefore, there will be only
one maximal cell per each sign. More details on this construction
can be found in Ref.~\cite{davidnode}.

\begin{figure}[!th]
\begin{center}
\includegraphics[width=0.8\columnwidth]{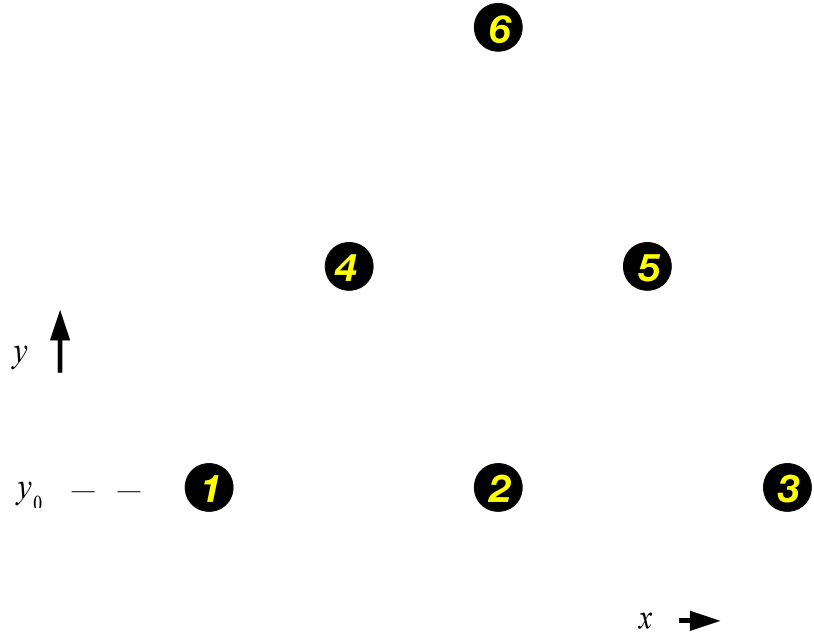}
\end{center}
\caption{
Positions of particles for proving the connectivity 
of six 2D harmonic fermions.}
\label{fig:sixparticles}
\end{figure}

We are now ready to show that six particles ($M=2$) in the harmonic 2D well also have
the ground state with the minimal nodal count of two. We position particles 1, 2 and 3
so that they have common $y-$coordinate, $y_0$. 
For $M=2$ the wave function reads
\begin{equation}
\Psi_{2D}(1,...,6)={\rm det}[1,x,y,x^2,xy,y^2]=\nonumber
\end{equation}
\begin{equation}
=\left| \begin{array}{cccccc}
1   &  1         &  1   & 1 & 1 & 1   \\
x_1 &  x_2       &  x_3 & x_4 & x_5 & x_6 \\
y_0 &  y_0       &  y_0 & y_4 & y_5 & y_6 \\
x_1^2 &  x_2^2   &  x_3^2 & x_4^2 & x_5^2 & x_6^2 \\
x_1y_0 & x_2y_0  &  x_3y_0 & x_4y_4 & x_5y_5 & x_6y_6 \\
y_0^2 &  y_0^2   &  y_0^2 & y_4^2 & y_5^2 & y_6^2 \\
\end{array} \right|,
\end{equation}
where the particle configuration which will enable 
to figure out the connectivity is given in Fig.\ref{fig:sixparticles}.
All elements containing
$y_0$ can be eliminated
by appropriate rearrangements and
the determinant factorizes as
\begin{equation}
\Psi_{2D}(1,...,6)=
\prod_{i=4}^6 (y_i-y_0)\Psi_{1D}(1,2,3)\Psi_{2D}(4,5,6).
\end{equation}
It is  now clear that we can rotate the triangle of particles 
4, 5, 6 without crossing the node since $\Psi_{2D}(4,5,6)$ remains constant and the wave
function does not change the sign. That implies that particles 4, 5, 6 are connected. However,
the wave function is rotationally invariant so the particles 1, 2, 4 must be connected, as well
as the particles 2, 3, 5. Therefore all six particles are connected and the wave function has 
only two nodal cells. 

With some effort it is possible to generalize these types of constructions
and show that harmonic particles are connected
for any closed-shell state 
with arbitrary $M \to \infty$  and for arbitrary $d$ as given elsewhere   
\cite{lubos_nodeprl}.

It seems that the harmonic oscillator might be somewhat special due to the simple 
structure of the wave function, since the nodes are zeros of antisymmetric polynomials
\cite{lubos_nodeprb}. 
In fact, {\em the minimal two-nodal cells property appears to be generic}. Let 
us present an example which is more realistic and does not reduce
to simple polynomials. 
Consider an atomic spin-polarized state  
$^6S(1s2s3p^3)$, i.e., 5 electrons in the Coulomb $-Z/r$ potential
for both  non-interacting and HF wave-functions.  
The wave function can be written as
\begin{equation}
\Psi_{at}(1,...,5)={\rm det}[\rho_{1s}^{*}(r),\rho_{2s}^{*}(r),x,y,z],
\end{equation}
where $\rho_{1s}^{*}(r)=\rho_{1s}(r)/\rho_{2p}(r)$ and
 $\rho_{2s}^{*}(r)=\rho_{2s}(r)/\rho_{2p}(r)$ since non-negative
$\rho_{2p}(r)$ is factorized out. The coordinates become dimensionless
by rescaling with the atomic number $Z$ 
as ${\bf r}\leftarrow Z{\bf r}$.
Let us position particle 1 at the origin and particles 2 to 5 on the surface
of a sphere with the radius $\eta_0$ equal to the radial node of
$\rho_{2s}(r)$ orbital,
i.e., $\rho_{2s}(\eta_0)=0$. For such configurations we obtain
\begin{equation}
\Psi_{at}(1,...,5)=
\rho_{1s}^{*}(\eta_0)\rho_{2s}^{*}(0)\, {\bf r}_{32} \cdot
({\bf r}_{42} \times {\bf r}_{52})
\end{equation}
Three-particle exchanges 2, 3, 4, 5 easily avoid
node-crossings by appropriate positioning and rotations on the sphere, e.g.,
using positions of the tetrahedron vertices. 
We can also show that particle 1 is connected by
the exchange $123\to 312$ which is parameterized
as ${\bf r}_1(t)=\eta_0[t,0,0]$; ${\bf r}_2(t)=\eta_0[c(t),
s(t), 0]$ and ${\bf r}_3(t)=\eta_0[0,1-t,0]$ where $t=0$ ($t=1$)
corresponds to the beginning (end) point of the exchange path 
while $c(t)=\cos(\pi t/2)$ and $s(t)=\sin(\pi t/2)$. Setting
 ${\bf r}_4=[0,0,\eta_0]$
and  ${\bf r}_5=[0,0,-\eta_0]$ we find that $\Psi_{at}$ is proportional to
\begin{equation}
\rho_{2s}^{*}(t\eta_0)c(t)(1-t)+ \rho_{2s}^{*}[(1-t)\eta_0]s(t)t>0
\end{equation}
The inequality holds for the whole path
$0\leq t\leq 1$ since  $\rho_{2s}^{*}(t\eta_0)>0$ for  $0\leq t <1$
for both non-interacting and HF cases.
Note that the demonstration is non-trivial since the HF wave function 
reflects the mean-field interactions and the orbitals can be found only by 
a numerical solution. Nevertheless,
qualitative features of the HF and non-interacting orbitals 
(symmetries, structure of radial nodes, etc) are basically identical and 
enable to verify the nodal cell count for both cases.
Note also that the wave functions is not a multi-variate polynomial
 like for the case of harmonic potentials.
The proof can be further extended to more shells with more electrons such
as $1s2s2p^33s3p^3$ and $1s2s2p^33s3p^33d^5$ \cite{lubos_nodeprb}.

Similar two-nodal cell demonstrations can be carried out for many paradigmatic 
quantum models such as 
2D and 3D homogeneous fermion gas, fermions on a sphere, etc  
\cite{lubos_nodeprb}. In addition, the same can be shown also for the 
temperature density matrices, for which one can define the nodes in similar
way as for the wave functions.
 
\subsection{Exact Nodal Surfaces}\label{nodessec:level2} 
We assume the usual electron-ion Hamiltonian and we first 
investigate a few-electron ions focusing on fermion nodes for
sub-shells of one-particle states with $s,p,d, ... $ symmetries using 
variable transformations, symmetry operations and explicit
expressions for the nodes.

\subsubsection{Two-Electron Atomic $^3P (p^2)$, $^3S(1s2s)$ and
Molecular $^3\Sigma_g(\pi^2)$ Triplet States} \label{nodessec:level22}
Apparently, the exact node of the atomic $^3P(p^2)$ 
state was derived in a different
context by Breit in 1930~\cite{Breit30,Bressanininew,darionew}.
Here we offer an independent
proof which enables us to apply the analysis
to molecular states with the same symmetries.
The state is the lowest within its symmetry class. It has even parity and
cylindric symmetry, say,  around $z$-axis. It is also odd
under rotation by $\pi$ around any axis orthogonal to the cylindric axis.
We denote the rotation by $\pi$ around, say, the $x-$axis, as  
$R(\pi x)$. 
In order to uncover the {\em exact} node
it is convenient to define new coordinates.
Let us denote
${\bf r}_{12}^+ ={\bf r}_1+{\bf r}_2, r_{12}^+=|{\bf r}_{12}^+| $,
together with the customary
${\bf r}_{12} ={\bf r}_1-{\bf r}_2, r_{12}=|{\bf r}_{12}|$.
We can now introduce the following
 coordinate map
\begin{equation}
({\bf r}_1, {\bf r}_2) \to
(r_{12}^+,r_{12},\cos\omega,\cos\beta,\varphi,\varphi'),
\end{equation}
where
$\cos\omega=
{\bf z}_0\cdot({\bf r}_1\times {\bf r}_2)/
|{\bf r}_1\times {\bf r}_2|$
with
${\bf z}_0$ being the unit vector along the $z$-axis,
$\cos\beta={\bf r}_{12}^+\cdot {\bf r}_{12}/(r_{12}^+ r_{12})$,
and
$\varphi'$ being the azimuthal angle of ${\bf r}_{1}\times  {\bf r}_{2}$.
Further, $\varphi$  is the azimuthal angle of ${\bf r}^+_{12}$
in the relative coordinate system with the $x$-axis unit vector
given by a projection of ${\bf z}_{0}$ into the plane
defined by ${\bf r}_1, {\bf r}_2$,
i.e., ${\bf e}_x= {\bf z}_{0p}/|{\bf z}_{0p}|$,
 ${\bf e}_z=({\bf r}_1\times {\bf r}_2)/|{\bf r}_1\times {\bf r}_2|$ and
 ${\bf e}_y={\bf e}_z\times {\bf e}_x$.
 The coordinate 
$\varphi'$ can be omitted from further 
considerations right away due to the cylindric symmetry so that the 
wave-function dependence simplifies to 
$\Psi(r_{12}^+, r_{12},\cos\beta,\cos\omega,\varphi)$.
We introduce an operator $Q$  which acts on coordinates and is given by 
 $Q=P_IP_{12}R(\pi x)$ where $P_I$ inverts around the origin and $P_{12}$
is the pair exchange of particles 1 and 2. Considering the state symmetries,
action of $Q$ reveals that the wave function is
invariant in the simultaneous change  $(\cos\beta,\varphi)$ $\to$
$(- \cos\beta, -\varphi)$. Combining action of $Q$ and $P_{12}$ we find
that
\begin{equation}
\Psi(...,-\cos\omega,...)=
-\Psi(...,\cos\omega,...)
\end{equation}
with the rest of variables unchanged.
The node is therefore given by $\cos\omega=0$, i.e., ${\bf z}_0\cdot
({\bf r}_1\cdot {\bf r}_2)=0$,
and it is encountered when both electrons happen to be 
on a plane which contains the $z$-axis. 
This exact node therefore agrees with the node of non-interacting or 
HF wave function $\Psi={\rm det}[\rho(r)x,\rho(r)y]$.

The derived node equation is applicable to any cylindric
 potential with $D_{\infty h}$ symmetry, e.g., 
equidistant homo-nuclear dimer, trimer, etc,
with one-particle orbitals
$\pi_x,\pi_y$ which couple into the triplet state
$^3\Sigma_g(\pi_x\pi_y)$.

Note that the parametrization given above
leads to the
exact node for the lowest two-electron triplet state,
$^3 S (1s2s)$ which has been known for some time~\cite{node1s2s_1,node1s2s_2}.
The spherical symmetry makes
 angles $\omega$ and $\varphi$  irrelevant and simplifies
the wave function dependence to distances $r_1,r_2, r_{12}$
or, alternatively, to
 $r_{12},r_{12}^+,\cos\beta$.
Applying $P_{12}$ 
leads to
\begin{equation}
-\Psi(r_{12},r_{12}^+,\cos\beta)=
\Psi(r_{12},r_{12}^+,-\cos\beta),
\end{equation}
so that the node is given by the condition $\cos\beta=0$, i.e., $r_1-r_2=0$, or,
equivalently,  $r_1^2=r_2^2$. 
The last equation clearly shows
that the node is generically a quadratic manifold, namely, a 5D hyperboloid in 6D
space (one recalls that in 3D space the 2D hyperboloid is given as $z^2 \pm {\rm const}
 =x^2-y^2$). Again, the nodal domain count is two and the domains correspond to $r_1>r_2$ or  $r_1<r_2$.

For the presented states the exact nodes are determined solely by
symmetry and are independent of interactions. That further implies that
all the excited states with the same symmetries
will contain the ground state node although they will exhibit also
additional nodal hypersurfaces. We will come back to this point in the following
subsection.

\subsubsection{Three-Electron Quartet $^4S (2p^3)$ State}\label{nodessec:level21}
Recently, we have found another case of exact node for a three-electron atomic
state, namely, the lowest spin quartet of $S$ symmetry and odd
parity, $^4S(2p^3)$. It is actually very easy to find the node by transforming
it  
 to the previously analyzed cylindric symmetry.
We can always rotate the system so that one of the electrons, say, the electron 3, is 
positioned on the 
$z-$axis so that the remaining symmetry is only cylindric. However, then the wave function
for the remaining two electrons is identical to the previous $^3P(2p^2)$ state
and we can directly write the nodal condition as 
${\bf r}_3\cdot({\bf r}_1\times {\bf r}_2)=0$. The wave function vanishes
when all three electrons appear on a plane which contains the origin. 

\begin{figure}[!t]
\centering
\begin{minipage}{\columnwidth}
\begin{tabular}{c c c }
\includegraphics[width=0.3\columnwidth]{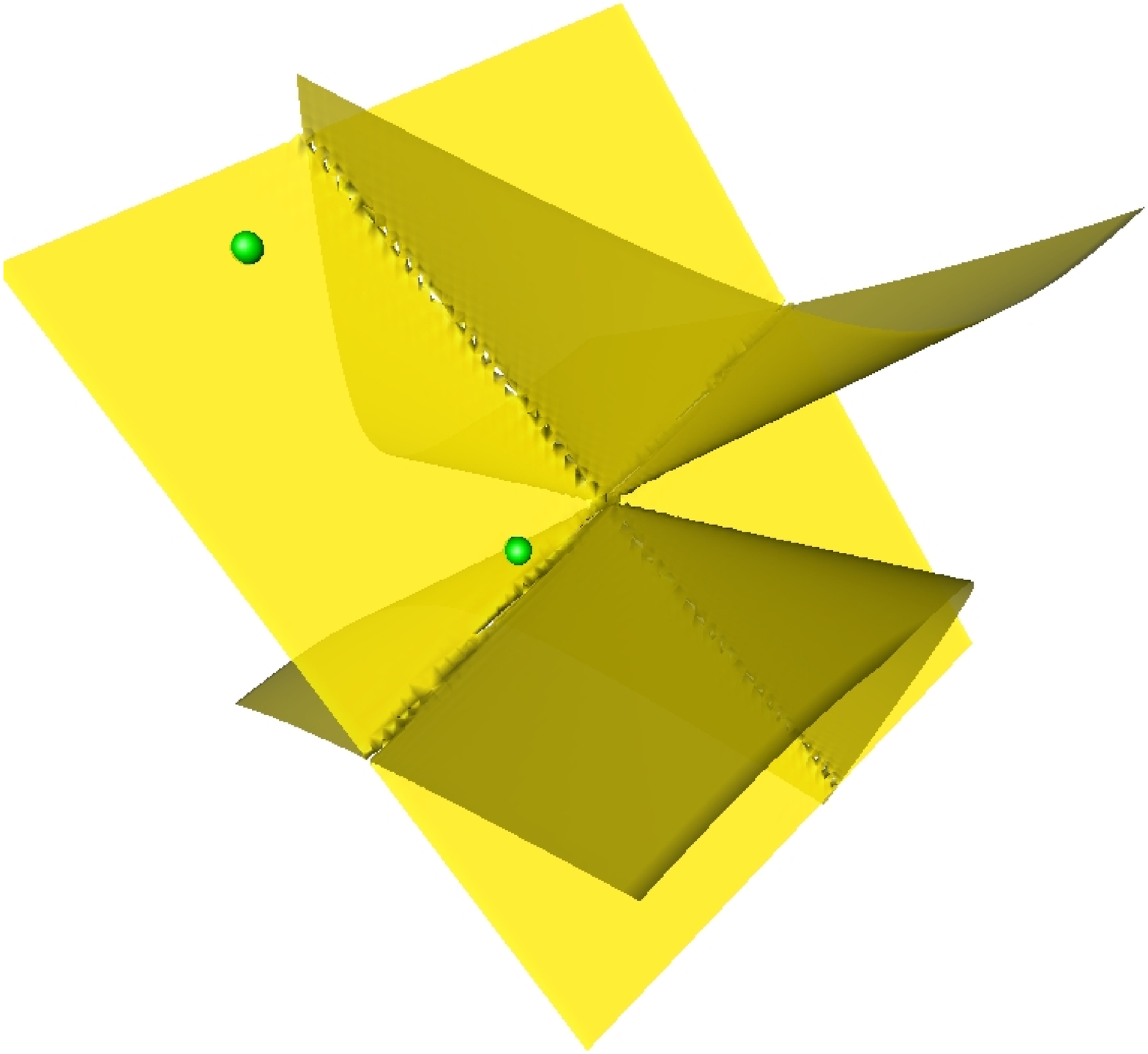} & 
\includegraphics[width=0.3\columnwidth]{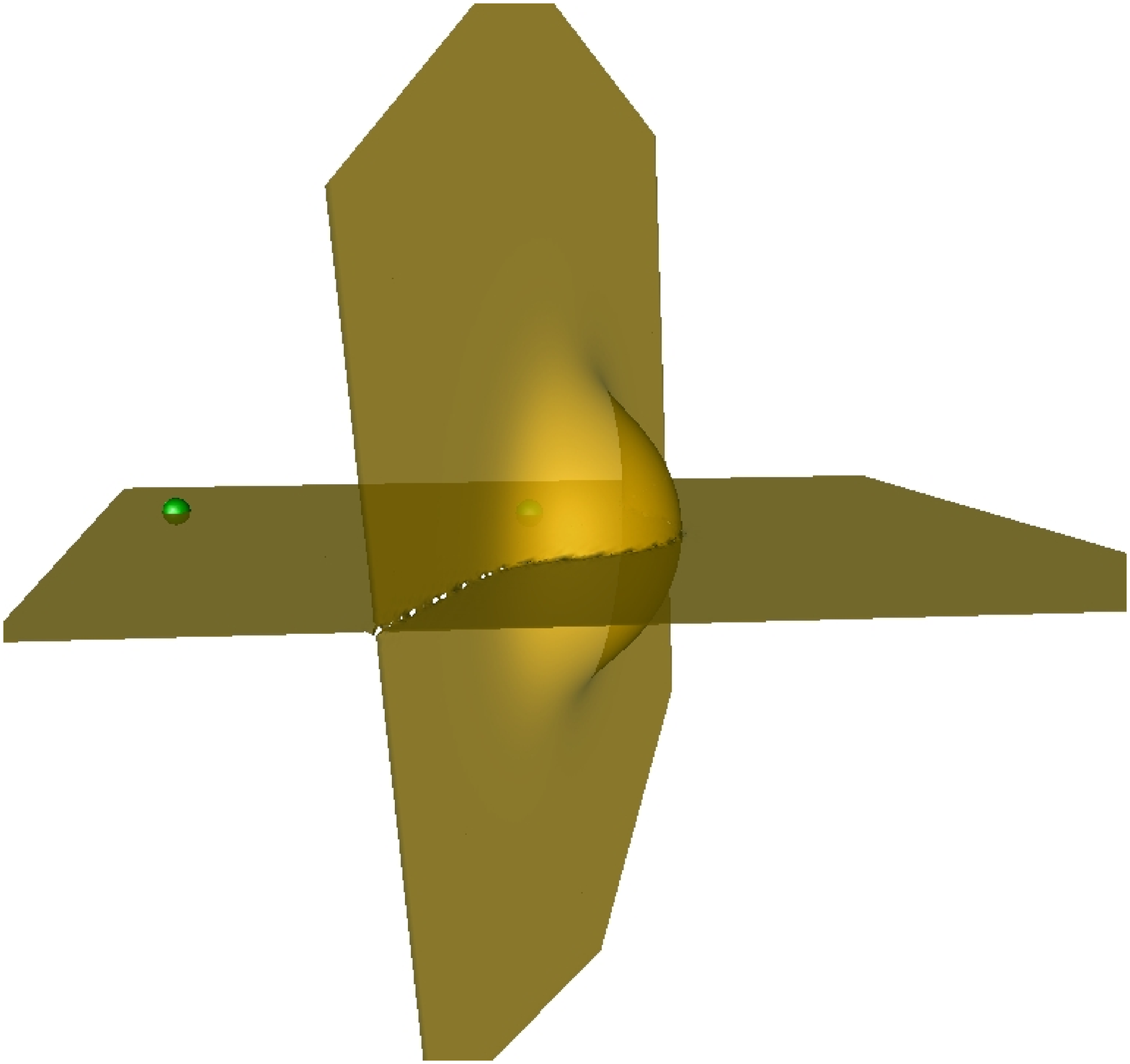}&
\includegraphics[width=0.3\columnwidth]{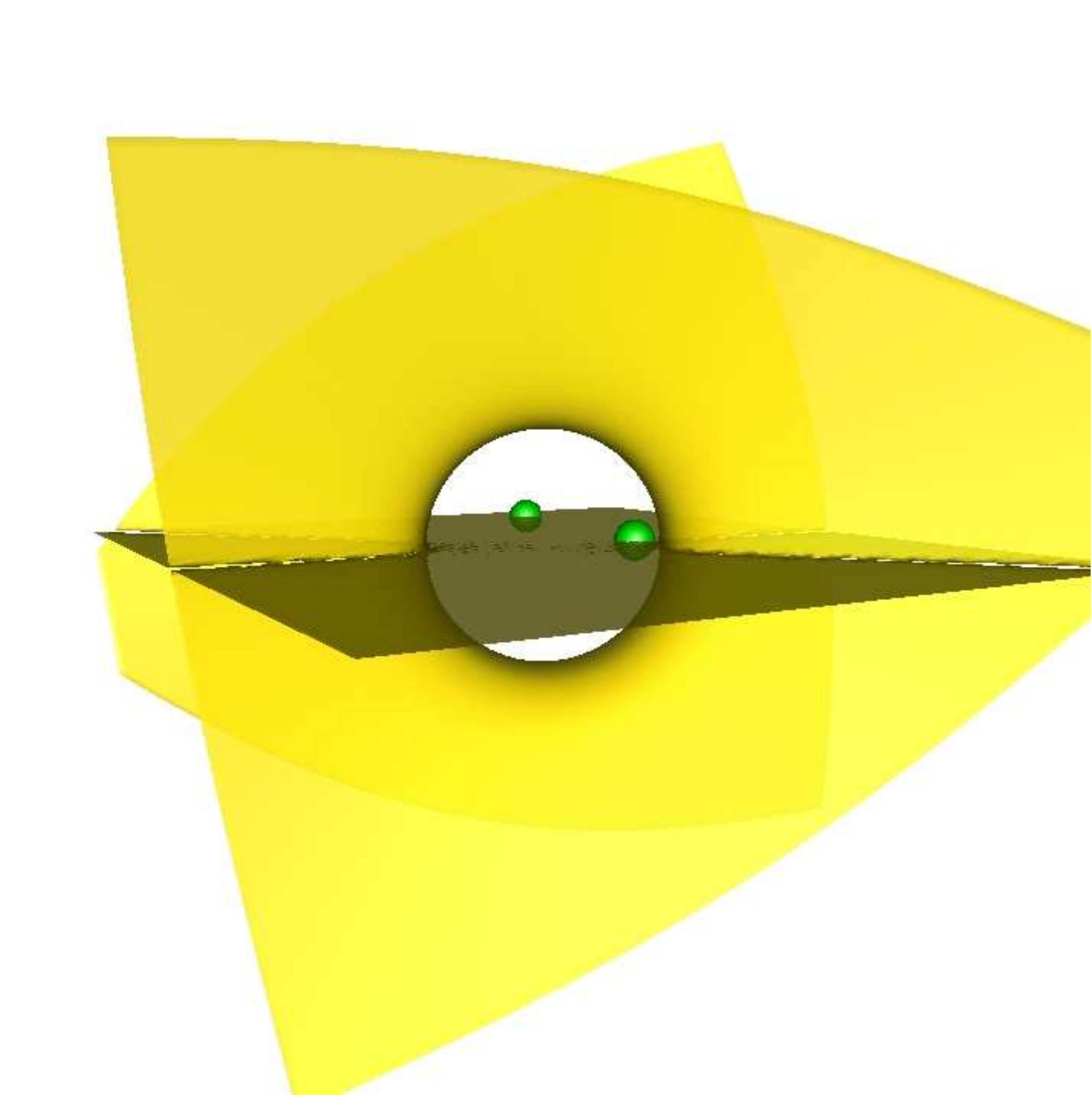}\\
{\Large $f^2$} & {\Large $pd^2$} & {\Large $pf^2$} \\
\end{tabular}
\end{minipage}
\caption{The 3D projected nodes of a few selected excitations
for the symmetry adopted CI expansion of the $^4S(p^3)$ ground state.
The exact planar node of the quartet ground state is also possessed
by all the excitations.
The small spheres show the fixed positions of two electrons
while the third one is scanning the nodal surface.
Labels indicate the types of excitations.}
\label{fig:p3_excit}
\end{figure}

One can demonstrate the same by defining an alternative coordinate map 
$\cos\alpha={\bf r}_3\cdot ({\bf r}_1\times {\bf r}_2)/r_3|{\bf r}_1\times {\bf r}_2|$,
$\cos\beta=[({\bf r}_1\times {\bf r}_2)\times {\bf r}_{12}^+]
\cdot  {\bf r}_{12}/
[|({\bf r}_1\times {\bf r}_2)\times {\bf r}_{12}^+|
|{\bf r}_{12}^+|]$
 and $\gamma$ being the azimuthal angle of ${\bf r}_3$ in the relative coordinate 
system given by the following unit vectors, ${\bf e}_x= {\bf r}_{12}^+/r_{12}^+$,
${\bf e}_z=[[({\bf r}_1\times {\bf r}_2) \times {\bf r}_{12}]\times {\bf r}_{12}^+]
/|[({\bf r}_1\times {\bf r}_2)\times {\bf r}_{12}]\times {\bf r}_{12}^+ |$
and ${\bf e}_y={\bf e}_z\times {\bf e}_x$. Besides these relative coordinates, 
the three Euler angles fix the orientation in the 
original coordinate system, however, the $S$ symmetry makes them irrelevant.
In the new coordinates the node comes out
by the action of $P_{12}$
on $\Psi(r_{12}^+, r_{12}, r_3, \cos\alpha,\cos\beta,\gamma)$ as
\begin{equation}
\Psi(...,-\cos\alpha,\cos\beta,\gamma)=
-\Psi(...,\cos\alpha,\cos\beta,\gamma).
\end{equation}
It is clear that $P_{23}$ and $P_{13}$ would lead to the same nodal 
condition, assuming one would redefine the relative coordinate
map accordingly.  The node of non-interacting and HF
wave function 
$\Psi_{HF}={\rm det} [\rho(r)x,
\rho(r)y, \rho(r)z]$ again agrees with the exact condition.
As mentioned before, the combination of symmetries is so restrictive
that interactions leave the node invariant and this is in fact the case
also for the excited states. Therefore {\em all\/} the excited states must
 contain the ground state fermionic node. This is illustrated in
Fig.~\ref{fig:p3_excit} which shows
the ground state (planar) node together with additional nodes, as expected for
excitations.

The excitations have therefore special symmetry structure 
which enables to show that the one-particle
$s$ channel is absent from the whole spectrum in this symmetry sector.
One can demonstrate this as follows.
We expand any state of this symmetry
in linear combinations of excited determinants 
\begin{equation}
|^4S(p^3)\rangle =\sum_{i,j,k}c_{ijk} {\rm det}[|n_il_im_i\rangle
|n_jl_jm_j\rangle|n_kl_km_k\rangle],
\end{equation}
where the sum is over the complete one-particle spectrum (including scattering
states). Due to the selection rules coefficients
$c_{ijk}$ vanish for
any excitation which contains one or more $l=0$ orbitals.
Consider first states with 
$l_i=0$. That implies that all determinants in this configuration
must have
$l_j=l_k$ otherwise the Clebsch--Gordan coefficients for $S$ symmetry vanish.
However, then the excitation would be of even parity and therefore such configurations
drop off. The only remaining possibility is that $l_i=l_j=0$. Such configurations then
require that also $l_k=0$, however,
such states with odd parity do not exist. Therefore the
one-particle channel with $s$ symmetry is ``switched-off'' completely.
This property was useful for showing that the fixed-node calculation of this state
provided the exact energy even when using nonlocal pseudopotential with $s$ non-locality
\cite{michal_prb}.

In order to illustrate how exact and approximate
nodal surfaces affect the fixed-node DMC results  
we studied the valence-electron states of
the spin-polarized nitrogen cation, some of which were analyzed above.
The core electrons were eliminated
by pseudopotentials~\cite{lester}. The trial wave function was
of the commonly used form with single HF determinant times the Jastrow correlation
factor~\cite{qmcrev}. As we explained above,
 the pseudopotential nonlocal $s-$channel does not couple
to either the odd parity $S$ state or the even parity $P(p^2)$ state so 
that the nonlocal contribution to these states vanishes exactly.

\begin{table}
\caption{Total energies (in Hartrees) of N$^{+}$, N$^{+2}$ and N$^{+3}$ ions with core electrons
eliminated by pseudopotentials~\cite{lester}. The energies are calculated
by HF, configuration
interaction (CI), variational
(VMC) and fixed-node diffusion Monte Carlo (DMC) methods.
}
\begin{center}
\begin{tabular}{l c c c c}
\hline
\hline
State  & HF  & CI  & VMC  & DMC \\
\hline
 $^3P(p^2)$  & -5.58528 &  -5.59491  &    -5.59491(2)  &  -5.59496(3)     \\
 $^4S(p^3)$  & -7.24716 &  -7.27566  &    -7.27577(1)  &  -7.27583(2)     \\
 $^5S(sp^3)$  & -8.98570 & -9.02027  &    -9.01819(4)  &  -9.01962(5)     \\
\hline
\hline
\end{tabular}
\end{center}
\label{tab_xx}
\end{table}

In order
to compare the fixed-node QMC calculations with an independent 
method
we have carried out also 
configuration interaction calculations 
with ccpV6Z basis~\cite{dunning} (with up to three
$g$ basis functions), which
generates more than 100 virtual orbitals in total. In the CI method the wave function
is expanded in excited determinants and we have included
all single, double  and triple excitations.
Since the doubles and triples include two- and three-particle correlations
exactly, the accuracy of the CI results is limited only by the size
of the basis set. By comparison with other two- and three-electron
 CI calculations we estimate that the order of magnitude of
the basis set CI bias is  $\approx$ 
0.01 mH for two electrons and  $\approx$ 0.1 mH.
and for three electrons (despite the large number of virtuals 
the CI expansion converges relatively slowly~\cite{kutz} in the maximum
angular momentum of the basis functions, in our case $l_{max}=4$).
The pseudopotentials we used were identical in both QMC and CI 
calculations.

The first two rows of Tab.~\ref{tab_xx} show the total energies of
variational and fixed-node DMC calculations 
with the trial wave functions with HF nodes together with
results from the CI calculations.
For $^3P(p^2)$ the energies agree within a few hundredths of mH
with the CI energy being slightly higher but within two standard
deviations from the fixed-node QMC result.  For  
$^4S (p^3)$ the CI energy is clearly above the fixed-node DMC by about 
0.17 mH which is the effect of the finite basis. In order to 
illustrate the fixed-node bias in the case
when the HF node is {\em not} exact we have also included calculations
for four electron state $^5S(sp^3)$ (for further discussion
of this Hartree--Fock node see Sec~\ref{nodessec:level33} below). For this case, 
we estimate that the CI energy 
is above the exact value by $\approx 0.3$ mH so that the fixed-node energy 
is significantly {\em higher} than both the exact and CI energies.
Using these results we estimate that the  
fixed-node error is  $\approx $ 1 mH, i.e., almost 3\% of the correlation energy. 

\subsection{Approximate Hartree--Fock Nodes}\label{nodessec:level3} 
It is quite instructive to investigate the nodes of half-filled 
sub-shells of one-particle states with higher angular momentum.
We will explore the nodal hypersurface projections into the 3D space
to understand what an individual electron ``sees'' during the imaginary time 
evolution. The study also helps to classify nodes for various cases of 
atomic and molecular systems.
For the next few subsections we will closely follow the derivations 
from our paper \cite{michal_prb}.

\begin{figure}[!t]
\begin{center}
\includegraphics[width=\columnwidth]{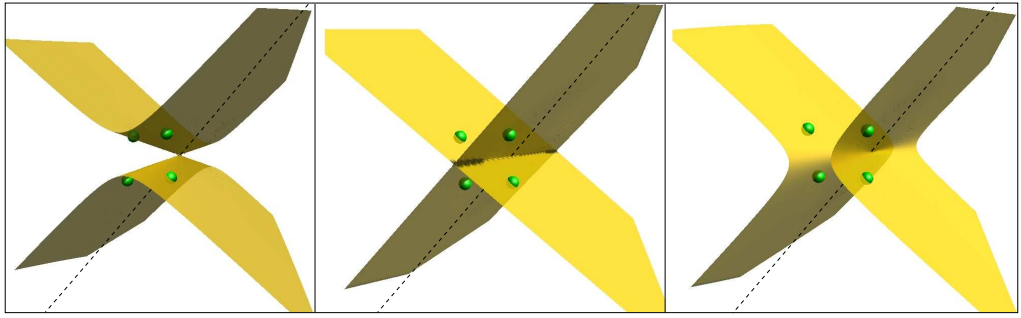}
\end{center}
\caption{ 
The 3D projected  Hartree--Fock node of $^6S(d^5)$ state, which is an elliptic cone
(left and right pictures). 
The picture in the center shows a configuration
of two pairs of electrons lying on two orthogonal
planes which contain the origin. This particular node is 
of lower dimension because of the additional constraint
on positions of the electrons. It appears
as a crossover between the elliptic cones with different
orientation (left 
and right pictures). The small spheres show the positions
of the four electrons while the line denotes the $z-$axis.}
\label{fig:d5}
\end{figure}

\subsubsection{Hartree--Fock Node of $^6S(d^5)$ State}\label{nodessec:level31}
The HF determinant wave function for $^6S(d^5)$
is given by
\begin{equation}
\Psi_{HF}= \prod_{i=1}^5 \rho(r_i){\rm det}[2z^2-x^2-y^2,x^2-y^2,xz,yz,xy],
\end{equation}
where $\rho(r_i)$ is the radial part of the $3d$-orbital. 
 Since the radial prefactor is irrelevant it can be omitted.
 The $S$ symmetry allows to rotate the system so that, say, electron 1 is
on the $z$-axis, and then the corresponding column in the Slater matrix
becomes $(2z_1^2,0,0,0,0)$. Assuming that $z_1\neq 0$ we 
can then write the nodal condition as
\begin{equation}
{\rm det}[x^2-y^2,xz,yz,xy] =0.
\end{equation}
Using one of the electrons as a 
{\em probe} (i.e., looking at the node from the
perspective of one of the electrons) we can find the projection 
of the node to the 3D space.
We denote the probe electron coordinates simply as
$(x,y,z)$ and expand the determinant so that we get
\begin{equation}\label{eq:5dHFstart}
(x^2-y^2)m_1 +xzm_2+yzm_3+xym_4=0
\end{equation}
where 
$m_i$ are the corresponding cofactors. We divide out $m_1$
assuming that it is non-zero 
(not a crucial assumption as clarified below).
We get 
\begin{equation}\label{eq:cone5d}
(x^2-y^2)+axz+byz+cxy=0,
\end{equation}
where  $\; a=m_2/m_1$,  $b=m_3/m_1$, $c=m_4/m_1$.
By completing the square
this can be further rearranged to 
\begin{equation}\label{eq:cone5d2}
(x-k_1y)(x-k_2y) +z(ax+by) =0
\end{equation}
with
$k_{1,2}=(-c\pm\sqrt{c^2+4})/2$. 
Let us define
rotated and rescaled coordinates 
\begin{align}
u^*&=-(ak_2-b)(x-k_1y)/(k_1-k_2)\\
v^*&=(ak_1-b)(x-k_2y)/(k_1-k_2)\\
w^*&=z[(ak_1-b)(ak_2-b)]/(k_1-k_2)^2
\end{align}
so we can write the Eq. (\ref{eq:cone5d}) as 
\begin{equation}\label{eq:5dHFgeneral}
u^*v^* +w^*u^* +w^*v^*=0.
\end{equation}
Note that this equation has a form which is
 identical to Eq. (\ref{eq:5dHFstart}) with 
$m_1=0$ so this representation is correct for any $m_1$.
After some effort one finds that Eq. (\ref{eq:5dHFgeneral}) is
a cone equation (i.e., $d_{z^2}$ orbital) as can be easily
verified by using the following identity
\begin{equation}
(2u^2-v^2-w^2)/8=u^*v^* +w^*u^* +w^*v^*,
\end{equation}
where  $ u=u^*+v^*+2w^*$, $ v=(-u^*+v^*+2w^*)$,
$w=(u^*-v^*+2w^*)$. 
 The 3D projected node is therefore an elliptic cone. 
Note that its equations is the same as rotated and rescaled angular part of the $d_{z^2}$ orbital, $2z^2-x^2-y^2$. 

Therefore the 3D projection of the $d^5$ HF node
is a family of cones given by the homogeneous 
second-order polynomial in
three variables with coefficients determined by the
positions of the four
electrons.
From the theory of quadrics
~\cite{rektorys}, one finds
that a general elliptic cone can possibly ``fit'' up to 
five 3D points/electrons. 
However, in our case there is an additional constraint 
of one electron being fixed 
on the $z$-axis.  That 
implies that the $z$-axis lies on the cone so that
the cone always 
cuts the $xy$ plane in two lines, which are orthogonal to each other.
The orthogonality can be verified by imposing
$z=0$ in Eq. (\ref{eq:cone5d2}) and checking that $k_1k_2=-1$. 
In addition, one can find
``degenerate'' configurations with two pairs of two electrons
lying on orthogonal planes (Fig.~\ref{fig:d5}).
This corresponds to the ``opening'' of the cone 
 with one of the elliptic radii becoming
infinite and the resulting node having a form of
 two orthogonal planes (Fig.~\ref{fig:d5}).
The condition is equivalent to
$A_{44}=b^2-a^2-abc=0$,  where
$A_{44}$ is one of the quadratic invariants~\cite{rektorys}.
There are more special cases of nodes with additional restrictions:
(a) when two electrons lie on a straight line going through
the origin;
(b) when three electrons lie on a plane
going through the origin; 
(c) when four electrons lie on a single plane.

%
%

\begin{figure}[!t]
\begin{center}
\includegraphics[width=\columnwidth]{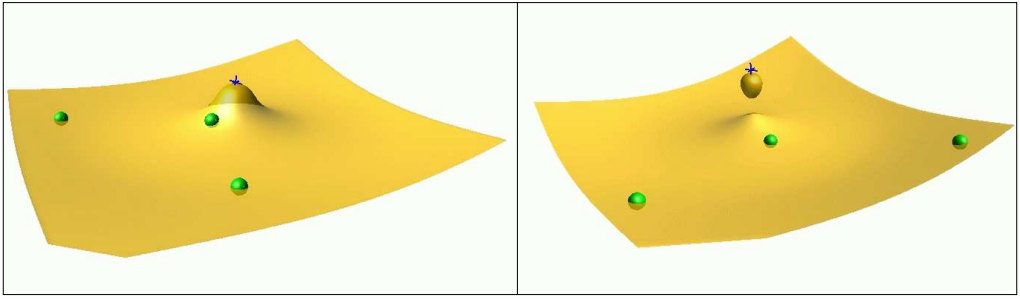}
\end{center}
\caption{
The 3D projection of the nitrogen cation $^5S(sp^3)$
Hartree--Fock node (the core electrons are eliminated by pseudopotentials). 
The projected node exhibits two
topologies. It is either
a planar surface deformed by the radial orbital functions 
at the nucleus or,
in certain configurations, the deformation forms a small bubble
detached form the surface (the picture on the right). 
The small cross is the location of the ion while the small spheres
denote positions of electrons.
}\label{fig:sp3}
\end{figure}

\subsubsection{Hartree--Fock Nodes of the $^5S(sp^3)$ Ion}\label{nodessec:level33}
The HF node for this spin-polarized 
state with $1s^2$ (He-core) states replaced by pseudopotentials can be
studied in a similar way as studied previously \cite{michal_prb}.
We can expand the determinant as in the previous cases
with a new feature that the radial parts will modify the shapes 
of the hypersurfaces. The system therefore corresponds, e.g., to the N$^+$ ion
from Tab.~\ref{tab_xx}.
Note that due to pseudized core the one-particle 
$s$ orbital has no radial node
although it corresponds to the physical $2s$ state.
By expanding the determinant in the column
of the probe electron with position $x,y,z$ 
the 3D node projection is simply given by
\begin{equation}
x+b'y+c'z+ d'\eta(r)=0,
\end{equation}
where $b',c',d'$ depend on ratios of cofactors 
and $\eta(r)=\rho_s(r)/\rho_p(r)$
is the ratio of radial parts of $s$ and $p$ orbitals. 
The probe electron will see a plane with a 
approximately bell-shape deformation in the area of the nucleus 
(see Fig.~\ref{fig:sp3}).
The shape of deformation depends on the ratio of $s$ and $p$ radial parts and the 
magnitudes and signs of the cofactors.
For certain configurations the deformation in a detached, separate
 ellipsoid-like bubble.
The bubble is caused by the radial dependence of $\eta(r)$ which  
for pseudized core is not a monotonic function and therefore
can create new topologies.
Note that despite the fact that the 3D projection shows a separated
region of space (the bubble) the complete node which is 11-dimensional has again 
the minimal number of two nodal cells.
 This can be checked by placing the four 
particles at the vertices of a regular tetrahedron with the
center at the ionic center. The wave function does not vanish
for such configuration as can be easily verified.
Rotations by $\pi/3$ around of each
of the four three-fold symmetry axes of the $T_d$ group then shows 
that all the particles are connected.

\subsubsection{Hartree--Fock Nodes of Spin-Polarized $p^3d^5$  
and $sp^3d^5$ Shells with $S$ Symmetry}\label{nodesec:level34}
\begin{figure}
\begin{center}
\includegraphics[width=\columnwidth]{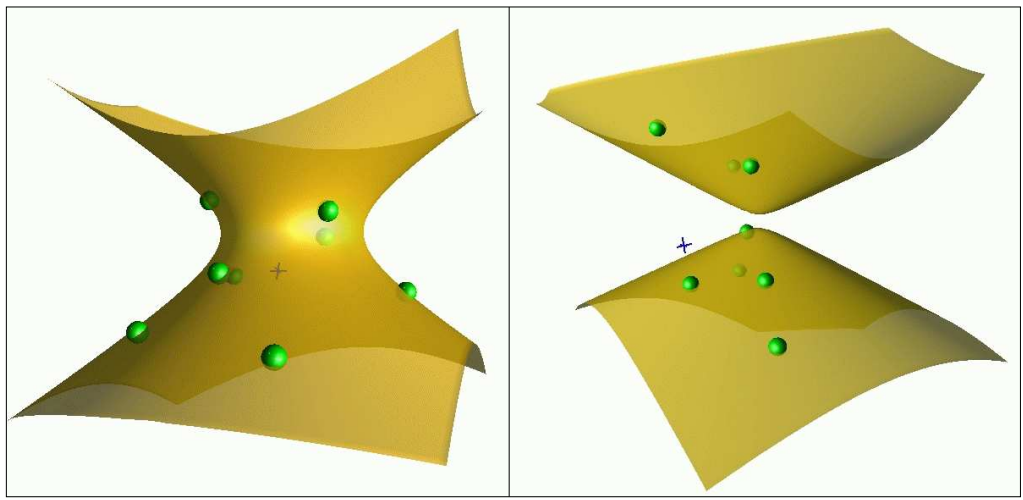}
\end{center}
\caption{
The 3D projection of the angular part of the $^{10}S(sp^3d^5)$ state
 Hartree--Fock node 
(with radial parts of orbitals identical for all $spd$ orbitals).  
The projection has a topology
of a single-sheet or double-sheet hyperboloid. The small cross shows the 
location of the nucleus while the spheres illustrate the electron positions. 
}\label{fig:sp3d5}
\end{figure}

\begin{figure}[!t]
\centering
\begin{minipage}{\columnwidth}
\begin{tabular}{c c}
\includegraphics[width=0.45\columnwidth]{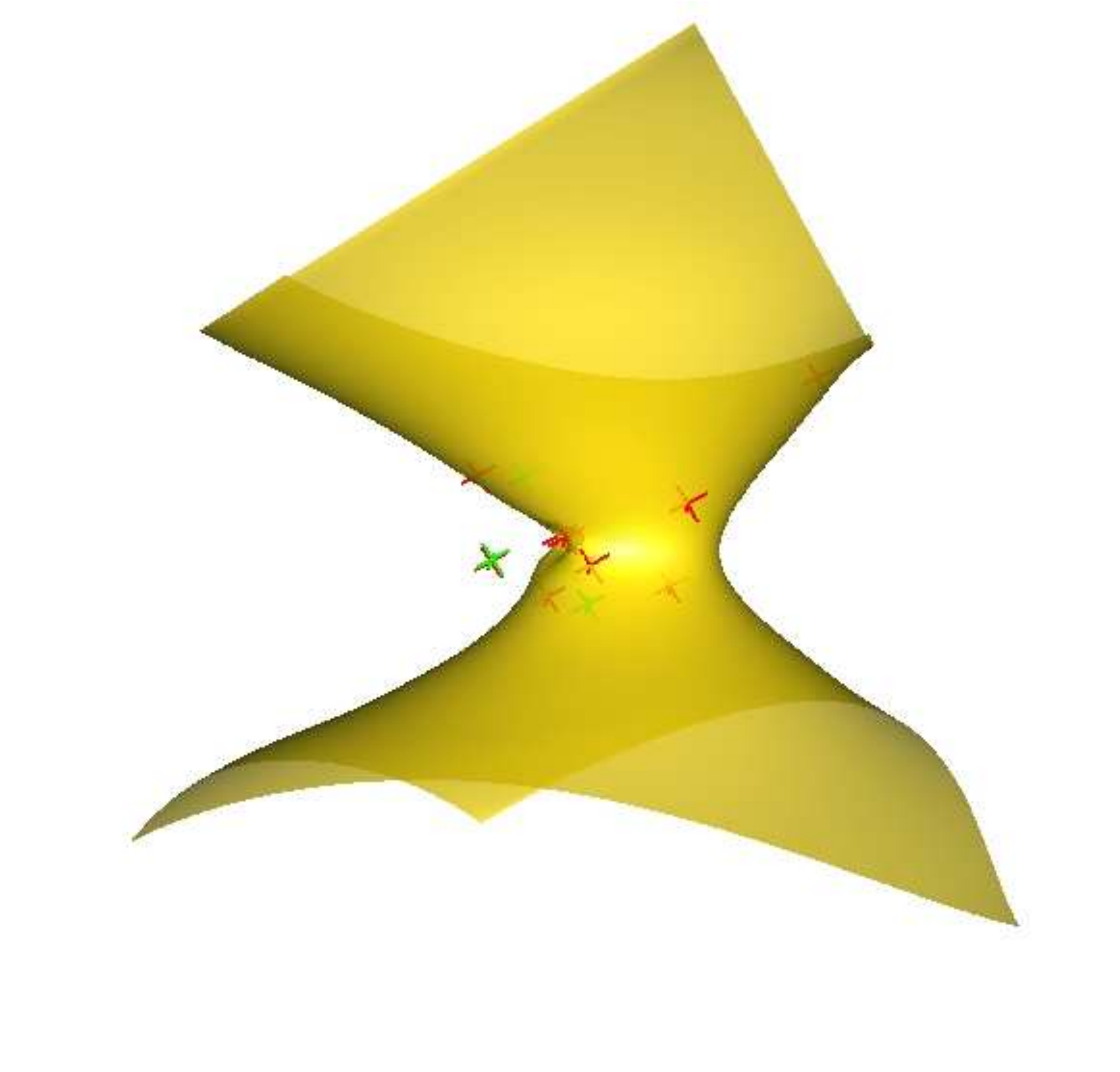} & 
\includegraphics[width=0.45\columnwidth]{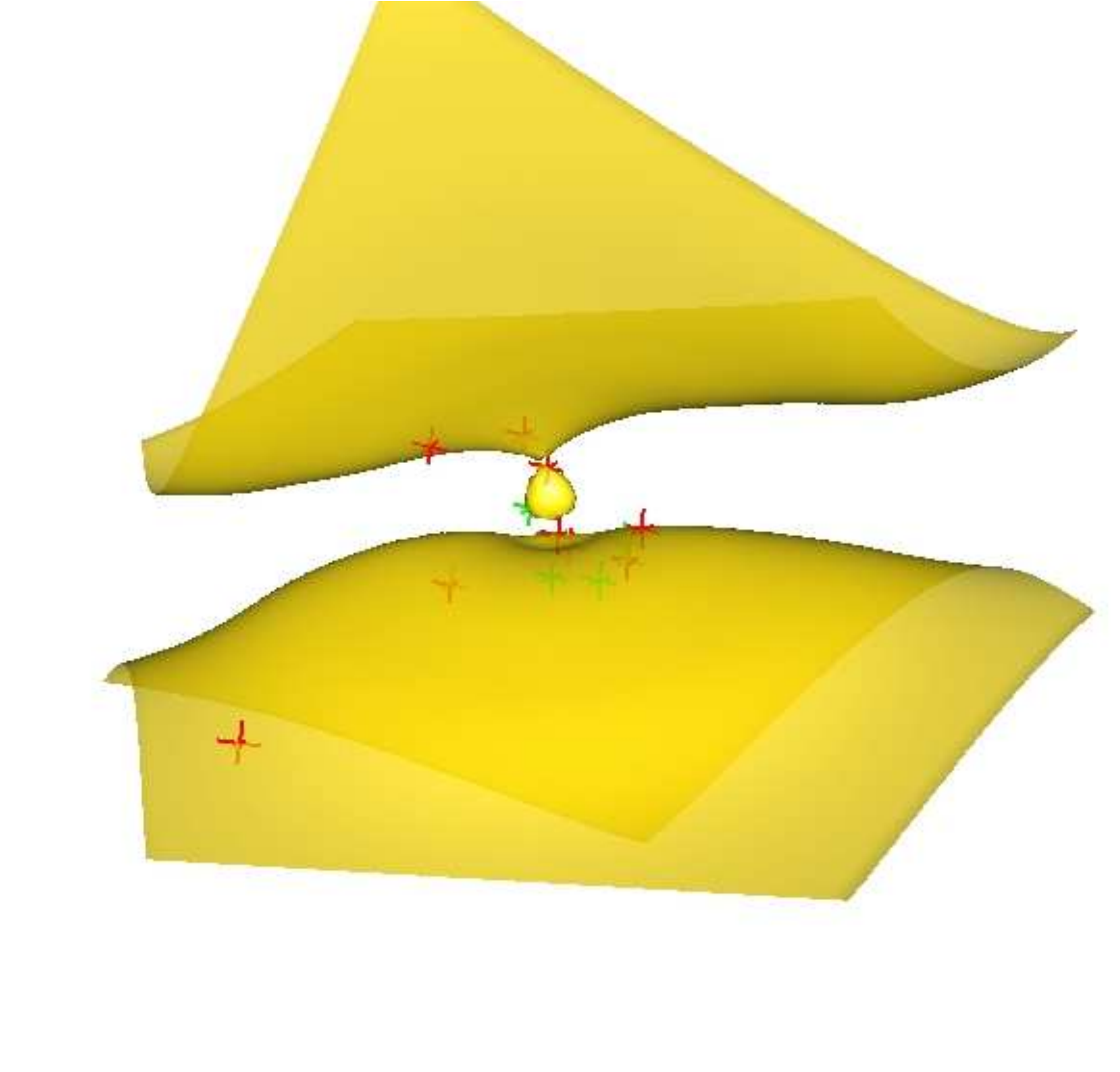}\\
\end{tabular}
\end{minipage}
\caption{ Projected Hartree--Fock node of
$^{10}S(sp^3d^5)$ of the majority spin valence electrons in
$\mathrm{Mn}^{+2}$ ion. The Ne-core electrons are eliminated 
by pseudopotentials. Note the deformations from the radial
parts of orbitals, including a  
small bubble detached from the rest of the surface (the right picture). 
The green cross indicates the nucleus and the red crosses the 
electron positions.}\label{fig:Mn}
\end{figure}

Let us for a moment assume a model wave function
in which the radial parts of $s,p,d$
orbitals are identical. Then, using the arrangements similar
to $d^5$ case, we can 
 expand the determinant of $p^3d^5$ in one column and 
for the 3D node projection we then get
\begin{equation}
2u^2-v^2 -w^2 +\alpha u +\beta v +\gamma w=0,
\end{equation}
where $u,v,w$ are appropriate linear combinations of $x,y,z$. 
This can be further
rewritten as 
\begin{equation}
2(u+\alpha/4)^2-(v-\beta/2)^2 -(w-\gamma/2)^2 +\delta_0 =0,
\end{equation}
where
$\delta_0=(-\alpha^2/2+\beta^2+\gamma^2)/4$.
It is clear that the 
 quadratic surface is offset
from the origin (nucleus)
by a vector normal to 
$\alpha u +\beta v +\gamma w =0$
plane.
Using the properties of quadratic surfaces one finds that
 for $(\alpha^2/(\alpha^2+\beta^2+\gamma^2))<2/3$ the node
is
a single-sheet hyperboloid with the radius $\sqrt{\delta_0}$; otherwise
it has a shape of
 a double-sheet hyperboloid. The double-sheet hyperboloid forms when
there is 
an electron located close to the origin.
A special case is a cone which corresponds to ($ \delta_0=0$).
The case of $sp^3d^5$ is similar, but with different $\delta_0$,
which now has a contribution from the $s$-orbital (see Fig.~\ref{fig:sp3d5}). 
Once we include also the correct radial parts of orbitals in the 
$s,p,d$ channels 
the coefficients of the quadratic form
depend on both cofactors and orbital radial functions.
The resulting
 nodal surface is deformed beyond an ideal quadric and shows some
more complicated structure around the nucleus (see Fig.~\ref{fig:Mn})
as illustrated on HF nodes of the majority spin electrons in Mn$^{+2}$ ion
(note that the Ne-core electrons were eliminated by pseudopotentials).

\subsection{Changes in Nodal Topologies From Interactions and Spin Correlations}\label{nodessec:level5}
We conjecture that a non-degenerate ground state of 
any given symmetry possesses only two maximal nodal cells. 
It was demonstrated for the considered fully spin-polarized systems
that the corresponding HF wave functions have the desired topology, i.e., two maximal nodal cells. 
On the other hand, for partially spin-polarized and unpolarized systems 
the corresponding HF wave functions of the form
\begin{equation}
\Psi_{HF}={\rm det}[\varphi_{\alpha}^{\uparrow}({\bf r}_i)]
{\rm det} [\varphi_{\beta}^{\downarrow}({\bf r}_j)]
\end{equation}
lead to at least {\em four} nodal cells due to the 
product of two determinants each with two cells.
As it turned out, this is correct only for special cases such as non-interacting 
systems with occupations in both spin channels and some separable models.
In general, such structure is an artefact. It appears since two different spins
and corresponding antisymmetry requirement on configuration subspaces is essentially 
a special non-uniform symmetry requirement on the wave function. We will see below
that requirements of additional or special symmetries could lead to the increase of nodal
counts even in the spin-polarized cases. Interestingly, generic interactions 
have the effect of smoothing out the artificial and redundant nodal structures 
so that the minimal two nodal cell property is restored. This is obvious since
it is not clear right away that interactions would drive the nodal topology changes, 
in particular, the changes of this type.
We will illustrate this on a few examples.

\begin{figure}[!t]
\begin{center}
\includegraphics[width=\columnwidth]{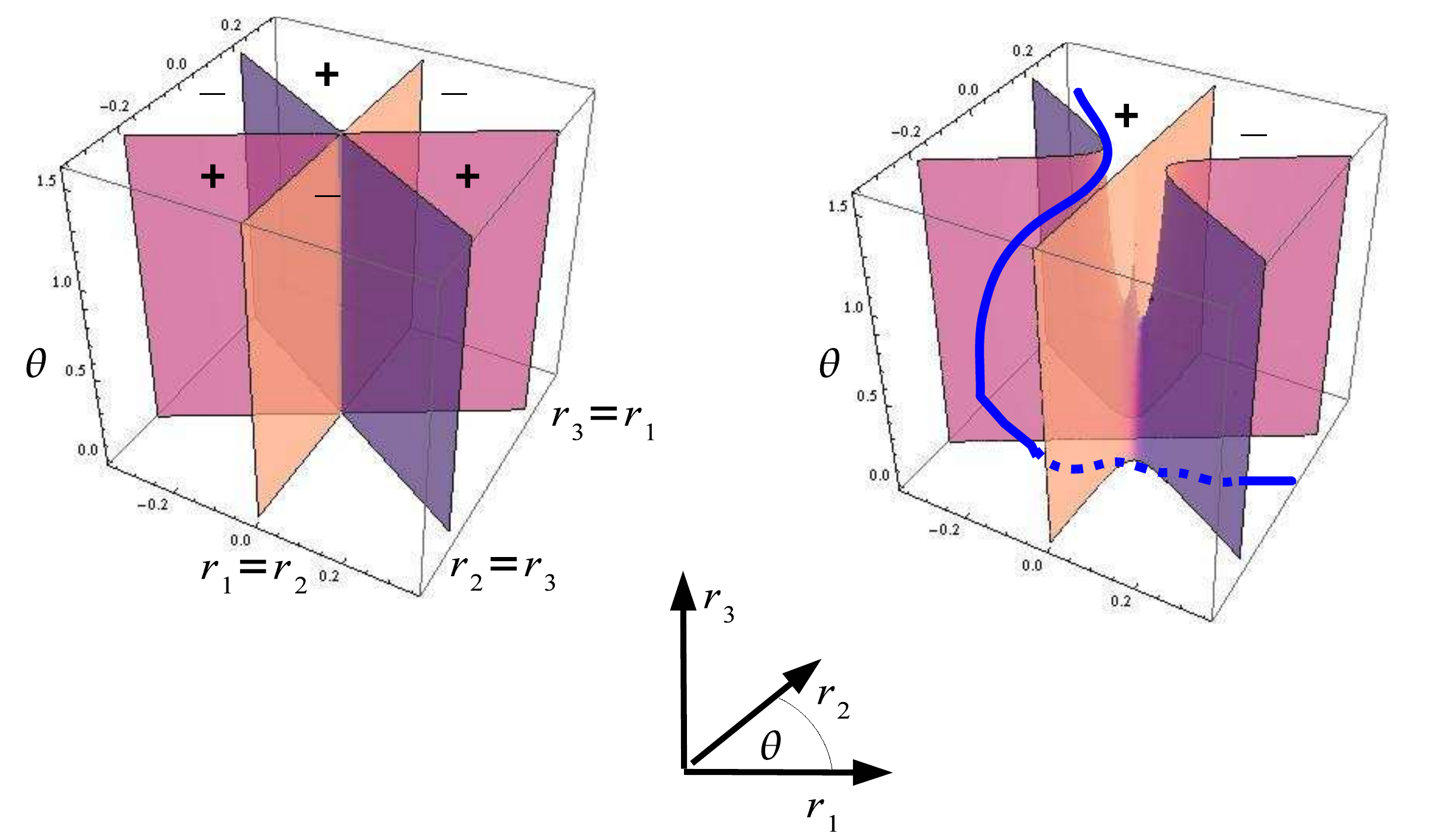}
\end{center}
\caption{
The HF node and correlated node for the $^4 S(1s2s3s)$ central potential state.
Note the change in topology after the inclusion of the correlation 
through the dominant additional configuration. The figure on the left plots
the {\em complete} HF node since it depends only on the three radii.
The figure on the right shows the node only on 3D subspace, there are more
nodal ``openings'' in other dimensions.
 }
\label{fig:quartetnode}
\end{figure}

Let us first analyze the simplest spin-polarized case of three electrons 
in a central, radially symmetric potential.
 Consider the lowest atomic
quartet of $S$ symmetry and even parity is $^4 S(1s2s3s)$. In the non-interacting
limit this state
 exhibits {\em six nodal cells}. The reason is
that all the single-particle orbitals depend only on radii and therefore
the nodes have effectively one-dimensional character. We will sketch how 
interactions change the topology and the number of nodal cells decreases
to minimal two. In the recent paper~\cite{lubos_nodeprl} we showed this numerically for the Coulomb
potential. For simplicity, we assume a simpler
model of harmonically confined electrons while using the same symmetries of
 states and one-particle orbitals as for the Coulomb case.
The harmonic oscillator has the advantage that
the exponential Gaussian prefactor is the same for all states and one-particle
orbitals irrespective of
the occupied sub-shells. As before, we  omit the prefactors in all expressions
since they are irrelevant for the nodes.
The non-interacting (or HF)
wave function then becomes simply
\begin{equation}
\Psi_{HF}({\bf r}_1, {\bf r}_2, {\bf r}_3)
=(r_2^2-r_1^2)(r_3^2-r_1^2)(r_3^2-r_2^2),
\end{equation}
since $\phi_{2s}$ and $\phi_{3s}$ are first
and second order polynomials in $r^2$,
respectively. It is clear that there are {\em six} nodal domains,
because $\Psi_{HF}$ vanishes whenever $r_i=r_j$ and
there are six permutations of $r_1,r_2,r_3$.
Now, we can switch-on the interactions and describe the resulting
correlations by expanding the wave function in
excitations. The excitations into
$s$-orbitals with higher quantum numbers do not change
the nodal structure
and their contributions to the exact wave function are actually very small.
The dominant excited configuration
is $^4S(1s,2p,3p)$, which captures angular correlations in the outer shells pair.
We write the correlated, two-configuration wave function as
\begin{align}
\Psi & = {\rm det}[\phi_{1s}, \phi_{2s}, \phi_{3s}]
\nonumber \\
& +\beta\{ {\rm det} [\phi_{1s}, \phi_{2px}, \phi_{3px}] + (x\to y)+(x\to z)\},
\end{align}
where $\beta$ is the expansion coefficient.
(This effect can be equivalently formulated also
as a Pfaffian with triplet pairing). Let us now try to carry out the triple
exchange. For this
purpose we position particles as follows: ${\bf r}_1=[0,r_a,0]$,
${\bf r}_2=[0,0,0]$ and ${\bf r}_3=[r_b,0,0]$ where $r_a>r_b>0$.
During the exchange the particles first move as follows:  $1\to 2$, $2\to3$
and 3$\to [r_a,0,0]$. The exchange is then finished by rotating particle
3 by $+\pi/2$ around the $z-$axis to the position $[0,r_a,0]$, completing thus
the repositioning
$3\to 1$. The particle $1$ moves
to the origin along the $y-$axis while particles 2, 3 move along the $x-$axis
while their radii are such that $r_3>r_2$ is maintained during the exchange.
Clearly, during the exchange the node of non-interacting wave function
will be crossed twice: when $r_1=r_3$ and again when $r_1=r_2$.
We will now show that for the correlated wave function the exchange
can be carried out
without crossing the node.
Using the expressions for the states of harmonic oscillator 
$\phi_{2px}=x$ and $\phi_{3px}=x(r^2-1)$ and substituting the coordinates
as outlined above,
one finds that the wave function becomes
\begin{equation}
\Psi = (r_3^2-r_2^2)[(r_2^2-r_1^2)(r_3^2-r_1^2)+\beta x_2x_3]
\end{equation}
We assume that $\beta>0$. 
(In the case $\beta <0$, it is straightforward
to modify the positions and exchange path
accordingly: the initial position of the particle 3 should be $[-r_b,0,0]$,
it moves to $[-r_a,0,0]$ and it is then rotated by $-\pi/2$
around the $z$-axis.)
Typical value of $\beta$ is of the order of $10^{-2}$ which reflects the fact that
the same spins (triplet) correlations are usually very  small.
After a little bit of analysis one finds that this function does not change
the sign provided the difference between the radii $r_2$ and $r_3$ when $r_1=
(r_2+r3)/2$ is sufficiently small. Then the following inequality holds
\begin{equation}
| r_2-r_3+ {\cal O}[(r_3-r_2)^2]| < \beta,
\end{equation}
which is easy to guarantee with some care when moving the particles
along the exchange path. The minimum of the
radial part appears essentially at the point $r_1=
(r_2+r_3)/2$ and the value is negative. However, the other term keeps
the wave function positive providing the negative term is not very large.
Since the term $| r_2-r_3+ {\cal O}[(r_3-r_2)^2]|$ can be made arbitrarily small by adjusting 
radii $r_2, r_3$ to be sufficiently
close to each other
one can easily fulfill the inequality.
The wave function then remains positive
along the whole exchange path. The nodes of uncorrelated and correlated
wave function are plotted in Fig.~\ref{fig:quartetnode}.

Similar topological change occurs in spin-unpolarized systems. The simplest 
possible case is the Be atom ground state with the electronic state $^1S[1s^22s^2]$. 
The wave-function can be written as a sum of HF and correlation components,
i.e.,  $\Psi=\Psi_{HF}+\alpha \Psi_{corr}$, where $\alpha$ is an expansion coefficient.
The HF part is given as 
\begin{align}
\Psi_{HF} & = {\rm det}^{\uparrow}[\phi_{1s}, \phi_{2s}]{\rm det}^{\downarrow}[\phi_{1s}, \phi_{2s}],
\end{align}
while the dominant correlating configuration is the $2s^2\to 2p^2$ excitation
\begin{align}
& \Psi_{corr}={\rm det}^{\uparrow}[\phi_{1s}, \phi_{2px}]{\rm det}^{\downarrow}[\phi_{1s}, \phi_{2px}]  + (x \to y)+( x \to z).
\end{align}
Clearly, the wave function has the required $S$ symmetry and it is
straightforward to
show that the wave function has only two nodal cells. Consider a plane passing through the origin
and configure
the two spin-up
electrons at antipodal points on a circle with radius $r_a>0$. 
Place the spin-down electrons on the same plane at antipodal points of 
a circle with different radius, say, $r_b>0$ and $r_b\neq r_a$.
In this configuration the HF wave function vanishes since particles lie on the
Hartree--Fock node. In general, however, the correlated wave function
 does not vanish
since we have
\begin{equation}
\Psi =C\,{\bf r}_{21}\cdot{\bf r}_{43},
\end{equation}
where $C$ is a symmetric non-zero factor.
 We can now rotate {\em all four particles} in the
common plane by $\pi$ and the wave function remains constant.
The rotation exchanges the two spin-up electrons 
and the two spin-down electrons simultaneously. This shows that the spin-up
and -down nodal {\em domains are connected} otherwise we would have encountered zeros.
Note that doing the rotation in one of  the spin-channels only, i.e., the two-particle
exchange for that spin, would necessarily result in hitting the node. The 
 correlation therefore enables the four-particle exchange to avoid the node and 
the example provides an interesting insight into the many-particle concerted exchange
effects.  Similar behavior can be found in homogeneous periodic gas and again can be generalized
to arbitrary sizes \cite{lubos_nodeprb}. In the periodic systems one can show that these effects 
enable to wind singlet pairs around the periodic cell without encountering the node providing 
the path is chosen appropriately and also that similar effects is impossible for HF 
wave functions  \cite{lubos_nodeprb}. 

 Multi-particle exchanges are closely related to quantum condensates such as 
Bose-Einstein condensation in bosonic systems or superconductivity in fermionic 
systems. The interpretation of condensates as systems which exhibit macroscopic chains 
of quantum exchanges goes back to Feynman. In bosonic systems this does not help much
and even looks ``trivial'' since there are no nodes to begin with and the emergence of 
condensates is driven by bosonic correlations. In fermionic systems the correct nodal 
topologies also appear only as a {\em necessary} but {\em not sufficient} condition
for the condensation to appear. However, from this point of view it is important that
the Bardeen--Cooper--Schrieffer wave functions, which will be described later, do exhibit
the correct nodal topologies as shown elsewhere \cite{lubos_nodeprl,lubos_nodeprb}. 

\begin{figure}[!ht]
\begin{center}
\includegraphics[width=\columnwidth]{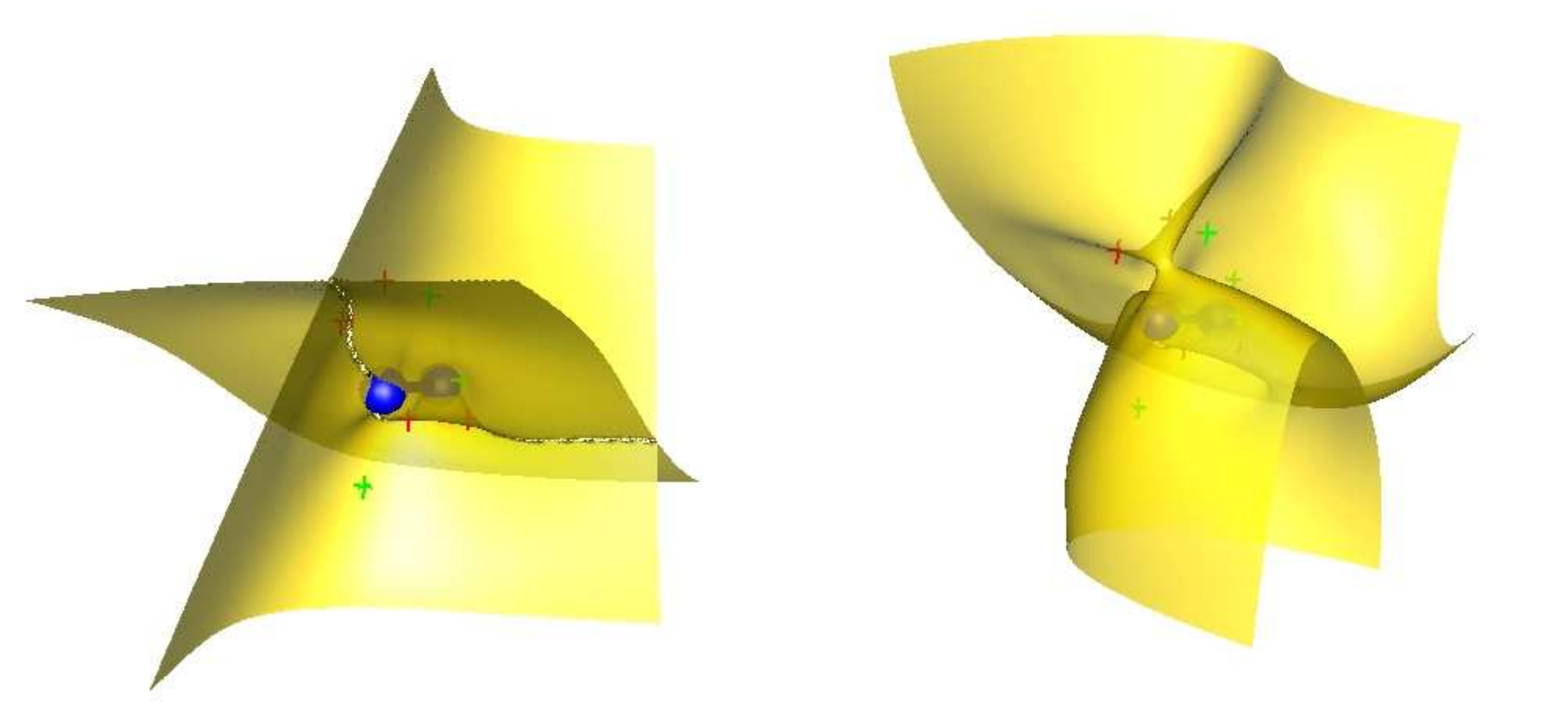}
\end{center}
\caption{The 3D projected nodes for N$_2$ dimer. 
The HF nodes (left) are bended and distorted due to spin correlations
into CI nodal surface (right) with just two nodal cells.
Sampling of nodes is performed with a pair of electrons with opposite spins, 
which are close to each other; positions of other electrons (crosses) 
are fixed; small spheres indicate ions.}
\label{fig:n2nodes}
\end{figure}

\subsubsection{HF and CI Nodes for N$_2$ Dimer State $^{1}\Sigma_g^+$} 
The projected nodal structure of the spin-unpolarized 
ground state $^{1}\Sigma_g^{+}$ for N$_2$ dimer
is shown in Fig.~\ref{fig:n2nodes}. The He-core electrons are eliminated by 
pseudopotentials as in previous calculations~\cite{lester}. The HF wave function
with the separation of electrons into independent spin-up and spin-down 
subspaces forms fours nodal cells. In the CI case we can see that 
the HF nodes have been distorted and bended to build up channels, 
which connect regions of the same sign with
resulting two maximal nodal cells.

In order to illustrate the accuracy of the constructed CI nodes we have 
carried out the  fixed-node DMC simulations with both HF and CI trial wave functions.
For the HF trial wave function 
 we have obtained the total energy 
-19.8395(7) H, which recovers $\approx$ 94\% of correlation energy.
For the CI trial function with more than 5000 determinants 
from double excitations into 45 virtual orbitals with  ccpV6Z basis
(with up to $f$ basis functions) the total energy 
reaches -19.870(5) H, recovering thus $\approx$ 98\% of 
correlation energy. 
The estimated exact energy is \mbox{-19.8822}~H. 
It is clear that despite the extensive determinantal expansion
the fixed-node error is still visible. It is just another demonstration that
convergence of such correlated wave functions is slow and more sophisticated approaches are needed
to reach beyond the limits of the determinantal expansions. 

\subsection{Conclusions}\label{nodessec:level6} 
We reviewed some of the recent advances in 
understanding the general properties of 
nodes of fermionic wave functions.
One of the key findings is that the nodal topologies of generic fermionic ground states 
appear to be remarkably simple. For $d>1$ 
the node {\em bisects } the configurations space with the resulting nodal domains being
connected. Intuitively, this is
not completely unexpected since presence of any node raises the wave function
curvature, i.e., the  kinetic energy. Therefore the energy minimization
leads to decreasing both the nodal ``volume'' and also the corresponding curvatures. 
We have also seen that by 
imposing additional symmetries one can increase the nodal domain counts. What is less 
obvious that even in such cases the interactions tend to smooth out the nodal surfaces 
and fuse the multiple domains into the minimal number.  This happens for both 
spin-polarized and unpolarized systems as we have shown on few-particle examples
with correlated wave functions.  

Comparison of bosonic and fermionic ground states from the
viewpoint of nodal topologies reveals an interesting perspective.
 It is well-known that generic bosonic 
ground states are node-less and so that one can say that they represent ``global S-waves''.
On the other hand, the bisection of the configuration space suggests
that generic fermionic ground states are ``global P-waves''. This is a remarkably
beautiful result of the interplay between continuous space and symmetries:
 it shows that Nature has chosen the simplest 
topologies to fulfill the fundamental symmetries. 

Furthermore, analysis of QMC data also reveals that much of 
the correlation in fermionic systems has actually bosonic
nature, i.e., particle correlations within the nodal 
domains such two-electron cusps, multi-particle even parity
exchanges, etc. This gives a rather prominent
role to the fixed-node DMC method which recovers such 
bosonic correlations exactly. Comparison
of fixed-node results with estimations of exact total 
energies (from other methods for small
number of electrons and from experiments for extended systems) 
show that this is indeed the case. Calculations
of atoms, molecules, clusters, solids, quantum liquids, etc, shows that 
the amount of recovered correlation
is similar essentially in all cases and reaches 90-95\% of the total correlation 
energy irrespective of the type of the system. This is very high 
when compared with other approaches based on explicitly correlated
wave functions. In fact, for extended systems, there is no other
method which can achieve such accuracy.
Indeed, this is a significant outcome of the QMC efforts 
over the years and it is also one of the key reasons why these approaches are becoming 
more common. The achieved level of accuracy now enables to predict many ``robust'' quantities,
i.e.,  which are larger than 0.1-0.2 eV, such as cohesions, optical gaps, etc.

On the other hand, the effects which are beyond the fixed-node
approximation still leave many important 
and exciting problems unsolved or unreachable. 
 Subtle effects such as magnetic ordering,
presence of new quantum phases, happen at scales of tens of meV or smaller
and, consequently,
can be significantly affected by  the fixed-node bias.
A limited understanding
of many of these aspects
is still the main reason why the progress has been very gradual albeit steady over the years.
Although the origin of these errors is exclusively in {\em inaccuracies 
of the nodal shapes} at present there are no clear-cut solutions which 
would enable to describe these in an efficient and general manner. 
Also how much and which types of nodal errors  
can influence interesting quantum phenomena is not 
yet understood. One possible path forward is to get
better handle of the nodal variational freedom using more effective nodal representations
or new constructions which will address this particular challenge. It seems that this 
is out of reach of traditional approaches:
common correlated wave function constructions, 
such as expansion in excited determinants, appear to be very slowly converging and
are 
therefore inefficient, limiting the applications to small sizes. Such constructions are essentially 
non-existent for extended
systems such as solids or surfaces.
Some of the recently explored ideas in this direction will be presented below. 
Another possibility to overcome this fundamental obstacle might come from better use of statistical 
methods to deal
with the nodal inaccuracies. In this respect, we mentioned some approaches in 
the previous section and there are ongoing efforts by us and others in 
exploring ideas in this research area as well.
Finally, ``brute force'' approaches, such as release node methods, are not necessarily useless
and could be useful for benchmarks in smaller and intermediate size systems.


%% file: pfaffianchapter/pfaffianchapter.tex
\section{Pfaffian Pairing Wave Functions}\label{ch:pfaffians}
\subsection{\label{sec:level1} Introduction}

As we mentioned, commonly used trial wave functions are based on
one-particle orbitals and Slater-Jastrow forms. 
In order to overcome this one-particle based constructions,
pair orbitals have been previously tried in various frameworks.
In condensed systems one 
such example is the Bardeen--Cooper--Schrieffer (BCS) wave function, which 
is an antisymmetrized product of singlet pairs. In quantum chemistry,
the pair orbital is sometimes referred to as geminal and the resulting wave function as 
the antisymmetrized geminal product (AGP).
It has been recently used  to calculate several atoms and molecules 
as well as  
superfluid ultracold Fermi gases~\cite{carlsonbcs,sorellabcs1,sorellabcs2}.
The results show promising gains, 
nevertheless, for partially
spin-polarized systems the improvements are
less systematic due to the lack of pair correlations in
the spin-polarized subspace~\cite{sorellabcs1,sorellabcs2}. 
The spin-polarized 
(triplet) pairing wave functions naturally lead to Pfaffians, 
 which have been previously applied to model systems~\cite{bouchaud2,Bhattacharjee,kevin}. 

In this section, we further develop ideas from Sec.~\ref{subsec:twf}, 
in which we have introduced 
the Pfaffian form. 
 Sec.~\ref{sec:pf:algebra} presents mathematical identities
and expressions associated with Pfaffians, some of them derived for the first time. 
In Sec.~\ref{sec:pf:pairingwf} we establish relationships 
between the Pfaffian, BCS and HF wave functions.
These forms are tested on atomic and molecular
systems in variational and fixed-node DMC
OA methods as described in Sec.~\ref{subsec:pf:spf}.
In Sec.~\ref{subsec:pf:mpf} 
we investigate generalizations to linear combinations 
of Pfaffians and compare the results with other methods in regard to 
to wave functions compactness.
the wave functions. In Sec.~\ref{subsec:pf:nodes}  
we analyze the topological differences between HF, Pfaffian and 
an essentially exact wave functions for a given test example.

\subsection{Algebra of Pfaffians}\label{sec:pf:algebra}
Here we will closely follow our exposition of Pfaffian algebra from Ref. \cite{pfaffianprb}. 
\subsubsection{Definitions}
First introduced by Cayley in 1852~\cite{cayley2},  the Pfaffian is named after 
German mathematician Johann Friedrich Pfaff.
Given a $2n\times 2n$ skew-symmetric matrix $A=\left[a_{i,j}\right]$, the 
Pfaffian of $A$ is defined as antisymmetrized product

\begin{align}\label{eq:pfdef}
{\rm pf}[A]&={\mathcal A}[a_{1,2}a_{3,4}\ldots a_{2n-1,2n}]\nonumber\\
&=\sum_{\alpha} {\rm sgn}(\alpha)\ a_{i_1, j_1}a_{i_2, j_2} \ldots a_{i_n, j_n},
\end{align}  
where the sum runs over all possible $(2n-1)!!$ pair partitions 
$\alpha=\{(i_1, j_1),(i_2, j_2),\ldots ,(i_n, j_n))\}$ of $\{1,2,\ldots,2n\}$ with $i_k < j_k$.
The sign of permutation associated with the partition $\alpha$ is denoted as
${\rm sgn}(\alpha)$. 
The Pfaffian for a matrix of odd order equals to zero. The following example gives Pfaffian
of a  $A(4 \times 4)$ skew-symmetric matrix
\begin{equation}
{\rm pf}\begin{bmatrix}
0 & a_{12} & a_{13} & a_{14} \\
-a_{12} & 0 & a_{23} & a_{24} \\
-a_{13} & -a_{23} & 0 &  a_{34} \\
-a_{14} & -a_{24} & -a_{34} & 0\\
\end{bmatrix}=a_{12}a_{34}-a_{13}a_{24}+a_{14}a_{23}.
\end{equation}   
It can be also evaluated recursively as 
\begin{align}\label{eq:pfrecurr}
{\rm pf}[A]&=\sum_{j=2}^{2n} a_{1,j}\sum_{\alpha_{1,j}}{\rm sgn}(\alpha_{1,j})\ a_{i_1, j_1}a_{i_2, j_2} \ldots a_{i_{n-1}, j_{n-1}}\nonumber\\ 
&\equiv \sum_{j=2}^{2n} a_{1, j} P_c(a_{1, j}) ,
\end{align} 
where $\alpha_{1,j}$ is partition with $i_k, j_k\neq 1, j$ and $P_c(a_{1, j})$ is 
defined as Pfaffian cofactor of $a_{1, j}$.
The cofactor for an element $a_{j,k}$ is given by a formula, 
\begin{equation}\label{eq:pfcof}
P_c(a_{j,k})=(-1)^{j+k+1}{\rm pf}[A(j,k;j,k)]  
\end{equation}
where the matrix $A(j,k;j,k)$ has the rank $2(n-1)\times 2(n-1)$  and is obtained  from  $A$ by eliminating
$j$ and $k$ rows and columns. 
\subsubsection{Calculation of a Pfaffian}
There exist several identities involving Pfaffians and determinants.
For any  $2n \times 2n$ skew-symmetric matrix $A$ and arbitrary matrices $B(2n \times 2n)$ 
and $M(n \times n)$ we have the following relations:
\begin{subequations}
\begin{align}
{\rm pf}[A^T]&=(-1)^n {\rm pf}[A] \label{eq:pfident1}\\
{\rm pf}[A]^2&={\rm det}[A] \label{eq:pfident2}\\
{\rm pf} \left[\begin{array}{cc}
A_1 & 0 \\
0 & A_2
\end{array} \right]&={\rm pf}[A_1]{\rm pf}[A_2] \label{eq:pfident5}\\
{\rm pf}[BAB^T]&={\rm det}[B]{\rm pf}[A]\label{eq:pfident4} \\
{\rm pf}\left[\begin{array}{cc}
0 & M \\
-M^T & 0
\end{array} \right]
&=(-1)^{{n(n-1)}\over{2}}{\rm det}[M].\label{eq:pfident3}
\end{align}
\end{subequations}
Let us comment on these identities and point out to respective proofs as appropriate:
\begin{itemize}
\item[(\ref{eq:pfident1})] Each permutation contains product of 
$n$ pairs resulting in an overall $(-1)^n$ factor.

\item[(\ref{eq:pfident2})] This is a well-known Cayley's relationship between
the Pfaffian and the determinant of a skew-symmetric
matrix. This is a well-known relationship which
 has been proved many times in variety of ways~\cite{nakahara,sss,cayley}.  

\item[(\ref{eq:pfident5})] Use the expansion by Pfaffian cofactors.

\item[(\ref{eq:pfident4})] By squaring (4d), using Eq.~(\ref{eq:pfident2}), and taking the
square root one finds
${\rm pf}[BAB^T]=\pm{\rm det}[B]{\rm pf}[A]$. Substituting the identity matrix $I$ for $B$ 
one finds $+$ to be the correct sign.

\item[(\ref{eq:pfident3})] Assume
\begin{align*}
B=\left(\begin{array}{cc}
M & 0\\
0 & I
\end{array}\right)
\quad {\rm and} \quad
A=\left(\begin{array}{cc}
0 & I\\
-I & 0
\end{array}\right)
\end{align*}
in Eq.~(\ref{eq:pfident4}). The overall sign is given by value of ${\rm pf}[A]$.
\end{itemize}

The identities listed above imply several important properties.
First, Eqs.~(\ref{eq:pfident4}) and (\ref{eq:pfident3}) show that every determinant 
can be written as a Pfaffian, but on the contrary, 
only the absolute value of Pfaffian can be given by determinant [Eq.~(\ref{eq:pfident2})].
The Pfaffian is therefore a generalized form of the determinant.
Second, by substituting suitable matrices~\cite{matrices} for $M$
in Eq.~(\ref{eq:pfident4}) one can verify the following three  
properties of Pfaffians~\cite{galbiati}, similar to the well-known
properties of determinants.
\begin{itemize}
\item[(a)] Multiplication of a row and a column by a constant
is equivalent to multiplication of the Pfaffian by the same constant.  
\item[(b)] Simultaneous interchange of two different rows 
and corresponding columns changes the Pfaffian sign.
\item[(c)] A multiple of a row and a corresponding column added to 
 another row and corresponding column leaves the Pfaffian value unchanged.
\end{itemize}
It is also clear that
any skew-symmetric matrix can be brought to the 
block-diagonal form by an orthogonal transformation.
Recursive evaluation [Eq.~(\ref{eq:pfrecurr})] implies that the Pfaffian of block-diagonal matrix is directly given by 
\begin{equation}\label{eq:pf:blockdiag}
{\rm pf}\begin{bmatrix}
0 & \lambda_1  & & & &  \\
-\lambda_1 & 0 & & & & 0 & \\
& & 0 & \lambda_2 & & \\
& & -\lambda_2 & 0 & & \\
& & & &  \ddots & \\
& 0 & & & & 0 & \lambda_n \\
& & & & & -\lambda_n & 0  \\
\end{bmatrix}=\lambda_1 \lambda_2 \ldots \lambda_n.
\end{equation}
Therefore by employing a simple Gaussian elimination technique with row pivoting (see Ref.~\cite{phdthesis}) 
we can transform any skew-symmetric matrix into block-diagonal form and obtain 
its Pfaffian value in $O(n^3)$ time.

However, in QMC applications, one often needs to 
evaluate the wave function after a single electron
update. Since Cayley~\cite{cayley} showed (for the proof see~\cite{pfaffianprb})
that  
\begin{align}\label{eq:cayley}
{\rm det}& \left[\begin{array}{ccccc}
0  & b_{12}  & b_{13} &\ldots &  b_{1n}\\
-a_{12}  & 0  & a_{23} & \ldots &  a_{2n}\\
-a_{13}  & -a_{23} & 0  & \ldots &  a_{3n}\\
 \vdots & \vdots & \vdots &  \ddots &  \vdots \\
-a_{1n} & -a_{2n} & -a_{3n} & \ldots &  0\\
\end{array}\right]\\
&={\rm pf} \left[\begin{array}{cccccc}
0  & a_{12}  & a_{13} &\ldots &  a_{1n}\\
-a_{12}  & 0  & a_{23} & \ldots &  a_{2n}\\
-a_{13}  & -a_{23} & 0  & \ldots &  a_{3n}\\
 \vdots & \vdots & \vdots &  \ddots &  \vdots \\
-a_{1n} & -a_{2n} & -a_{3n} & \ldots &  0\\
\end{array}\right]\,
{\rm pf} \left[\begin{array}{ccccc}
0  & b_{12}  & b_{13} &\ldots &  b_{1n}\\
-b_{12}  & 0  & a_{23} & \ldots &  a_{2n}\\
-b_{13}  & -a_{23} & 0  & \ldots &  a_{3n}\\
 \vdots & \vdots & \vdots &  \ddots &  \vdots \\
-b_{1n} & -a_{2n} & -a_{3n} & \ldots &  0\\
\end{array}\right],\nonumber
\end{align}
we can relate the Pfaffian of original matrix ${\rm pf}[A]$ to  
the Pfaffian of a matrix with updated first row and column ${\rm pf}[B]$ using the inverse 
matrix $A^{-1}$ in only $O(n)$ operations by 
\begin{equation}\label{eg:inverseupdate}
{\rm pf}[B]=\frac{{\rm det}[A]\sum_j b_{1j}A^{-1}_{j1}}{{\rm pf}[A]}={\rm pf}[A]\sum_j b_{1j}A^{-1}_{j1}.
\end{equation}
The second part of Eq.~(\ref{eg:inverseupdate}) was obtained by 
taking advantage of the identity in Eq.~(\ref{eq:pfident2}).
Slightly more complicated relation between ${\rm pf}[A]$ and ${\rm pf}[B]$ 
can be derived if one considers simultaneous change of two separate rows and columns,
which represents the two electron update of a wave function.  

\subsubsection{Gradient and Hessian of Pfaffian}
In optimization routines, it is often favorable to calculate the gradient 
and Hessian of wave function with respect to its variational parameters. 
Given the matrix elements $A$ dependent on a set of parameters $\{c\}$, 
one can derive the following useful relations:
\begin{equation}\label{eq:gradient}
\frac{1}{{\rm pf}[A]}\frac{\partial {\rm pf}[A]}{\partial c_i}=\frac{1}{2} {\rm tr}
\left[ A^{-1}\frac{\partial A}{\partial c_i}\right]
\end{equation}
and
\begin{align}\label{eq:hessian}
\frac{1}{{\rm pf}[A]}\frac{\partial^2{\rm pf}[A]}{\partial c_i \,\partial c_j}=&\frac{1}{4}
{\rm tr}
\left[A^{-1}\frac{\partial A}{\partial c_i}\right] \, {\rm tr}\left[ A^{-1}\frac{\partial A}{\partial c_j}\right] 
\\ \nonumber
&-\frac{1}{2} {\rm tr} \left[  A^{-1}\frac{\partial A}{\partial c_i}A^{-1}\frac{\partial A}{\partial c_j} \right]
+\frac{1}{2} {\rm tr} \left[  A^{-1}\frac{\partial^2 A}{\partial c_i\partial c_j} \right]
\end{align}
where 
$A^{-1}$ is again the inverse of $A$. 

\subsection{\label{sec:level2} Pairing Wave Functions}\label{sec:pf:pairingwf}


Let us now consider the generalization of the one-particle orbital
to a two-particle (or pair) orbital
$\tilde{\phi}(i,j)$, where tilde again denotes dependence on both
spatial and spin variables. The simplest antisymmetric wave function 
for $2N$ electrons constructed from the pair orbital is a
{\em Pfaffian} 
\begin{equation}\label{eq:generalpairingwf}
\Psi={\mathcal A}[\tilde{\phi}(1,2),\tilde{\phi}(3,4) \ldots\tilde{\phi}(2N-1,2N)]={\rm pf}[\tilde{\phi}(i,j)].
\end{equation}
The antisymmetry is guaranteed by the definition (\ref{eq:pfdef}), since the signs 
of pair partitions alternate depending on the parity of the corresponding
permutation. 
The important difference from Slater determinant is that in the simplest case only {\em one} 
pair orbital is necessary. (This can be generalized, of course, as will be shown later.)
If we further restrict our description to systems with collinear spins, 
the pair orbital $\tilde{\phi}({\bf r}_i, \sigma_i; {\bf r}_j, \sigma_j)$
for two electrons in positions ${\bf r}_i$ and ${\bf r}_j$ 
and with spins projections $\sigma_i$ and $\sigma_j$ can be expressed as 
\begin{align}\label{eq:generalpair}
\tilde{\phi}({\bf r}_i, \sigma_i; {\bf r}_j, \sigma_j)&=
\phi(i,j)
\langle \sigma_i\sigma_j|[|\uparrow \downarrow\rangle -|\downarrow\uparrow\rangle]/\sqrt{2}\\ \nonumber
&+\chi^{\uparrow \uparrow}(i,j)\langle \sigma_i\sigma_j|\uparrow\uparrow\rangle\\ \nonumber
&+\chi^{\uparrow \downarrow}(i,j)
\langle \sigma_i\sigma_j|[|\uparrow \downarrow\rangle +|\downarrow\uparrow\rangle]/\sqrt{2}\\ \nonumber
&+\chi^{\downarrow \downarrow}(i,j)\langle \sigma_i\sigma_j|\downarrow\downarrow\rangle.
\end{align}
Here  
$\phi(i,j)=\phi({\bf r}_i,{\bf r}_j)$ is even while $\chi^{\uparrow \uparrow}$, $\chi^{\uparrow \downarrow}$ and $\chi^{\downarrow \downarrow}$ are 
odd functions of spatial coordinates. 
In the rest of this section we will discuss special cases of the wave function~(\ref{eq:generalpairingwf}).

\subsubsection{Singlet Pairing Wave Function}\label{subsec:singlet}
Let us consider the first $1,2, ...,N$ electrons to be spin-up and the rest $N+1, ...,2N$ 
electrons to be spin-down and allow only $\phi({\bf r}_i,{\bf r}_j)$ in $\tilde{\phi}({\bf r}_i, \sigma_i; {\bf r}_j, \sigma_j)$ to be non-zero. 
Using the Pfaffian identity [Eq.~(\ref{eq:pfident3})], we can write 
the wave function for $N$ singlet pairs, also known as the 
Bardeen--Cooper--Schrieffer (BCS) wave function (or AGP), in the following form
\begin{equation}\label{eq:bcs1} 
\Psi_{BCS}={\rm pf}\begin{bmatrix}
0 & 
{\boldsymbol \Phi}^{\uparrow\downarrow}\\
-{\boldsymbol \Phi}^{\uparrow\downarrow T}&
0 \\
\end{bmatrix}={\rm det}[\boldsymbol \Phi^{\uparrow\downarrow}],
\end{equation}
which is simply a determinant of the $N\times N$ matrix $\boldsymbol \Phi^{\uparrow\downarrow}=\left[\phi (i,j)\right]$ as was shown
previously~\cite{bouchaud1,bouchaud3}.

Let us comment on the BCS wave function which has its origin on the theory
of superconductivity \cite{bcs}. The original form is actually for variable number 
of particles, since the theory is formulated in the grand canonical ensemble---the number of particles adjusts to the chemical potential. 
Hence, if the grand canonical form is projected onto the state with
{\em fixed } number of particles, its form is then given by Eq.~(\ref{eq:bcs1}).

It is straightforward to show that the BCS wave function contains the 
restricted HF wave function as a special case.
Let us define the Slater matrix $C=\left[\varphi_i(j)\right]$ where $\{\varphi_i\}$ is a set of HF occupied
orbitals. Then we can write 
\begin{equation}
\Psi_{HF}={\rm det}[C]{\rm det}[C]=
{\rm det}[CC^T]={\rm det}[{\boldsymbol \Phi}_{HF}^{\uparrow\downarrow}],
\end{equation}
where 
\begin{equation}
( \boldsymbol \Phi_{HF}^{\uparrow\downarrow})_{i,j}=\phi_{HF}(i,j)=\sum_{k=1}^{N}\varphi_k(i)\varphi_k(j).
\end{equation}
On the other hand, we can think of the BCS wave function as a natural generalization
of the HF one. To do so we write the singlet pair orbital as
\begin{equation}\label{eq:phi}
\phi(i,j)=\sum_{k,l}^{M>N}S_{k,l}\varphi_k(i)\varphi_l(j)=
{\boldsymbol \varphi}(i)\,{\bf S}\,{\boldsymbol \varphi}(j),
\end{equation}
where the sum runs over all $M$ (occupied and virtual) single-particle orbitals and ${\bf S}$ is some symmetric matrix. 
Therefore, we can define one-particle orbitals, which diagonalize this matrix and call them 
\emph{natural orbitals of a singlet pair}.


The BCS wave function is efficient for describing
systems with single-band correlations such
as Cooper pairs in conventional BCS
superconductors, where pairs form from
one-particle states close to the Fermi level.

\subsubsection{Triplet Pairing Wave Function}\label{subsec:triplet}
Let us assume, in our system of $2N$ electrons, that the first $M_1$ electrons are spin-up 
and remaining $M_2=2N-M_1$ electrons are spin-down. Further, we restrict $M_1$ and $M_2$ to be even numbers.  
Then by allowing only $\chi^{\uparrow \uparrow}(i,j)$ and $\chi^{\downarrow \downarrow}(i,j)$
in (\ref{eq:generalpair}) to be non-zero, we obtain from expression~(\ref{eq:generalpairingwf})
by the use of Eq.~(\ref{eq:pfident5})
\begin{equation}
\Psi_{TP}={\rm pf}\begin{bmatrix}
 {\boldsymbol \xi}^{\uparrow\uparrow} & 
0\\
0 & {\boldsymbol \xi}^{\downarrow\downarrow}\\
\end{bmatrix}={\rm pf}[{\boldsymbol \xi}^{\uparrow\uparrow}]{\rm pf}[{\boldsymbol \xi}^{\downarrow\downarrow}],
\end{equation}
where we have introduced $M_1\times M_1(M_2\times M_2)$ matrices 
${\boldsymbol \xi}^{\uparrow\uparrow(\downarrow\downarrow)}=\left[\chi^{\uparrow\uparrow(\downarrow\downarrow)}(i,j)\right]$.
This result was previously found in a weaker form
in Ref.~\cite{bouchaud1,bouchaud3}.

The connection to a restricted HF wave function for the above state can be again 
established as follows.
In accord with what we defined above,
 ${\rm det}[(C)^{\uparrow(\downarrow)}]$ are
spin-up(-down) Slater determinants of some HF orbitals $\{\varphi_i\}$.
Then, by taking advantage of  Eq.~(\ref{eq:pfident3}) we can write
\begin{align}
\Psi_{HF}&={\rm det}[C^{\uparrow}]{\rm det}[C^{\downarrow}]\\ \nonumber
&=\frac{{\rm pf}[C^{\uparrow} A_1 {C^{\uparrow}}^T]{\rm pf}[C^{\downarrow} A_2 {C^{\downarrow}}^T]}{{\rm pf}[A_1]{\rm pf}[A_2]},
\end{align}
given $A_1$ and $A_2$ are some skew-symmetric non-singular matrices.
In the simplest case, when $A_1$ and $A_2$ have block-diagonal form~(\ref{eq:pf:blockdiag}) with all values $\lambda_i=1$,
one gets
\begin{equation}
\Psi_{HF}={\rm pf}[\boldsymbol \xi_{HF}^{\uparrow\uparrow}]{\rm pf}[\boldsymbol \xi_{HF}^{\downarrow\downarrow}].
\end{equation}
The pair orbitals can be then expressed as
\begin{align}
(\boldsymbol \xi_{HF}^{\uparrow\uparrow(\downarrow\downarrow)})_{i,j}&=\chi_{HF}^{\uparrow\uparrow(\downarrow\downarrow)}(i,j)\\ \nonumber
&=\sum_{k=1}^{M_1(M_2)/2}(\varphi_{2k-1}(i)\varphi_{2k}(j)-\varphi_{2k-1}(j)\varphi_{2k}(i)).
\end{align} 
Similarly to the singlet pairing case, 
 the triplet pairing wave function appears
as a natural generalization of the HF one. We can write the triplet pair orbitals as
\begin{align}\label{eq:chi}
\chi(i,j)^{\uparrow\uparrow(\downarrow\downarrow)}&=\sum_{k,l}^{M>M_1(M_2)}A^{\uparrow\uparrow(\downarrow\downarrow)}_{k,l}\varphi_k(i)\varphi_l(j) \nonumber \\
&={\boldsymbol \varphi}(i)\,{\bf A}^{\uparrow\uparrow(\downarrow\downarrow)}\,{\boldsymbol \varphi}(j),
\end{align}
where again the sum runs over all $M$ (occupied and virtual) 
single-particle orbitals and ${\bf A}^{\uparrow\uparrow(\downarrow\downarrow)}$ are some skew-symmetric
matrices. Therefore we can define one-particle orbitals which block-diagonalize these matrices and call them 
\emph{natural orbitals of a triplet spin-up-up (down-down) pair}. 

\subsubsection{Generalized Pairing Wave Function}
Let us now consider a partially spin-polarized system with
unpaired electrons. In order to introduce both types 
of pairing we allow $\chi^{\uparrow \uparrow}(i,j)$, $\chi^{\downarrow \downarrow}(i,j)$
and $\phi(i,j)$ in (\ref{eq:generalpair}) to be non-zero. 
However, we omit the $\chi^{\uparrow \downarrow}(i,j)$ term. 
Then our usual ordered choice of electrons labels with all spin-up electrons first 
and remaining electrons spin-down enables us to directly write from (\ref{eq:generalpairingwf}) the
singlet-triplet-unpaired (STU) orbital Pfaffian wave function~\cite{pfaffianprl}
\begin{equation}\label{eg:stu}
\Psi_{STU}=
{\rm pf}\begin{bmatrix}
{\boldsymbol \xi}^{\uparrow\uparrow} & 
{\boldsymbol \Phi}^{\uparrow\downarrow} & 
{\boldsymbol\varphi}^{\uparrow} \\
-{\boldsymbol \Phi}^{\uparrow\downarrow T} &
{\boldsymbol \xi}^{\downarrow\downarrow} &
{\boldsymbol \varphi}^{\downarrow} \\
-{\boldsymbol\varphi}^{\uparrow T} &
-{\boldsymbol\varphi}^{\downarrow T} &
0 \;\; \\
\end{bmatrix},
\end{equation} 
where the bold symbols are block matrices or vectors of
corresponding orbitals as defined in Sections~\ref{subsec:singlet} and \ref{subsec:triplet} and $T$ denotes
transposition. For a spin-restricted STU wave function the
pair and one-particle orbitals of spin-up and -down channels
would be identical. 

The Pfaffian form can accommodate
both singlet and triplet pairs as well as one-particle
unpaired orbitals into a single, compact wave function.
The correspondence of STU, Pfaffian wave function to HF wave function can 
be established in a similar way for the pure singlet and triplet pairing cases.

It is instructive to write down some illustrative 
examples of the introduced form for
a few-electron atomic states. For example, for the Li atom with $^2S(1s^22s)$ we
can write 
\begin{equation}\label{eg:stuli}
\Psi_{STU}=
{\rm pf}\begin{bmatrix}
0 &
 \xi^{\uparrow\uparrow}(1,2) &
 \phi^{\uparrow\downarrow}(1,3) &
\varphi^{\uparrow}_{2s}(1) \\
 \xi^{\uparrow\uparrow}(2,1) &
0&
\phi^{\uparrow\downarrow}(2,3) &
\varphi^{\uparrow}_{2s}(2) \\
\phi^{\uparrow\downarrow}(3,1) &
\phi^{\uparrow\downarrow}(3,2) &
0 &
\varphi_{1s}^{\downarrow}(3) \\
-\varphi^{\uparrow}_{2s}(1) &
-\varphi^{\uparrow}_{2s}(2) &
-\varphi_{1s}^{\downarrow}(3)&
0 \;\; \\ \nonumber
\end{bmatrix}=
\end{equation}
\begin{equation}\label{eg:stuc}
=  \xi^{\uparrow\uparrow}(1,2)\varphi_{1s}^{\downarrow}(3) 
- \phi^{\uparrow\downarrow}(1,3)\varphi^{\uparrow}_{2s}(2)
+ \phi^{\uparrow\downarrow}(2,3)\varphi^{\uparrow}_{2s}(1). 
\end{equation}
It is easy to check that the expression has the correct antisymmetry. 
Also note that on can  get
the HF wave function by substitution 
\begin{equation}\label{eg:stuli2}
\phi^{\uparrow\downarrow}(i,j)=\varphi^{\uparrow}_{1s}(i)\varphi^{\downarrow}_{1s}(j), \;\;\;
\xi^{\uparrow\uparrow}(i,j)=0
\end{equation}
 or, alternatively,  by setting
\begin{equation}\label{eg:stuli3}
\phi^{\uparrow\downarrow}(i,j)=0, \;\;\;
 \xi^{\uparrow\uparrow}(i,j)=\varphi^{\uparrow}_{1s}(i)\varphi^{\uparrow}_{2s}(j)
-\varphi^{\uparrow}_{1s}(j)\varphi^{\uparrow}_{2s}(i).
\end{equation}
For the valence part 
of the carbon atom (we skip the He-core), 
we can write the state as $^3P(2s^22p^2)$ 
with the corresponding wave function
\begin{equation}\label{eg:stuli}
\Psi_{STU}=
{\rm pf}\begin{bmatrix}
0 &
 \xi^{\uparrow\uparrow}(1,2) &
 \xi^{\uparrow\uparrow}(1,3) &
 \phi^{\uparrow\downarrow}(1,4) \\
 \xi^{\uparrow\uparrow}(2,1) &
0&
 \xi^{\uparrow\uparrow}(2,3) &
\phi^{\uparrow\downarrow}(2,4) \\
 \xi^{\uparrow\uparrow}(3,1) &
 \xi^{\uparrow\uparrow}(3,2) &
0 &
\phi^{\uparrow\downarrow}(3,4) \\
-\phi^{\uparrow\downarrow}(4,1) &
-\phi^{\uparrow\downarrow}(4,2) &
-\phi^{\uparrow\downarrow}(4,3) &
0 \;\; \\
\end{bmatrix}.
\end{equation}
Finally, for the nitrogen atom valence electrons the quartet state given by $^4S(2s^22p^3)$ and the wave function can be written as
\begin{equation}\label{eg:stun}
\Psi_{STU}=
{\rm pf}\begin{bmatrix}
0 &
 \xi^{\uparrow\uparrow}(1,2) &
 \xi^{\uparrow\uparrow}(1,3) &
 \xi^{\uparrow\uparrow}(1,4) &
 \phi^{\uparrow\downarrow}(1,5) &
 \varphi^{\uparrow}(1) \\
 \xi^{\uparrow\uparrow}(2,1) &
0&
 \xi^{\uparrow\uparrow}(2,3) &
 \xi^{\uparrow\uparrow}(2,4) &
\phi^{\uparrow\downarrow}(2,5) &
 \varphi^{\uparrow}(2) \\
 \xi^{\uparrow\uparrow}(3,1) &
 \xi^{\uparrow\uparrow}(3,2) &
0 &
 \xi^{\uparrow\uparrow}(3,4) &
\phi^{\uparrow\downarrow}(3,5) &
 \varphi^{\uparrow}(3) \\
 \xi^{\uparrow\uparrow}(4,1) &
 \xi^{\uparrow\uparrow}(4,2) &
 \xi^{\uparrow\uparrow}(4,3) &
0 &
 \phi^{\uparrow\downarrow}(4,5) &
 \varphi^{\uparrow}(4) \\
-\phi^{\uparrow\downarrow}(5,1) &
-\phi^{\uparrow\downarrow}(5,2) &
-\phi^{\uparrow\downarrow}(5,3) &
-\phi^{\uparrow\downarrow}(5,4) &
0 &
\varphi^{\downarrow}(5)\\
-\varphi^{\uparrow}(1) &
-\varphi^{\uparrow}(2) &
-\varphi^{\uparrow}(3) &
-\varphi^{\uparrow}(4) &
-\varphi^{\downarrow}(5)&
0 \;\; \\
\end{bmatrix}.
\end{equation}

\subsection{Applications of Pairing Wave Functions  in QMC}\label{sec:pf:results}
In the remainder of this section, we would like to illustrate the applications of Pfaffian paring wave functions in QMC methods.
As we have mentioned before in Sec.~\ref{subsec:twf}, the trial wave function used in QMC calculations
by variational and fixed-node diffusion Monte Carlo is a product of an antisymmetric part $\Psi_A$
times the Jastrow correlation factor:
\begin{equation}
\Psi_{T} ({\bf R}) = \Psi_{A}({\bf R}) \exp[U_{corr}(\{r_{ij}\},\{r_{iI}\},\{r_{jI}\})],
\end{equation}
where  $U_{corr}$ depends on
electron-electron, electron-ion and, possibly, on 
electron-electron-ion combinations of distances.
The antisymmetric part can be represented as $\Psi_A=\Psi_{HF}$ and $\Psi_A=\Psi_{STU}$
or $\Psi_A=\Psi_{BCS}$.
Further, the pair orbitals $\chi(i,j)$ and $\phi(i,j)$ are expanded in products of a one-particle orbital
basis~\cite{sorellabcs1} according to Eqs.~(\ref{eq:phi}) and (\ref{eq:chi}). 
This approach was particularly convenient due to availability of one-particle atomic and molecular orbitals from either Hartree--Fock 
or CI correlated calculations. However, expansion of this type might not be the most optimal for homogeneous or extended periodic systems 
and other forms were also successfully employed~\cite{carlsonbcs}. For the applications below, we 
typically used about 10 virtual orbitals. The natural orbitals (see e.g. Ref.~\cite{szabo}) produced better and more systematic
results than the HF ones. The pair orbital expansion coefficients were then optimized
in VMC by minimizations of energy, variance or a combination
of energy and variance (for details, see Sec.~\ref{subsec:opt3}). 
The optimization procedure required the calculation of gradient and the Hessian of the wave function 
according to Eqs.~(\ref{eq:gradient}) and (\ref{eq:hessian}).

\subsubsection{Single Pfaffian Calculations}\label{subsec:pf:spf}
The performance of single Pfaffian pairing wave functions in the STU form~[Eq.~(\ref{eg:stu})] was tested 
on several first row atoms and dimers~\cite{pfaffianprl}. The pseudopotentials~\cite{lester,stevens} 
were used to eliminate the atomic core electrons, except in Be atom case. 

\begin{figure}[!ht]
\centering
\includegraphics[width=\columnwidth]{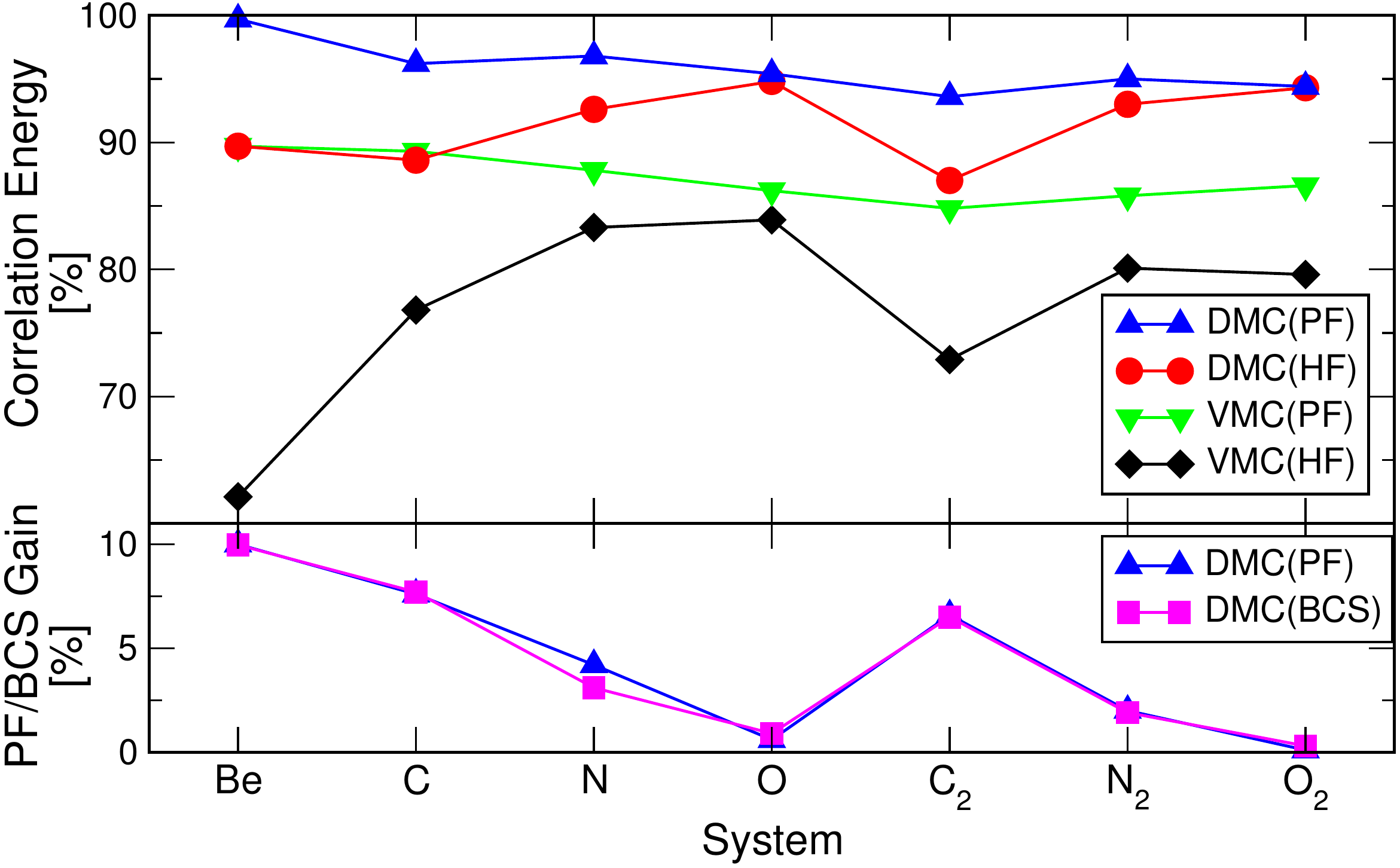}
\caption{
Correlation energies obtained by QMC methods
with the different trial wave functions: VMC
and fixed-node DMC with HF nodes (HF) and STU Pfaffian nodes (PF).
The lower plot shows the fixed-node DMC correlation energy gains
over HF nodes for BCS and STU Pfaffian
wave functions.
The statistical error bars are of the symbol sizes or smaller.
Except for the Be atom all the calculations used the same pseudopotentials~\cite{lester,stevens}.
\label{fig:singlepfresults}}
\end{figure}

The most important result is that STU wave functions  
systematically achieve higher percentage of recovered
correlation energy than single determinant wave functions 
in the fixed-node DMC method (see Table \ref{energies:1} and Fig.~\ref{fig:singlepfresults}).
Another significant result is that in general the triplet contribution for
these single Pfaffian STU wave functions are small, perhaps the only
exception being the nitrogen atom, where we see a
gain of  additional  1\% in correlation energy when compared
to a trial wave function without triplet pairs.
We believe, this is due to the fact, that  
ground state of nitrogen atom has a quartet spin state and 
therefore the highest spin polarization from all studied cases. 
Overall, the single Pfaffian form is 
capable of capturing near-degeneracies
and mixing of excited states for both
spin-polarized and unpolarized systems.

\begin{table}[!t]
\caption{ Total energies for C, N and O atoms and their dimers with 
amounts of the correlation energy recovered in VMC and DMC methods with wave functions 
as discussed in the text. Unless noted otherwise, 
the numbers in parentheses are the statistical errors in the last digit from corresponding QMC calculation.
Energies are in Hartree atomic units.
For C, N, O atoms we used the correlation energies by Dolg~\cite{dolgcpl}(0.1031, 0.1303, 0.1937 H).
For the estimation of correlation energies of dimers we needed accurate HF energies at experimental distances~\cite{NIST-JANAF}
and the estimated exact total energies. 
Each exact total energy was estimated as a sum of total energies of constituent atoms minus experimental binding energy~\cite{NIST-JANAF,UBJ,HH}
adjusted for experimental zero-point energy~\cite{HH}.} 
\begin{minipage}{\columnwidth}
\renewcommand{\thefootnote}{\alph{footnote}}
\renewcommand{\thempfootnote}{\alph{mpfootnote}}
\centering
\begin{tabular}{l c c c c c c }
\hline
\hline
 \multicolumn{1}{l}{Method/WF}&\multicolumn{1}{c}{C}&\multicolumn{1}{c}{E$_{corr}$[\%]} & \multicolumn{1}{c}{N} &\multicolumn{1}{c}{E$_{corr}$[\%]}&\multicolumn{1}{c}{O}& \multicolumn{1}{c}{E$_{corr}$[\%]}\\
\hline 
HF                & -5.31471   &   0     & -9.62892   & 0       & -15.65851   &  0      \\[-0.2em]
VMC/HF            & -5.3939(4) & 76.8(4) & -9.7375(1) & 83.3(1) & -15.8210(6) & 83.9(3) \\[-0.2em]
VMC/BCS           & -5.4061(2) & 88.6(2) & -9.7427(3) & 87.3(2) & -15.8250(3) & 86.0(2) \\[-0.2em]
VMC/STU           & -5.4068(2) & 89.3(2) & -9.7433(1) & 87.8(1) & -15.8255(3) & 86.2(2) \\[-0.2em]
DMC/HF            & -5.4061(3) & 88.6(2) & -9.7496(2) & 92.6(2) & -15.8421(2) & 94.8(1) \\[-0.2em]
DMC/BCS           & -5.4140(2) & 96.3(2) & -9.7536(2) & 95.7(2) & -15.8439(4) & 95.7(2) \\[-0.2em]
DMC/STU           & -5.4139(2) & 96.2(2) & -9.7551(2) & 96.8(1) & -15.8433(3) & 95.4(2) \\[-0.2em]
Est./Exact        & -5.417806  & 100     & -9.759215  & 100     & -15.85216   & 100     \\
\hline 
\multicolumn{1}{l}{Method/WF} &  \multicolumn{1}{c}{C$_2$} &  \multicolumn{1}{c}{E$_{corr}$[\%]} &  \multicolumn{1}{c}{N$_2$} &  \multicolumn{1}{c}{E$_{corr}$[\%]} &  \multicolumn{1}{c}{O$_2$} &  \multicolumn{1}{c}{E$_{corr}$[\%]} \\
\hline 
HF                & -10.6604    & 0       & -19.4504   &  0       & -31.3580    & 0      \\[-0.2em]
VMC/HF            & -10.9579(4) & 72.9(1) & -19.7958(5)&  80.0(1) & -31.7858(6) & 79.6(1)\\[-0.2em]
VMC/BCS           & -11.0059(4) & 84.7(1) & -19.8179(6)&  85.0(1) & -31.8237(4) & 86.7(1)\\[-0.2em]
VMC/STU           & -11.0062(3) & 84.8(1) & -19.821(1) &  85.8(2) & -31.8234(4) & 86.6(1)\\[-0.2em]
DMC/HF            & -11.0153(4) & 87.0(1) & -19.8521(3)&  93.0(1) & -31.8649(5) & 94.3(1)\\[-0.2em]
DMC/BCS           & -11.0416(3) & 93.5(1) & -19.8605(6)&  94.9(1) & -31.8664(5) & 94.6(1)\\[-0.2em]
DMC/STU           & -11.0421(5) & 93.6(1) & -19.8607(4)&  95.0(1) & -31.8654(5) & 94.4(1)\\[-0.2em]
Est./Exact\footnotemark[3]  & -11.068(5)\footnotemark[1]  & 100.0(10)  & -19.8825(6)\footnotemark[2]   &  100.0(1)     & -31.8954(1)\footnotemark[2]    & 100.0(1)    \\
\hline
\hline
\end{tabular}
\footnotetext[1] { 
There is rather large discrepancy in the experimental values of C$_2$ binding energy
($141.8(9)$~\cite{NIST-JANAF}, $143(3)$~\cite{HH} and $145.2(5)$ kcal/mol~\cite{UBJ}). 
For the estimation of exact energy we have taken the average value of $143(3)$ kcal/mol.}
\footnotetext[2] {Experimental binding energies taken from ref.~\cite{NIST-JANAF}.}
\footnotetext[3] {The error bars on estimated exact total energies are due to experiment.}
\end{minipage}
\label{energies:1}
\end{table}


\subsubsection{Multi-Pfaffian Calculations}\label{subsec:pf:mpf}
In order to capture the correlation energy missing from the single STU Pfaffian wave function and to 
test the limits of the Pfaffian functional form, we have also proposed to expand the $\Psi_A$ in 
the linear combination of STU Pfaffian wave functions~\cite{pfaffianprl, pfaffianprb}.  

Following the approach adopted from the CI correlated calculations, accurate $\Psi_A$ can be expressed as 
a linear combination of reference state $\Psi_0$ and single $\Psi_i^k$ and double$\Psi_{ij}^{kl}$  excitations:
\begin{align}
  \Psi_{CISD} =c_0 \Psi_0 +\sum^N_i \sum^M_k  c_i^k \Psi_i^k + \sum^N_{ij} \sum^M_{kl} c_{ij}^{kl} \Psi_{ij}^{kl}, 
\end{align} 
where $i$-th and $j$-th electrons are being excited into $k$ and $l$ virtual orbitals. Note that the 
the number of determinants in the expansion will be of the order
of $N^2\times M^2$. 
Analogously, we postulate the multi-Pfaffian (MPF) wave function as a linear combination of STU Pfaffians:
\begin{align}
  \Psi_{MPF} = c_0 \Psi_0 +\sum_{i}^N{\rm pf}[\tilde\phi_{i}]+\sum_{ij}^N{\rm pf}[\tilde\phi_{ij}], 
\end{align} 
where each $\tilde\phi_{ij}$ is the generalized paring orbital [Eq.~(\ref{eq:generalpair})] containing 
all possible $M^2$ excitations of $i$ and $j$ electrons. The resulting wave function
will in general consist of only $N^2$ Pfaffians. 
Further, if the reference state $\Psi_0$ is the most dominant state (i.e., $c_0 \gg c_{ij}^{kl}$), 
it is possible to show by expanding $\tilde\phi_{ij}$ into orders of $c_{ij}^{kl}/c_0$ that 
\begin{align}
  \Psi_{MPF} =  \Psi_{CISD} + {\mathcal O}\left(({c_{ij}^{kl}}/c_0)^2\right).
\end{align} 
In fact, the only difference between the $\Psi_{MPF}$ and $\Psi_{CISD}$ wave functions is in presence of the higher order 
excitations, which are approximately present in the $\Psi_{MPF}$. 

The mapping of the MPF wave functions onto equivalent CISD wave functions was also verified numerically 
in the variational and FN-DMC methods using the above first row atoms and molecules~\cite{pfaffianprl, pfaffianprb}.
The results in Table~\ref{energies2} show that for the atomic systems the MPF wave functions are able to
recover close to  99\% of correlation energy---very similarly to CISD wave functions, while requiring order less terms. 
The results for diatomic cases (see Table~\ref{energies3}) show very similar behavior. 
The correlation energy recovered is on the order 98\% with MPF wave functions closely matching the CISD wave functions
---despite much richer electronic structure than in atomic cases.
The comparison with the CI results therefore demonstrates that it is possible to obtain similar quality wave functions 
with corresponding improvements of the fermion nodes at much smaller calculational cost.
\begin{table}
\caption{Percentages of correlation energies recovered 
for C, N and O atoms by VMC and DMC methods 
with wave functions as discussed in the text.
The corresponding number of Pfaffians or determinants $n$ for each
wave function is also shown. For details, see caption of Table~\ref{energies:1}.}
\begin{minipage}{\columnwidth}
\renewcommand{\thefootnote}{\alph{footnote}}
\renewcommand{\thempfootnote}{\alph{mpfootnote}}
\centering
\begin{tabular}{l c c c c c c}
\hline
\hline
\multicolumn{1}{l}{Method/WF}&
\multicolumn{1}{c}{$n$}&\multicolumn{1}{c}{C} &\multicolumn{1}{c}{$n$}&\multicolumn{1}{c}{N}& \multicolumn{1}{c}{$n$}&\multicolumn{1}{c}{O}\\
\hline 
VMC/MPF & 3   & 92.3(1) & 5   & 90.6(1) &  11  & 92.6(3) \\ 
VMC/CI\footnotemark[1]  & 98  & 89.7(4) & 85  & 91.9(2) & 136  & 89.7(4) \\ 
DMC/MPF & 3   & 98.9(2) & 5   & 98.4(1) &  11 &  97.2(1) \\ 
DMC/CI\footnotemark[1]  & 98  & 99.3(3) & 85  & 98.9(2) & 136  & 98.4(2) \\
\hline
\hline
\end{tabular}
\footnotetext[1]{The determinantal weights were taken directly from CI calculation without re-optimization in VMC.}
\end{minipage}
\label{energies2}
\end{table}

\begin{table}
\caption{Total energies for C$_2$ and N$_2$ dimers with 
amounts of correlation energy recovered in VMC and DMC methods
with wave functions as discussed in the text. Energies are in Hartree atomic units.
The corresponding number of Pfaffians or determinants $n$ for each
wave function is also shown. For details, see caption of Table~\ref{energies:1}.}
\begin{minipage}{\columnwidth}
\renewcommand{\thefootnote}{\alph{footnote}}
\renewcommand{\thempfootnote}{\alph{mpfootnote}}
\centering
\begin{tabular}{l c c c c c c}
\hline 
\hline
\multicolumn{1}{l}{Method/WF} & \multicolumn{1}{c}{$n$} & \multicolumn{1}{c}{C$_2$} &  \multicolumn{1}{c}{E$_{corr}$[\%]} &  \multicolumn{1}{c}{$n$} &  
\multicolumn{1}{c}{N$_2$} & \multicolumn{1}{c}{E$_{corr}$[\%]}\\
\hline  
VMC/MPF\footnotemark[3] & 11   &  -11.0402(1) & 93.1(1)   &  16 &  -19.8413(6)   &  90.5(1)\\
VMC/CI\footnotemark[1]  & 404  &  -11.0409(3) & 93.3(1)   & 535  &  -19.8487(6)   & 92.2(1) \\
DMC/MPF\footnotemark[3] &  11   &  -11.0574(5) & 97.3(1)  &  16   & -19.8670(8)  &  96.4(2) \\
DMC/CI\footnotemark[1]  & 404  &  -11.0593(6)\footnotemark[2]  &  97.8(2) & 535  &  -19.8713(5)    &  97.4(1)\\
\hline
\hline
\end{tabular}
\footnotetext[1]{The determinantal weights were re-optimized in the VMC method.}
\footnotetext[2]{Recently, Umrigar~{\it et.al.}~\cite{umrigarC2} published very accurate 
DMC result for fully optimized CI wave function with up to 500 determinants for C$_2$ molecule. 
The resulting well-depth of his calculation is $6.33(1)$ eV, which is only $0.03$ eV form estimated exact value of Ref.~\cite{bytautas}.
The well-depth resulting from our DMC/CI energy of $-11.0593(6)$ H equals to $6.08(3)$ eV.}
\footnotetext[3]{one type of triplet-like excitation not included.}
\end{minipage}
\label{energies3}
\end{table}

\subsubsection{Nodal Properties}\label{subsec:pf:nodes}
\begin{figure}[!t]
  \centering
  \begin{tabular}{c}
   \mbox{
       {\resizebox{0.3\columnwidth}{!}{\includegraphics{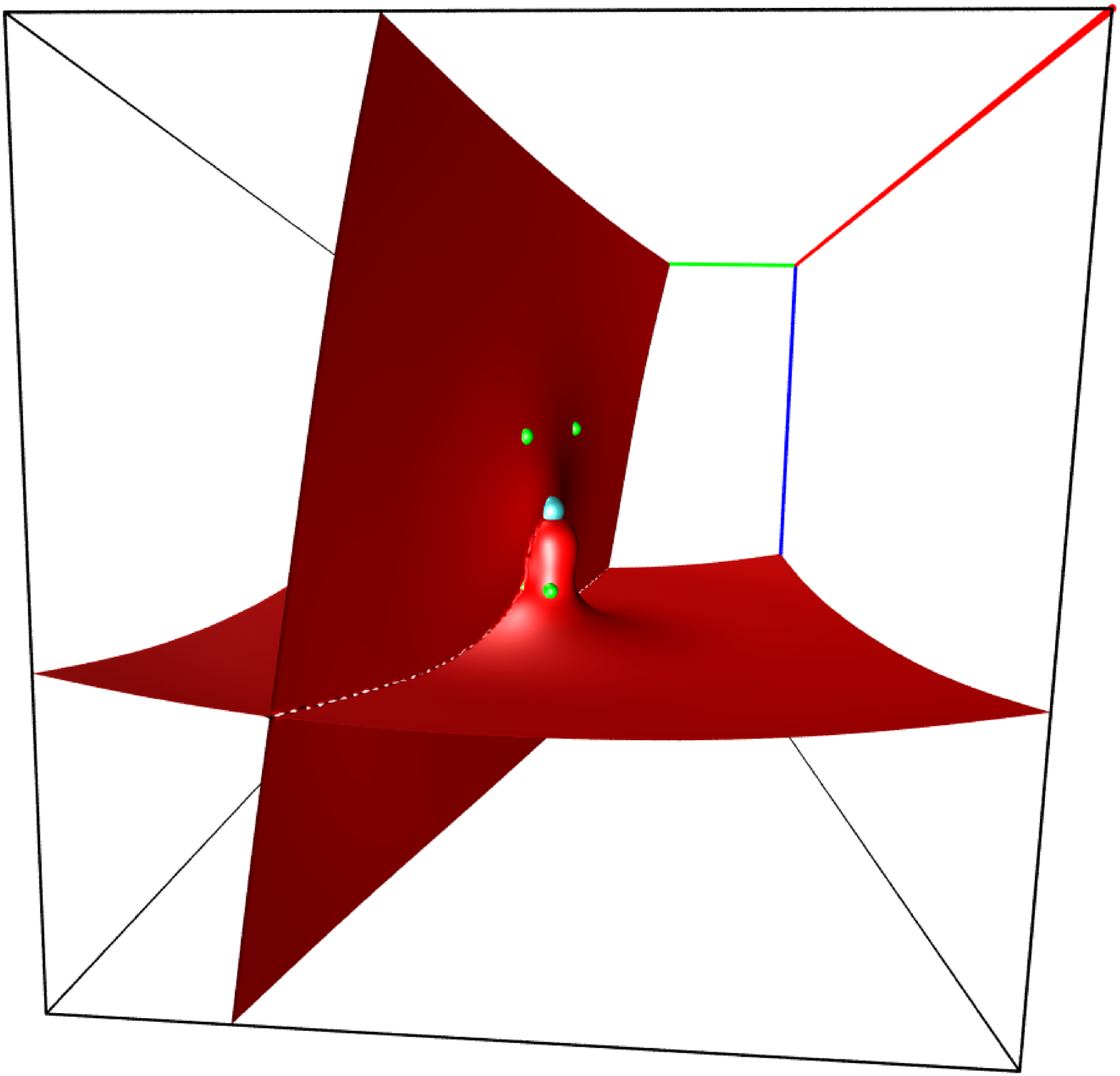}}}
       {\resizebox{0.3\columnwidth}{!}{\includegraphics{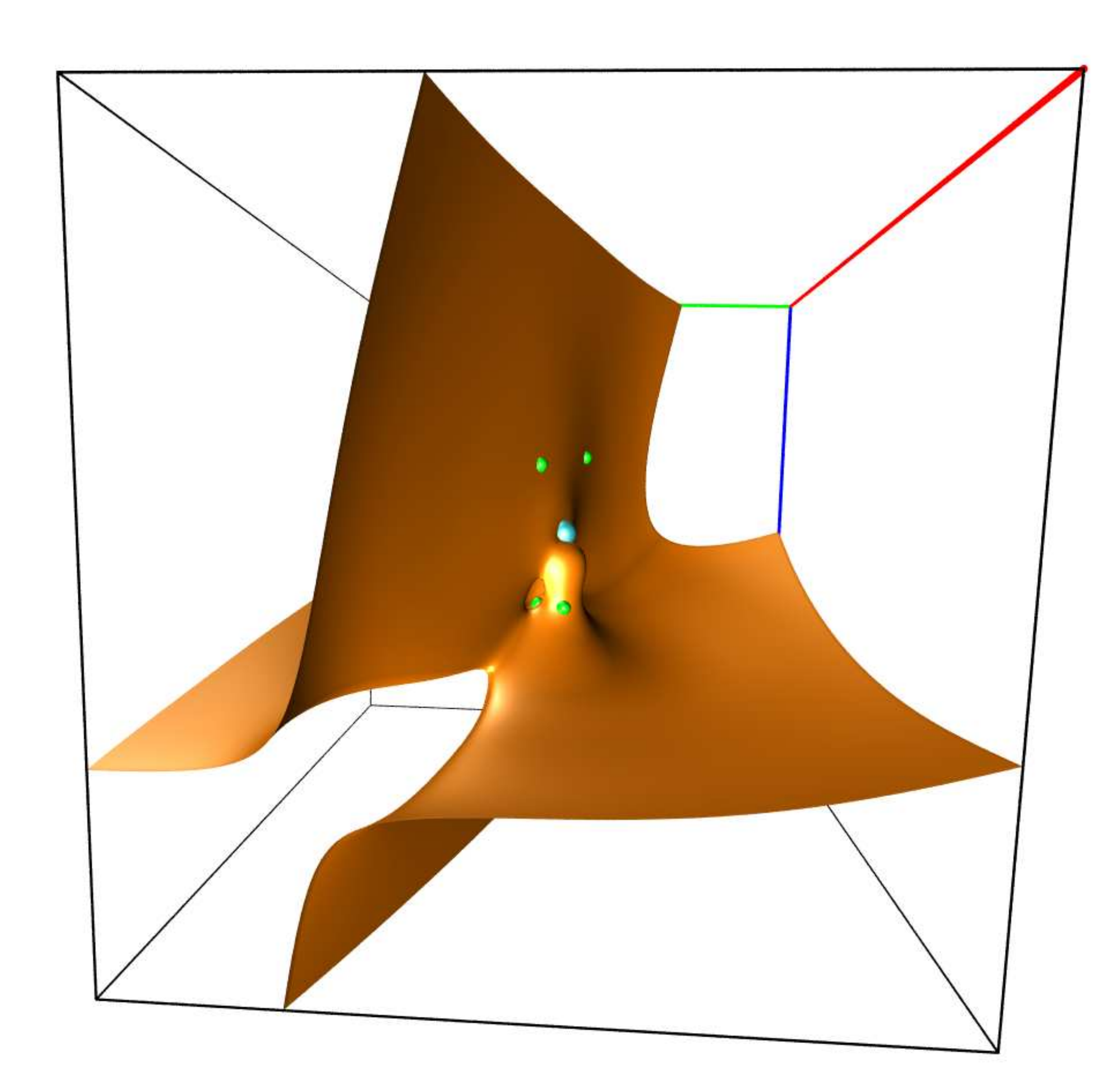}}} 
       {\resizebox{0.3\columnwidth}{!}{\includegraphics{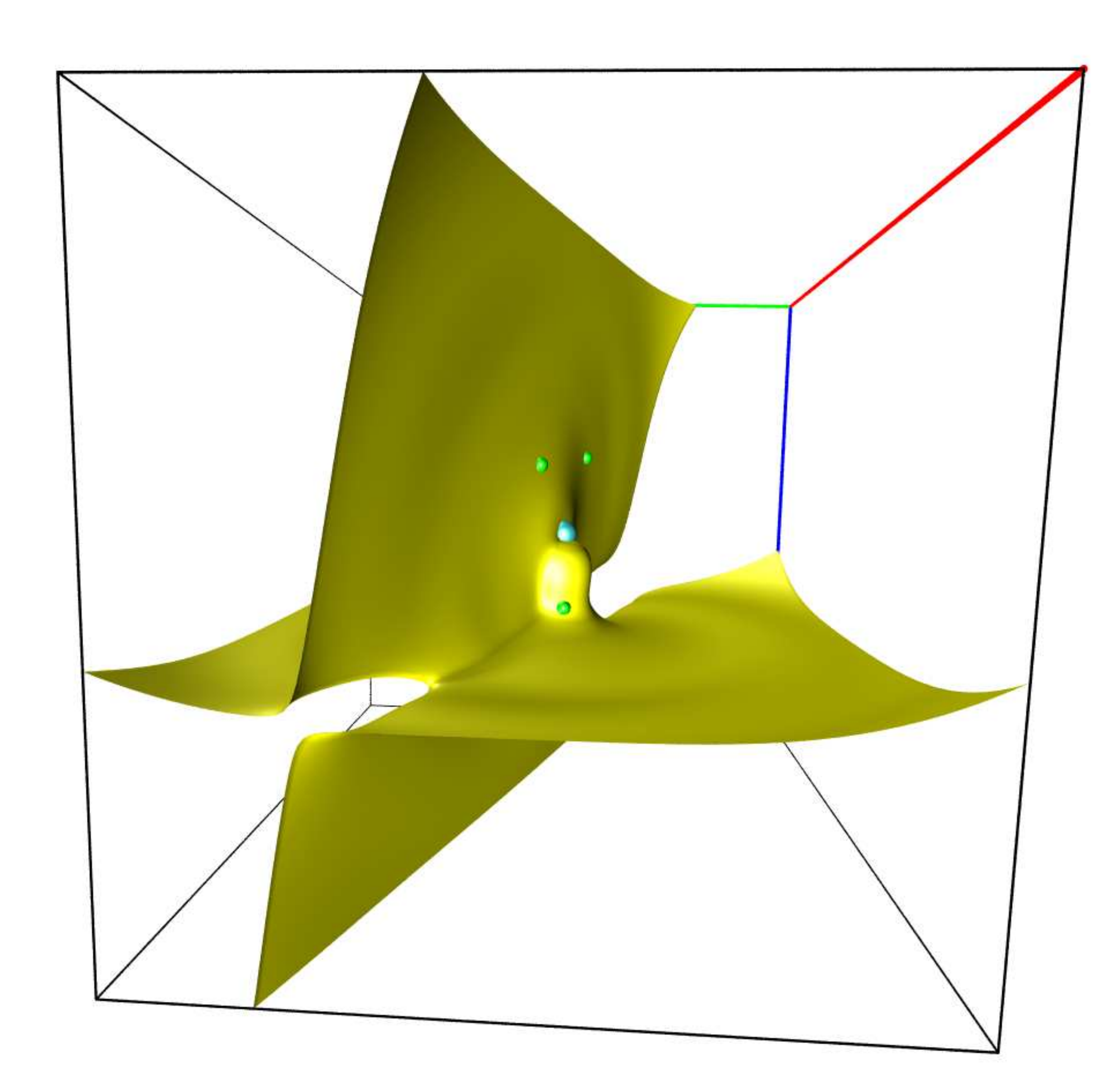}}}
      } 
 \end{tabular}
\caption{
A three-dimensional cut through the fermion node hypersurface of oxygen atom 
obtained by scanning the wave function with a spin-up and -down (singlet)
pair of electrons at the equal positions, while keeping the 
rest of electrons at a given VMC snapshot positions (small green spheres). 
Nucleus is depicted in the center of the cube by the blue 
sphere. The three colors (from left to right) show nodes of: Hartree--Fock (red/dark gray);
Multi-Pfaffian nodes (orange/medium gray); and the nodes of the CI
wave function (yellow/light gray).
The CI nodal surface is very close to the exact one (see text).
The HF node clearly divides the space into four
nodal cells while Pfaffian and CI wave functions
partitioning leads to the minimal number of two nodal cells.
The changes in the nodal topology occur on the appreciable spatial 
scale of the order of 1 a.u.}
\label{fig:nodes2}
\end{figure}
As we have already extensively discussed in Sec.~\ref{ch:nodes}, 
the fermion node is a complicated hypersurface defined by an implicit equation $\Psi_A(R)=0$. 
Due to its complexity, we approximate the exact fermion node with the node of the best available trial wave function. 
However, any deviations from the exact node lead to the fixed-node DMC errors. 
It is therefore quite valuable to compare the nodes of a different accuracy.

The Fig.~\ref{fig:nodes2} shows the example of nodal structure of oxygen atom 
($1s^2$ electrons were removed by pseudopotential). Here we compare the nodal surfaces of 
HF (no pairing), MPF Pfaffian (STU pairing) and a high accuracy CI wave function with more than 3000 determinants, which gives 
essentially exact fermion nodes (i.e., $99.8(3)\%$ of correlation energy in fixed-node DMC).
From the figure, it is clear that the changes in the nodal surfaces are significant---the most 
important one being the elimination of artificial four nodal cells resulting from the independence of spin-up and -down
channels in HF. The Pfaffian smooths-out the crossings and fuses
the compartments of the same sign into the single ones. These topology changes
therefore lead to the minimal number of two nodal cells (for details see Sec.~\ref{ch:nodes}).
However, the MPF Pfaffian nodes are noticeably different close to nucleus and in the tails and 
the amount of missing correlation energy is still non-negligible. 

\subsection{\label{sec:level5} Conclusions}
The proposed Pfaffians with
singlet and triplet pairs together with unpaired orbitals provides
a variationally rich and compact wave functions.
They offer significant and systematic improvements
over commonly used Slater determinant-based wave functions
and can be calculated as easily as determinants using the 
presented formulas. 
We have also shown relationships between HF, BCS(AGP) and Pfaffian
wave functions. Clearly,
Pfaffian pairing wave functions are able to capture a large fraction of
missing correlation energy with consistent treatment 
of both spin-polarized and unpolarized pairs.
We have explored Pfaffian expansions which enabled us to capture additional correlation
and map them onto the equivalent CISD wave functions.
Finally, the gains in correlation energy for Pfaffians come from
improved fermion nodes, which are very close to the exact ones and exhibit 
the correct topology with the minimal number of two nodal cells.

%% file: backflow/backflow.tex
\section{Backflow Correlations in Slater and Pfaffian Wave Functions}\label{ch:bf}
\subsection{Introduction}
Another approach for improvement of the nodal accuracy of variational trial wave functions 
is to employ backflow correlations. They were first introduced by Feynman and Cohen~\cite{feynman}
as a simple ansatz to describe correlations and excitations in liquid $^4$He. 
Since then, a number of authors~\cite{schmidt_bf,panoff,moskowitz,kwon1,kwon2,kwon3}
showed that the backflow correlations are useful also in fermionic systems. 
Recently, the backflow transformation of electron coordinates has been applied to 
chemical (e.i., inhomogeneous) systems~\cite{drummond_bf,rios_bf, gurtubay_bf, brown_bf}. 
This section contains detailed description of inhomogeneous backflow transformation for multi-determinantal and Pfaffian wave functions 
and their applications. 

Let us try to briefly elucidate where the motivation for the backflow comes from.
Consider the trial function of the familiar form which we used a number of times before
$\Psi_T({\bf R})={\det}[...]e^{U_{corr}}$, and let us apply the projector $\exp(-\tau{\mathcal H})$ as given by  
\begin{align}
\exp(-\tau {\mathcal H})\Psi_T({\bf R})= \Psi_T({\bf R}) -\tau{\mathcal H}\Psi_T({\bf R})
+{\mathcal O}(\tau^2)
\end{align}
In order to analyze the first order term we re-arrange the terms in the Hamiltonian
\begin{align}
{\mathcal H} =T +V_{ext}+V_{ee}= T+(V_{ext}+V_{mf}) + (V_{ee}-V_{mf}),
\end{align}
where $V_{mf}$ is a one-particle mean-field approximation for $V_{ee}$ which corresponds 
to the theory used for generation of orbitals.
A closer look on the first order term then leads to the following
\begin{align}
{\mathcal H} {\det}[...]e^{U_{corr}}=&e^{U_{corr}}(T+V_{ext}+V_{mf}) {\det}[...] 
+{\det}[...] (T+V_{ee}-V_{mf}) e^{U_{corr}} \nonumber \\
&-\nabla e^{U_{corr}} \cdot \nabla  {\det}[...].
\end{align}
Note that the first two terms should be close to $\Psi_T({\bf R})E_{loc}({\bf R})$,
 where $E_{loc}({\bf R})$ is some function which is not very far from a constant. The reason is 
that in the first term the determinant solves the
corresponding one-particle Hamiltonian, while in the second term the correlation is optimized to cancel out
the difference $V_{ee}-V_{mf}$. 
 The third term appears to be ``spurious'' since it is not compensated by any potential.
It has a form of the dot product of two ``fluxes'', one resulting from $V_{ext}+V_{mf}$ while the other
from the electron-electron interactions. These terms need to be ``cancelled'' out what
requires modifying the wave function or including additional terms.
Which types of terms would be effective? We consider two limiting cases. The first one corresponds to
\begin{align}
|\nabla  {\det}[...]| \gg | \nabla e^{U_{corr}}  |
\end{align}
with that the ``flux'' from the determinant having much larger amplitude so that the system is 
strongly inhomogeneous---typical behaviour for chemical systems and solid state materials. For these cases the 
excited determinants, Pfaffians, etc, appear as the most obvious, if not the most efficient, way
to cancel out the spurious component.
On the other hand, if the inequality goes the other way
\begin{align}
|\nabla  {\det}[...]| \ll  | \nabla e^{U_{corr}}  |
\end{align}
then the system has density close to a constant or only mildly varying so that the electron-electron term
dominates. This is typical for homogeneous electron gas, quantum liquids such $^4$He and $^3$He, and similar
systems. Feynman and Cohen showed that in this case the backflow coordinates can partially compensate for the spurious
term, basically by slightly displacing a given particle to reflect the influence of particles which happen to be nearby.

\subsection{Backflow Wave Function Form}
As already discussed in Sec.~\ref{subsec:twf}, our trial wave function has the form
\begin{align}
\Psi_T({\bf R})=\Psi_A({\bf X})\times \exp[U_{corr}({\bf R})],
\end{align}
where ${\bf X}=({\bf x}_1,\ldots,{\bf x}_N)$ represents some quasi-particle coordinates 
dependent on all $N$ electron positions ${\bf R}$. Further, $\Psi_A$
is either (multi)-determinantal or Pfaffian wave function and $U_{corr}$
is the Jastrow correlation factor both defined in the previous sections. 

The quasi-coordinate of $i$th electron at position ${\bf r}_i$ is given as 
\begin{align}
{\bf x}_i&={\bf r}_i+{\boldsymbol \xi}_i({\bf R}) \nonumber \\
&={\bf r}_i+{\boldsymbol \xi}_i^{en}({\bf R})+{\boldsymbol \xi}_i^{ee}({\bf R})+{\boldsymbol \xi}_i^{een}({\bf R}),
\end{align}
where ${\boldsymbol \xi}_i$ is the $i$th electron's backflow displacement 
divided to the contributions from one-body (electron-nucleus), two-body (electron-electron) 
and three-body (electron-electron-nucleus) terms. 
They can be further expressed as 
\begin{align}\label{eg:bfterms}
{\boldsymbol \xi}_i^{en}({\bf R})&=\sum_I \chi(r_{iI}) {\bf r}_{iI} \\
{\boldsymbol \xi}_i^{ee}({\bf R})&=\sum_{j\ne i} u(r_{ij}) {\bf r}_{ij} \label{eg:bfterms2} \\
{\boldsymbol \xi}_i^{een}({\bf R})&=\sum_I \sum_{j\ne i} [w_1(r_{ij},r_{iI},r_{jI}) {\bf r}_{ij} + w_2(r_{ij},r_{iI},r_{jI}) {\bf r}_{iI}],
\end{align}
where ${\bf r}_{ij}={\bf r}_i-{\bf r}_j$ and ${\bf r}_{iI}={\bf r}_i-{\bf r}_I$. The $\chi$, $u$ and $w_1$ with $w_2$
terms are similar to one, two and three-body Jastrow terms and are further expanded 
as 
\begin{align}
\chi(r)&=\sum_k c_k a_k(r), \\
u(r)&=\sum_k d_k b_k(r), \\
w_{1,2}(r_{ij},r_{iI},r_{jI})&=\sum_{klm} g_{klm} a_k(r_{iI})a_l(r_{jI})b_m(r_{ij}). 
\end{align}
The one dimensional basis functions $\{a\}$ and $\{b\}$ are chosen as Gaussians with the center in origin or as polynomial Pad\' e functions (Sec.~\ref{appendix:functions}) to 
preserve the electron-electron [Eqs.~(\ref{eq:cupsupup}) and~(\ref{eq:cupsupdown})] and electron-nucleus [Eq.~(\ref{eq:necusp})] cusp conditions.
The set of variational parameters $\{c\}$, $\{d\}$ and $\{g\}$ is minimized with respect to 
mixture of energy and variance (for details, see Sec.~\ref{subsec:opt3}). In addition, all electron-electron 
coefficients ($\{d_k\}$ and $\{g_{klm}\}$ with fixed $k$ and $l$) are allowed to be different for spin-like and for spin-unlike electron pairs. 

\subsection{Applications of Backflow Correlated Wave Functions}
In the remainder of this section we present VMC and DMC results obtained with above implementation of backflow 
correlations for determinant and Pfaffian wave functions.

\subsubsection{Homogeneous electron gas}
We benchmark our implementation of the backflow correlations on the homogeneous electron gas (HEG) system. 
For comparison purposes, we choose to study the HEG system of 54 unpolarized electrons in the simple cubic simulation cell with periodic boundary conditions 
(the system was studied before several times~\cite{kwon4,markus,rios_bf}). Due to the homogeneity of the problem, 
the backflow displacement in Eq.~(\ref{eg:bfterms2}) has only one non-zero component: ${\boldsymbol \xi}_i^{ee}$. 
We conveniently choose $u(r)$ to be expanded in the basis of several polynomial Pad\' e functions to preserve cusp conditions 
with cutoff equal to half of the simulation cell. Finally, our Jastrow correlation factor contains only electron-electron 
terms.

In Table ~\ref{table:heg} we show our results for the following three electron densities of $r_s=1$, $5$ and $20$. 
The First important result is that at the normal density ($r_s=1$), the size of correlation (e.i., $E_{DMC}-E_{HF}$) is relatively small, 
but at lower densities (larger $r_s$), the correlation energy is larger fraction of total energy and becomes more important. 
Second, we observe that the HF and Slater--Jastrow (SJ) fixed-node DMC energies are in very good agreement with  previous results~\cite{kwon4,markus,rios_bf}.
However, due to the omission of higher order correlations from Jastrow factor and also from backflow displacement, we obtain 
the expected higher VMC energies and variances for SJ and backflow displaced SJ (SJBF) trial wave functions. 
Nevertheless, our fixed-node DMC energies for SJBF trial wave functions closely match the results of Kwon {\it et al.\/}\cite{kwon4}, and 
only slightly deviate at $r_s=20$ from results of Rios {\it et al.\/}\cite{rios_bf}.
 
\begin{table}
\centering
\caption{VMC and fixed-node DMC energies per electron and variances of local energies for various 
trial wave functions (S, Slater; SJ, Slater--Jastrow; SJBF, backflow correlated SJ) for 3D unpolarized HEG of 54 electrons.}
\begin{tabular}{l l l c c }
\hline 
\hline
\multicolumn{1}{l}{$r_s$} & \multicolumn{1}{l}{Method} & \multicolumn{1}{l}{WF}& \multicolumn{1}{c}{$E/N$[a.u./electron]} & \multicolumn{1}{c}{$\sigma^2$[a.u.$^2$]}\\
\hline
1.0 & HF  & S    &  0.56925(2)  &  19.3(1)   \\
    & VMC & SJ   &  0.53360(4)  &  1.26(4)   \\
    &     & SJBF &  0.53139(4)  &  0.81(4)   \\
    & DMC & SJ   &  0.53087(4)  &  -        \\ 
    &     & SJBF &  0.52990(4)  &  -        \\ 
\hline            
5.0 & HF  & S    &  -0.056297(7) &  0.776(4)  \\
    & VMC & SJ   &  -0.075941(6) &  0.0636(1) \\
    &     & SJBF &  -0.078087(4) &  0.0236(1) \\
    & DMC & SJ   &  -0.07862(1)  &  -         \\ 
    &     & SJBF &  -0.07886(1)  &  -         \\    
\hline            
20.0 & HF  & S   &  -0.022051(2) &  0.0482(1)  \\
     & VMC & SJ  &  -0.031692(2) &  0.000711(4) \\
     &     & SJBF&  -0.031862(1) &  0.000437(1) \\
     & DMC & SJ  &  -0.031948(2) &  -        \\ 
     &     & SJBF&  -0.032007(2) &  -        \\    
\hline
\hline
\end{tabular}
\label{table:heg}
\end{table}

\begin{table}
\caption{VMC and DMC energies and variances of local energy 
for Slater--Jastrow (SJ), Pfaffian--Jastrow (PF) and CI-Jastrow (CI) trial wave functions with backflow (BF) correlations 
for C atom. Notation for backflow parameters: 2B, electron-nucleus and electron-electron terms; 3B, all electron-electron-nucleus terms;
23B for all terms together.}
\begin{minipage}{\columnwidth}
\renewcommand{\thefootnote}{\alph{footnote}}
\renewcommand{\thempfootnote}{\alph{mpfootnote}}
\begin{tabular}{l c c c c c  c c c c c  }
\hline
\hline
\multicolumn{1}{l}{Method} &\multicolumn{1}{c}{WF} & \multicolumn{1}{c}{$N_\mu$}  & \multicolumn{1}{c}{$N_\eta$} & \multicolumn{1}{c}{$N_{\theta_1}$}  & \multicolumn{1}{c}{$N_{\theta_2}$} &
 \multicolumn{1}{c}{N$_p$} & \multicolumn{1}{c}{E [a.u.]} &  \multicolumn{1}{c}{$\sigma^2$ [a.u.$^2$]} & \multicolumn{1}{c}{E$_{corr}$[\%]} \\
\hline 
HF  & S & -& - & - & -& -& -5.31471 & - & 0.0 \\
\hline 
VMC & SJ      & -  & -  & -   & -   & -   & -5.3990(1) & 0.0677 & 81.8(1)\\
    & SJBF2B  & 11 & 22 & -   & -   & 33  & -5.4011(2) & 0.0544 & 83.8(2)\\ 
    & SJBF3B  & -  & -  & 128 & 128 & 256 & -5.4023(3) & 0.0504 & 85.0(3)\\
    & SJBF23B & 4  & 8  & 128 & 128 & 268 & -5.4020(2) & 0.0498 & 84.7(2)\\
    & PF      & -  & -  & -   & -   & -   & -5.4083(2) & 0.0626 & 90.8(2)\\
    & PFBF2B  & 11 & 22 & -   & -   & 33  & -5.4097(1) & 0.0427 & 92.1(1)\\
    & PFBF23B & 4  & 8  & 128 & 128 & 268 & -5.4107(1) & 0.0411 & 93.1(1)\\
    & CI\footnotemark[1]      & -  & -  & -   & -   & -   & -5.4127(1) & 0.0447 & 95.0(1)\\
    & CIBF2B  & 11 & 22 & -   & -   & 33  & -5.4131(3) & 0.0427 & 95.4(3)\\
    & CIBF23B & 4  & 8  & 128 & 128 & 268 & -5.4140(1) & 0.0342 & 96.3(1)\\
\hline 
DMC & SJ      & -  & -  & -   & -   & -   & -5.4065(3) & - & 89.0(3)\\
    & SJBF2B  & 11 & 22 & -   & -   & 33  & -5.4090(3) & - & 91.5(3)\\
    & SJBF3B  & -  & -  & 128 & 128 & 256 & -5.4085(3) & - & 91.0(3)\\
    & SJBF23B & 4  & 8  & 128 & 128 & 268 & -5.4094(3) & - & 91.8(3)\\
    & PF      & -  & -  & -   & -   & -   & -5.4137(3) & - & 96.0(3)\\
    & PFBF2B  & 11 & 22 & -   & -   & 33  & -5.4145(3) & - & 96.8(3)\\
    & PFBF23B & 4  & 8  & 128 & 128 & 268 & -5.4152(3) & - & 97.5(3)\\
    & CI      & -  & -  & -   & -   & -   & -5.4178(1) & - & 100.0(1)\\
    & CIBF2B  & 11 & 22 & -   & -   & 33  & -5.4177(3) & - & 99.9(3) \\
    & CIBF23B & 4  & 8  & 128 & 128 & 268 & -5.4174(2) & - & 99.6(2) \\
\hline
Est.& Exact & - & -& - & - & - & -5.417806 & - & 100.0 \\
\hline
\hline
\end{tabular}
\footnotetext[1] {Wave function consists of 100 determinants re-optimized in VMC.}
\end{minipage}
\label{table:bf:C}
\end{table}

\begin{table}
\caption{Slater--Jastrow (SJ), Pfaffian--Jastrow (PF)  and CI--Jastrow (CI) wave functions with backflow (BF) correlations for C dimer.
Notation is the same as in Table~\ref{table:bf:C}.}
\begin{minipage}{\columnwidth}
\renewcommand{\thefootnote}{\alph{footnote}}
\renewcommand{\thempfootnote}{\alph{mpfootnote}}
\centering
\begin{tabular}{l c c c c c  c c c c c  }
\hline
\hline
\multicolumn{1}{l}{Method} &\multicolumn{1}{c}{WF} & \multicolumn{1}{c}{$N_\mu$}  & \multicolumn{1}{c}{$N_\eta$} & \multicolumn{1}{c}{$N_{\theta_1}$}  & \multicolumn{1}{c}{$N_{\theta_2}$} &
 \multicolumn{1}{c}{N$_p$} & \multicolumn{1}{c}{E [a.u.]} &  \multicolumn{1}{c}{$\sigma^2$ [a.u.$^2$]} & \multicolumn{1}{c}{E$_{corr}$[\%]} \\
\hline 
HF & S & -& - & - & -& -& -10.6604 & - & 0.0 \\
\hline 
VMC & SJ\footnotemark[1]     & -  & -  & -   & -   & -   & -10.9936(4) & 0.179 & 81.7(1)\\
    & SJBF2B                 & 11 & 22 & -   & -   & 33  & -11.0012(3) & 0.144 & 83.5(1)\\
    & SJBF23B                & 4  & 8  & 128 & 128 & 268 & -11.0014(2) & 0.141 & 83.6(1)\\
    & PF\footnotemark[2]     & -  & -  & -   & -   & -   & -11.0171(2) & 0.160 & 87.4(1)\\
    & PFBF2B                 & 11 & 22 & -   & -   & 33  & -11.0223(3) & 0.123 & 88.7(1)\\
    & PFBF23B                & 4  & 8  & 128 & 128 & 268 & -11.0223(2) & 0.128 & 88.7(1)\\
    & CI\footnotemark[3]     &  - & -  & -   & -   & -   & -11.0420(4) & 0.112 & 93.6(1)\\
    & CIBF2B                 & 11 & 22 & -   & -   & 33  & -11.0440(3) & 0.100 & 94.0(1)\\
    & CIBF23B                & 4  & 8  & 128 & 128 & 268 & -11.0438(3) & 0.123 & 94.0(1)\\
\hline 
DMC & SJ      & -  & -  & -   & -   & -   & -11.0227(2) & - & 88.8(1)\\
    & SJBF2B  & 11 & 22 & -   & -   & 33  & -11.0269(4) & - & 89.9(1)\\
    & SJBF23B & 4  & 8  & 128 & 128 & 268 & -11.0280(3) & - & 90.1(1)\\
    & PF      & -  & -  & -   & -   & -   & -11.0419(9) & - & 93.5(2)\\
    & PFBF2B  & 11 & 22 & -   & -   & 33  & -11.0443(6) & - & 94.1(2)\\
    & PFBF23B & 4  & 8  & 128 & 128 & 268 & -11.0447(3) & - & 94.2(1)\\
    & CI      &  - & -  & -   & -   & -   & -11.0579(5) & - & 97.5(1)\\
    & CIBF2B  & 11 & 22 & -   & -   & 33  & -11.0580(4) & - & 97.5(1)\\
    & CIBF23B & 4  & 8  & 128 & 128 & 268 & -11.0585(5) & - & 97.7(1)\\
\hline
Est.& Exact & - & -& - & - & - & -11.068(5) & - & 100.0 \\
\hline
\hline
\end{tabular}
\footnotetext[1] {Slater determinant contains PBE DFT orbitals.}
\footnotetext[2] {Same PBE DFT orbitals are used also in PF wave function.}
\footnotetext[3] {Uses natural orbitals with weights of the 148 determinants re-optimized in VMC.}
\end{minipage}
\label{table:bf:C2}
\end{table}

\subsubsection{Carbon atom and dimer}
The backflow correlations in single-determinant Slater--Jastrow (SJBF)~\cite{drummond_bf,rios_bf,gurtubay_bf}
as well as in multi-determinant Slater--Jastrow (CIBF)~\cite{brown_bf} trial wave functions were recently 
applied to chemical systems. We extend these studies by including the backflow 
into the Pfaffian--Jastrow (PFBF) pairing wave functions. Below is a brief discussion of our 
implementation and results for carbon atom and dimer systems.

We employ the inhomogeneous backflow given by Eq.~(\ref{eg:bfterms}), with the functions $u$ and $w_{1,2}$ are allowed to be spin-dependent.
The $\chi$ and $u$ functions are expanded in the 11 Gaussian basis functions, while the three-body 
functions $w_{1,2}$ were limited to a product of $4\times4\times4$ Gaussians.

The numerical results are summarized in Tables~\ref{table:bf:C} and~\ref{table:bf:C2}. 
The backflow correlations are able to capture additional few percent of the correlation energy 
for both Slater--Jastrow and Pfaffian--Jastrow wave functions. Another important feature of backflow is 20-30\% decrease in 
variances of local energy with respect to the wave functions without backflow correlations.
We find that for the fully optimized backflow, the spin-unlike electron-electron functions are almost order of magnitude larger than spin-like ones as well as electron-nucleus functions. The gains are systematic with increasing number of parameters, however we do not 
find the three-body terms as important as reported in previous study~\cite{rios_bf}. 
This difference can be attributed to two main reasons---we use a different basis to expand the three-body 
functions $w_{1,2}$ and we also eliminate atomic cores by pseudopotentials. It is plausible
that for systems with core electrons the three-body correlations are more important due to the 
strong variations of orbitals close to the nucleus. 

Finally, let us discuss the difference between the two systems with respect to missing correlation energy.
For the C atom, we have shown previously~\cite{pfaffianprl} that less than 100 determinants gives more than $99\%$ of correlation energy ($E_{corr}$).
The C dimer's fixed node errors are more pronounced, since the 148 determinants with re-optimized weights
give only 97.5(1)\% in a close agreement with recent calculations by Umrigar {\it et al.\/}\cite{umrigarC2}.
Employing backflow correlations for our 148 determinant CI-Jastrow wave function gives no apparent gain in $E_{corr}$
except for decrease in the variance of local energy.
The improvement for the Pfaffian-Jastrow wave function is also very modest (less than 1\%).
Our results suggest that to reach beyond 99\% of correlation one still needs 
complicated multi-reference wave functions, even after including quite general forms
of the backflow correlations.

\subsection{Conclusions}
We have benchmarked our backflow correlated wave functions on the HEG system and achieved high level 
of agreement with previous results. Further, we have applied the Pfaffian pairing wave functions to chemical systems.
The results for to testing cases of carbon pseudo atom and its dimer show promising gains in correlations energies, 
decreases in variances and improvements in the nodal structure. 

\section{Example of Large Scale Application: FeO Solid}\label{ch:feo}

As an example of calculations, which are feasible at present using
available both software and hardware resources, we briefly report 
our recent study of the FeO solid.

FeO solid is a very interesting
system which has been calculated a number of times before by several approaches.
Nevertheless, it is difficult to claim that the problems and research questions
have been all ``solved'' and, in fact, our understanding of this and many other
transition metal compounds is still rather limited. FeO belongs to the group of
other paradigmatic transition metal (TM) oxides such as MnO, CoO and NiO.
All of these have a rather simple rocksalt (NaCl type) structure denoted as B1,
with very small deformations at lower temperatures resulting from effects 
such as magnetostriction.
Since the transition elements in these systems have open d-shells, such solids should
be nominally metals.
However, for a long time it has been known and experiments have shown that these systems
are
{\em large gap insulators}. In addition, they exhibit antiferromagnetic ordering
with the Neel temperatures of the order of a few hundred Kelvins.
Interestingly, the gap remains present
even at high temperatures when there is no long-range magnetic order.
How is it possible? Using qualitative classification introduced by
Zaanen and coworkers\cite{zaanen}, there are basically two relevant and somewhat competing
 mechanisms: Mott-Hubbard and charge transfer. In the Mott-Hubbard picture,
the gap opens because of large Coulomb repulsion associated with the double occupancy
of the strongly localized d-states. 
The spin minority bands are pushed up in the energy, leading to gap opening.    
 In the charge transfer mechanism,  4s electrons of the transition metal atom  
 fill the unoccupied
p-states of oxygen.  The electronic structure has an almost ionic TM$^{+2}$O$^{-2}$ character, 
with resulting gap opening as well.
 In the real materials both of these mechanisms are present and therefore
the systems exhibit insulating behavior. The antiferromagnetism happens on the
top of this and is caused by weak super-exchange interactions between the
moments of neighbouring transition metal atoms mediated by bridging oxygens.
We will not go into further details, suffice it to say that the electron correlations
are very large and crucially affect many properties of these materials. 

If one
applies standard DFT approaches to FeO, the results are highly unsatisfactory.
First, the atomic structure which comes out is not correct. Instead of the rocksalt
antiferromagnet another structure, so-called iB8, appears to have the lowest energy at
equilibrium conditions. Interestingly, iB8 is actually a high pressure phase of FeO.
  This discrepancy is surprising since DFT methods typically provide correct equilibrium structures
and very reasonable geometric parameters such as lattice constants and others.
 Another problem appears in the electronic structure since for the
correct geometry DFT predicts a {\em metallic} state. More sophisticated methods beyond DFT 
have been applied
to this system in order to reconcile some of the results with experiments,
nevertheless, a number of questions remain unanswered. For example, at high 
pressures, FeO undergoes a structural transition into the iB8 phase, however, the 
value of the transition pressure which agrees with experiment
 is difficult to obtain using the mainstream approaches. Similar transition
appears in MnO, where recent bechmarking of several DFT approaches provided
transition pressure estimates between 65 and 220 GPa, i.e., more than 300\% 
spread in the predictions
\cite{kasinathan}. Large
discrepancies with expriment exist also for the CoO crystal and more complicated
transition metal compounds.

We have carried out QMC calculations of FeO using supercells with periodic
boundary conditions to model the infinite solid~\cite{feo_prl}. Several supercell sizes were calculated 
and  k-point sampling of the Brillouin zone
was carried out by the so-called twist averaging with
the purpose of eliminating finite size effects. 
The core electrons were replaced by pseudopotentials for both Fe (Ne-core) and
O (He-core). The largest simulated supercells had more than 300 valence 
electrons and the total energies were sizeable due to the presence of
``semicore'' 3s and 3p states of Fe in the valence space, what have made the calculations rather
demanding. The wave function had the Slater-Jastrow form 
and the orbitals were obtained from unrestricted (spin-polarized) calculations within 
DFT with hybrid functionals and HF.

In Tab.~\ref{con_feo_table} are shown the QMC calculated equilibrium
parameters. Note that QMC identifies the correct equilibrium structure 
and provides very accurate value of the cohesive energy.
Cohesive energies are very difficult to calculate since the DFT methods show typical bias of 
15-30\%. At present, there is basically no other method besides QMC
which can get this level of accuracy. Note also good agreement with
experiments for the other quantities including the band gap.
In QMC the band gap is calculated as a difference of two total energies---
ground state and  excited state, where the excited state is formed by promoting an electron
from the top valence band into the conduction band.

\begin{figure}[!ht]
\centering
\includegraphics[width=\columnwidth]{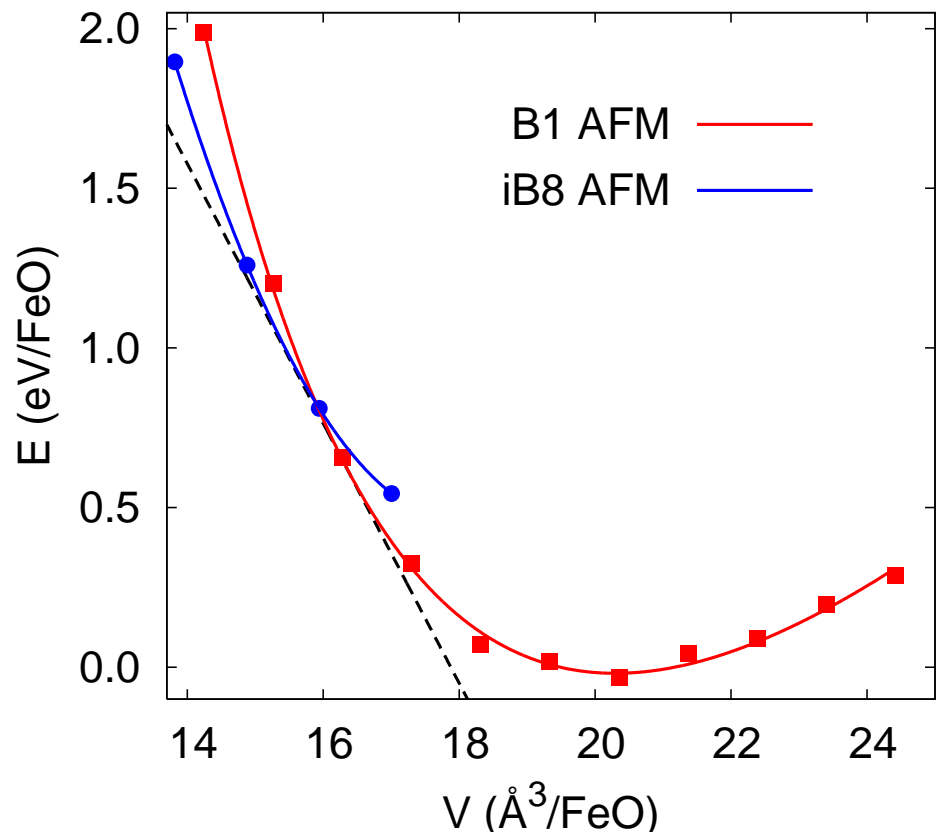}
\caption{FN-DMC Energy as a function of volume for FeO for B1 (red squares) and iB8 (blue circles) phases. 
Lines are fits with Murnagham equation of state.}
\label{fig:feo}
\end{figure}

In addition, the equations of state have been calculated for both
the equilibrium structure and also for the high pressure phase,
see Fig.~\ref{fig:feo}. The estimated transition pressure
is 65(5) GPa, at the lower end of the experimental range 70-100 GPa~\cite{fang1999,zhang2000}.  

Finally, considering that only a simple trial wave function was employed, the
results are remarkable and very encouraging. Note that the calculations do not have any 
free non-variational parameters. It is simply the best possible solution
within the trial function nodes and the given Hamiltonian.

\begin{table}
\caption{Comparision of the calculated structural properies of FeO solid in DFT and in the fixed-node DMC
with experimental data. The energy difference E$_{iB8}$-E$_{B1}$ is evaluated at
the experimental lattice constant 4.334 \AA. }
\label{con_feo_table}
\begin{minipage}{\columnwidth}
\renewcommand{\thefootnote}{\alph{footnote}}
\renewcommand{\thempfootnote}{\alph{mpfootnote}}
\centering
\begin{tabular}{l c c c}
\hline
\hline
Method/quantity & DFT/PBE & FN-DMC & Exper. \\
\hline
E$_{iB8}$-E$_{B1}$ [eV] & -0.2~\cite{fang1998} & 0.5(1) & $>0$ \\
cohesion energy [eV] & $\sim$ 11 & 9.66(4) & 9.7~\cite{feo_exp} \\
lattice constant [\AA] & 4.28~\cite{fang1999} & 4.324(6) & 4.334\cite{mccammon1984} \\
bulk modulus [GPa] & 180~\cite{fang1999}  & 170(10) & $\sim$ 180~\cite{zhang2000} \\
band gap [eV] &  $\sim$ 0 & 2.8(3) & 2.4~\cite{bowen1975}  \\
\hline
\hline
\end{tabular}
\end{minipage}
\end{table}

%% file: qwalk/qwalk.tex
\section{QWalk: A Quantum Monte Carlo Program for Electronic Structure}\label{ch:qwalk}
The solution of the electronic structure of many electron systems in quantum Monte Carlo
requires the use of computers and appropriate software tools. 
For the practitioners of the QMC methods, the computational aspects 
such as the use of efficient algorithms, flexible code organization and parallel efficiency 
are important, since they influence which types of calculations and which types of
physical effects are feasible to study.  
 This section is therefore aimed to introduce a reader to one of the existing
QMC packages, Open Source  QWalk code available under the GPL licence   
\cite{qwalk} (www.qwalk.org). 

This  section is organized as follows. In Sec.~\ref{ch:qwalk:sec1} we overview the program's organization and structure.
The description of the implementation of wave function optimization is in  Sec.~\ref{subsec:opt3}. 

\subsection{Organization and Implementation}\label{ch:qwalk:sec1}
\begin{figure*}
\includegraphics[width=\columnwidth]{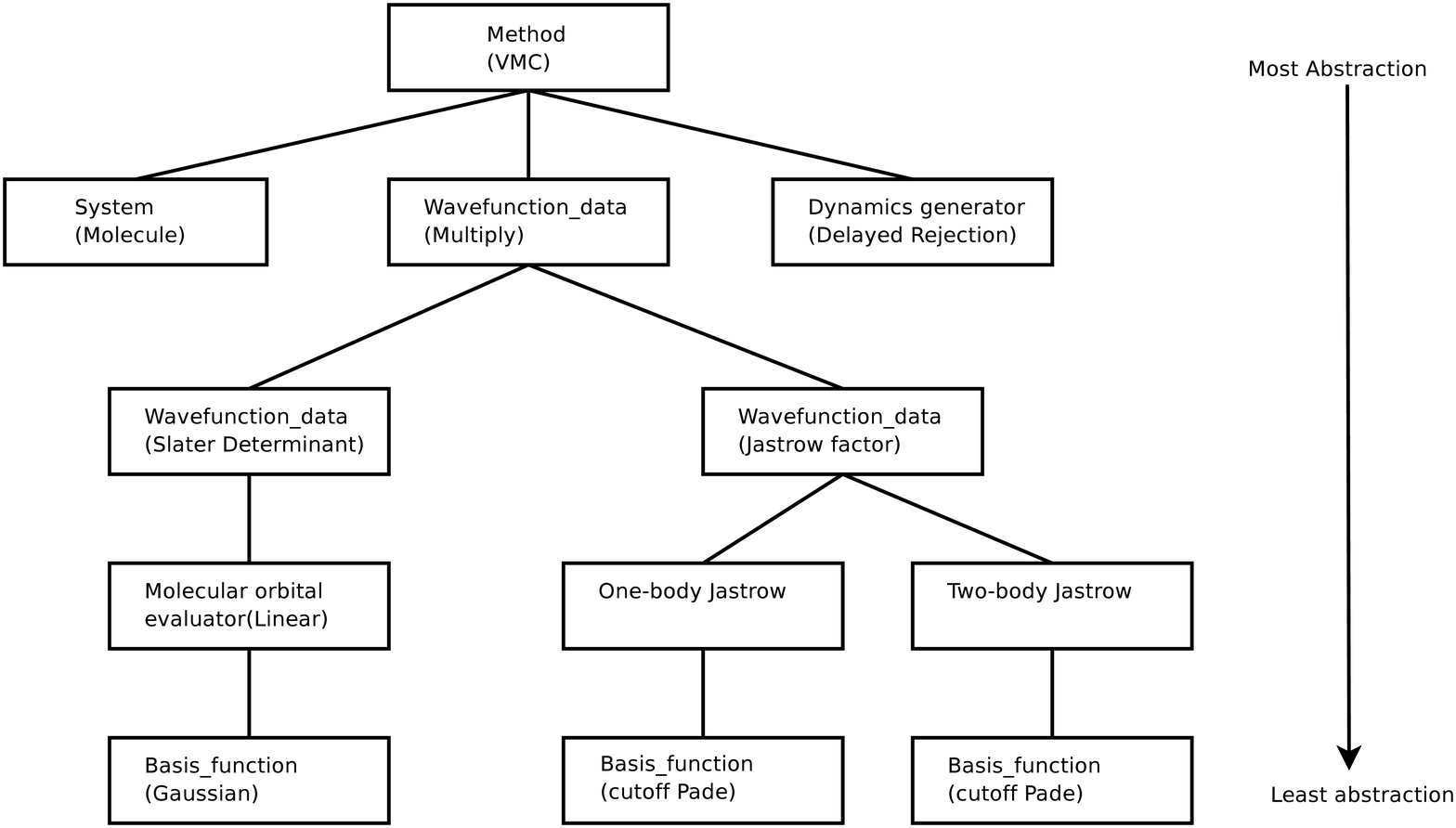}
\caption{Calculation structure for the VMC method on a molecule using a 
Slater-Jastrow wave function. }
\label{fig:tree}
\end{figure*}
Qwalk (``Quantum Walk'') is a computational package for QMC calculations of electronic structure
systems such as molecules and
solids. However, others systems can be calculated such as homogeneous electron gas, BCS-BEC
condensates, 1D rigns with persistent currents, etc.
 The code is written in a combination of object-oriented and procedural techniques. 
The object-oriented approach is coarse-grained, creating independent sections of code
that are written efficiently in a procedural fashion. It is extremely modular; 
almost every piece can be removed and replaced with another. For example, 
a contributor of a new  module only has to change one line in the main code to allow its use. 
This allows for flexibility while keeping the code base relatively simple and separable.  
The modular structure also allows for partial rewrites of the code without worrying 
about other parts.  In fact, each major module has been rewritten several times in this 
manner as we add new features and re-optimize the code flexibility and performance.

\begin{table}
\caption{The central objects of the code and their physical correspondents}
\label{table:correspondence}
\begin{center}
\begin{tabular}{ll}
\hline
\hline
{\bf Module name} & {\bf Mathematical object} \\
\hline
System & parameters and form of the Hamiltonian \\
Sample point & $\bR$, the integration variables \\
Wave function type  & Wave function ansatz \\
Wave function & $\Psi_T(\bR)$, $\nabla\Psi_T(\bR)$, $\nabla^2\Psi_T(\bR)$ \\
Dynamics generator & Metropolis trial move \\
                   & (Green's function) \\
\hline
\hline
\end{tabular}
\end{center}

\end{table}

\begin{figure}[!t]
\begin{center}
\includegraphics[width=0.8\columnwidth]{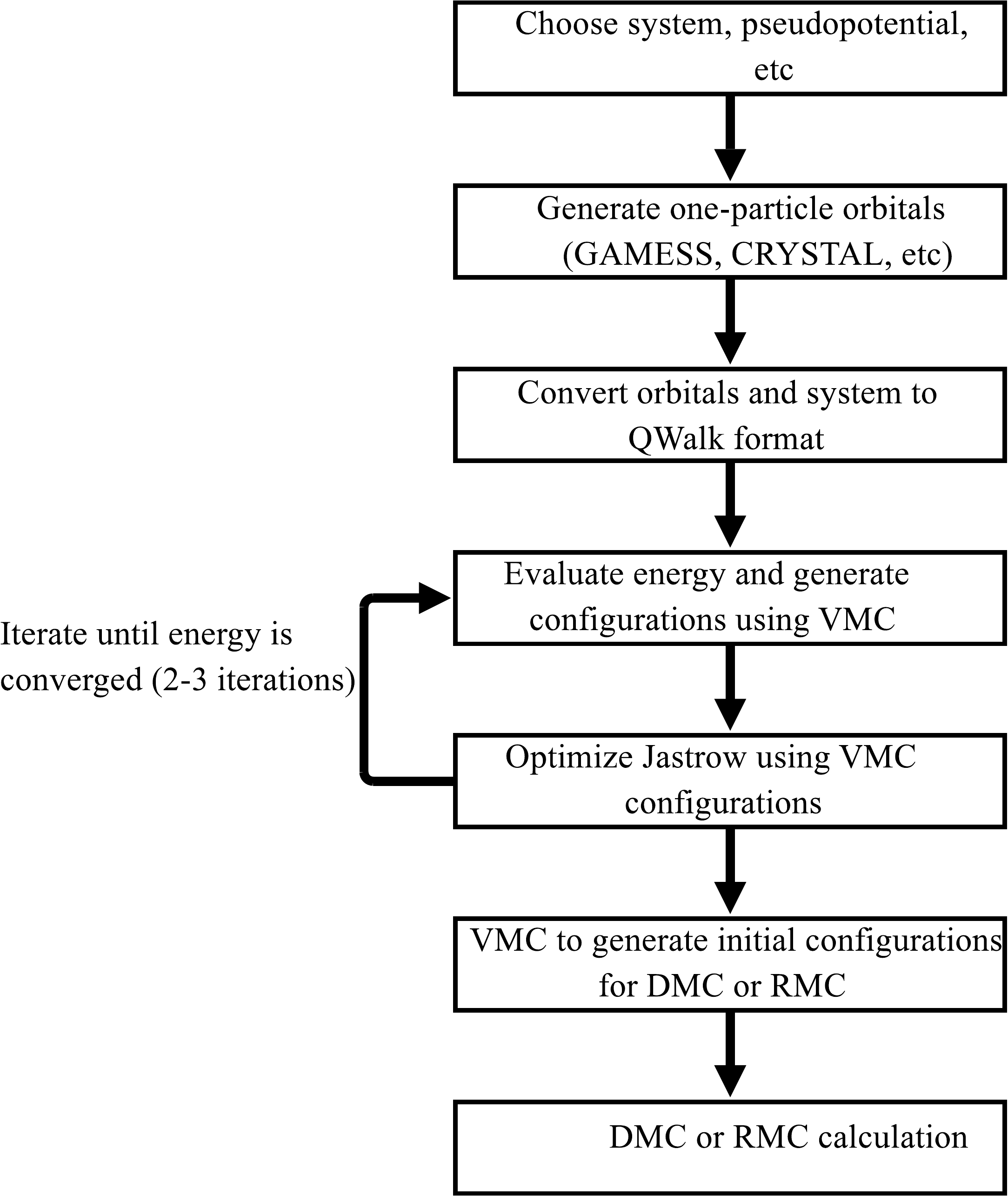}
\end{center}
\caption{Flow of a QMC calculation}
\label{fig:flowchart}
\end{figure}

The modules form a tree of successive abstractions (Fig.~\ref{fig:tree}).  At the top of the tree
is the QMC method, VMC in this case.  It works only in terms of the objects directly
below it, which are the concepts of System, Wave function data, etc. (see Table~\ref{table:correspondence}).
These in turn may have further abstractions below them.
 The highest wave function object is of type ``Multiply'', which uses two 
wave function types to create a combined wave function.  In this case, it multiplies 
a Slater determinant with a Jastrow correlation factor to form a Slater-Jastrow
function.  Since various wave functions can be plugged-in, the Slater determinant can be
replaced with any antisymmetric function. The same applies to the Jastrow factor.  
The type is listed along with the specific instant of that type in parenthesis.  At 
each level, the part in parenthesis could be replaced with another module
of the same type. The example of the flowchart of the whole program is given in Fig.~\ref{fig:flowchart}.

%
%
%
%
%


\subsection{Optimization Methods in QWalk}\label{subsec:opt3}
Historically, the first implementation of minimization method in QWalk code was the OPTIMIZE method, 
based on the quasi-Newton minimizer. 
It has proved effective for variance minimization of linear and nonlinear Jastrow parameters, 
however, the energy minimizations require very large configuration samples. 

This deficiency was partially fixed by the OPTIMIZE2 method, 
which  uses modified Hessians and gradients of variance and energy of Ref.~\cite{cyrus2}
together with the Levenberg-Marquardt minimization algorithm.
The analytical gradients and Hessians 
with respect to selected variational parameters were implemented 
(determinantal and orbital coefficients, but also Pfaffian and pair orbital coefficients).  
OPTIMIZE2 has proved to be very effective method and has been used for most of recent calculations.

In an attempt to remove the dependence of OPTIMIZE2 method on the fixed MC sample, 
the NEWTON\_OPT method was implemented as well. Its principal advantage is that the
energy and variance are obtained from a small MC run performed after each optimization step. 

Besides the variance and energy minimizations, it is possible
to minimize several other {\em objective or cost functions}.
Their incomplete list can be found in Table~\ref{table:cost functions}.
\begin{table}[!t]
\caption{List of some useful cost functions for wave function optimizations
with indicated implementations in QWalk~\cite{qwalk}.}
\begin{center}
\begin{tabular}{l r c c c}
\hline
\hline
\multicolumn{1}{l}{Cost Function} & \multicolumn{1}{c}{Minimized quantity} & \multicolumn{3}{c}{QWalk implementation} \\
\multicolumn{1}{l}{} & \multicolumn{1}{c}{} & \multicolumn{1}{c}{{\small OPTIMIZE}} & \multicolumn{1}{c}{{\small OPTIMIZE2}} & \multicolumn{1}{c}{{\small NEWTON\_OPT}}\\
\hline
Variance  & $\langle(E_L-\bar{E})^2\rangle$ & $\surd$ & $\surd$ & $\surd$\\
Energy & $\bar{E}$  & $\surd$ & $\surd$ & $\surd$\\
Mixed & $x\bar{E} +(1-x)\langle(E_L-\bar{E})^2\rangle$  & $\surd$ & $\surd$ & $\surd$\\
Absolute value & $ \langle |E_L-\bar{E}| \rangle$  & $\surd$ & & \\
Lorentz & $\langle \ln(1+(E_L-\bar{E})^2/2) \rangle$ & $\surd$ & & \\
Ratio  & $\frac{\langle(E_L-\bar{E})^2\rangle}{\bar{E}}$ & & &  \\
Overlap & $\frac{\int \Psi_T\Psi(\tau\to\infty)}{\int \Psi_T^2}$ & & & \\
\hline
\hline
\end{tabular}
\end{center}
\label{table:cost functions}
\end{table}
As was recently pointed out by Umrigar and Filippi~\cite{cyrus2}, 
the wave functions optimized by minimization of a mix of energy and variance 
produce almost as low energy as energy optimized wave functions 
but with lower variances. 

\begin{figure}[!h]
\centering
\begin{minipage}{\columnwidth}
\centering
\includegraphics[width=0.9\columnwidth]{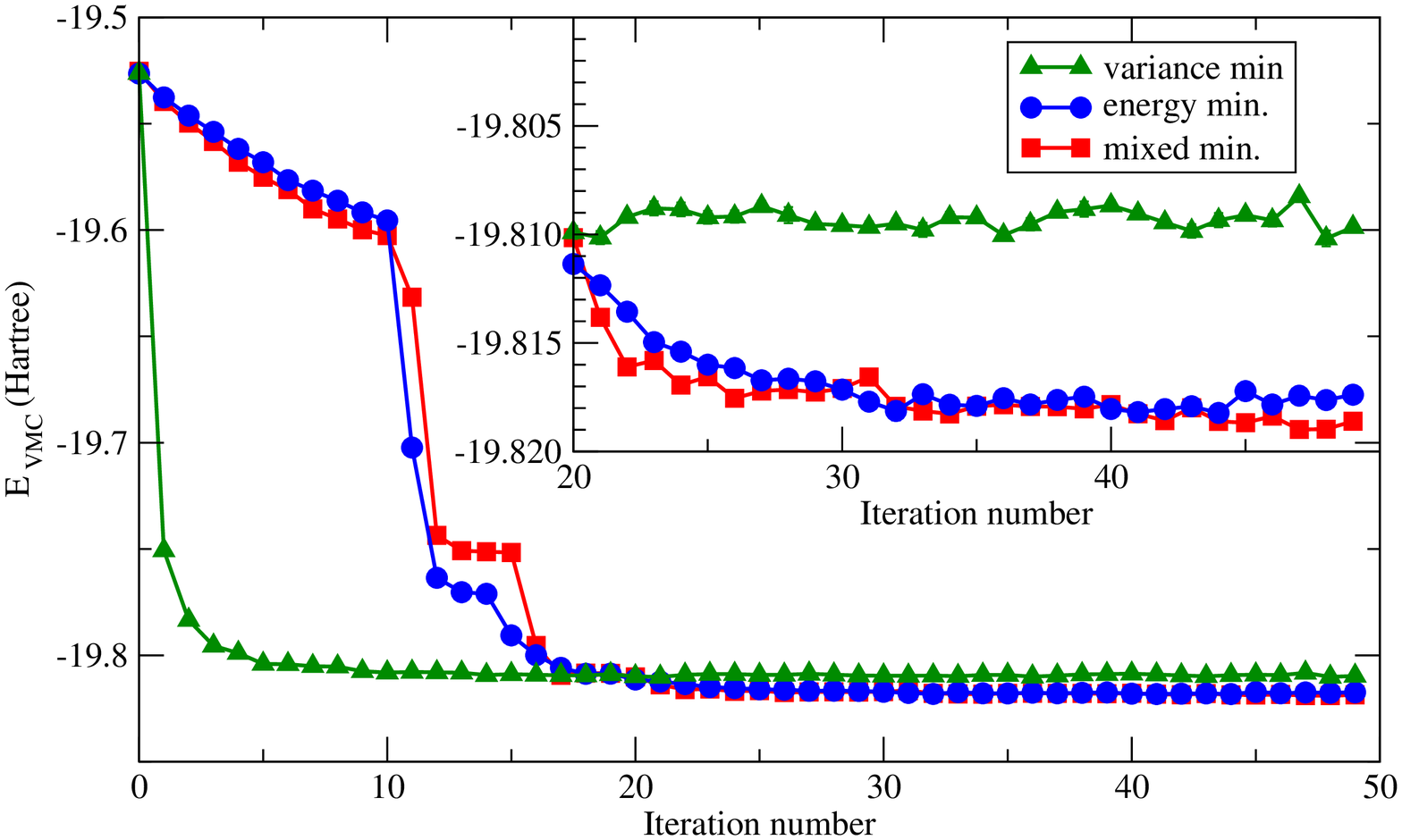}
\includegraphics[width=0.9\columnwidth]{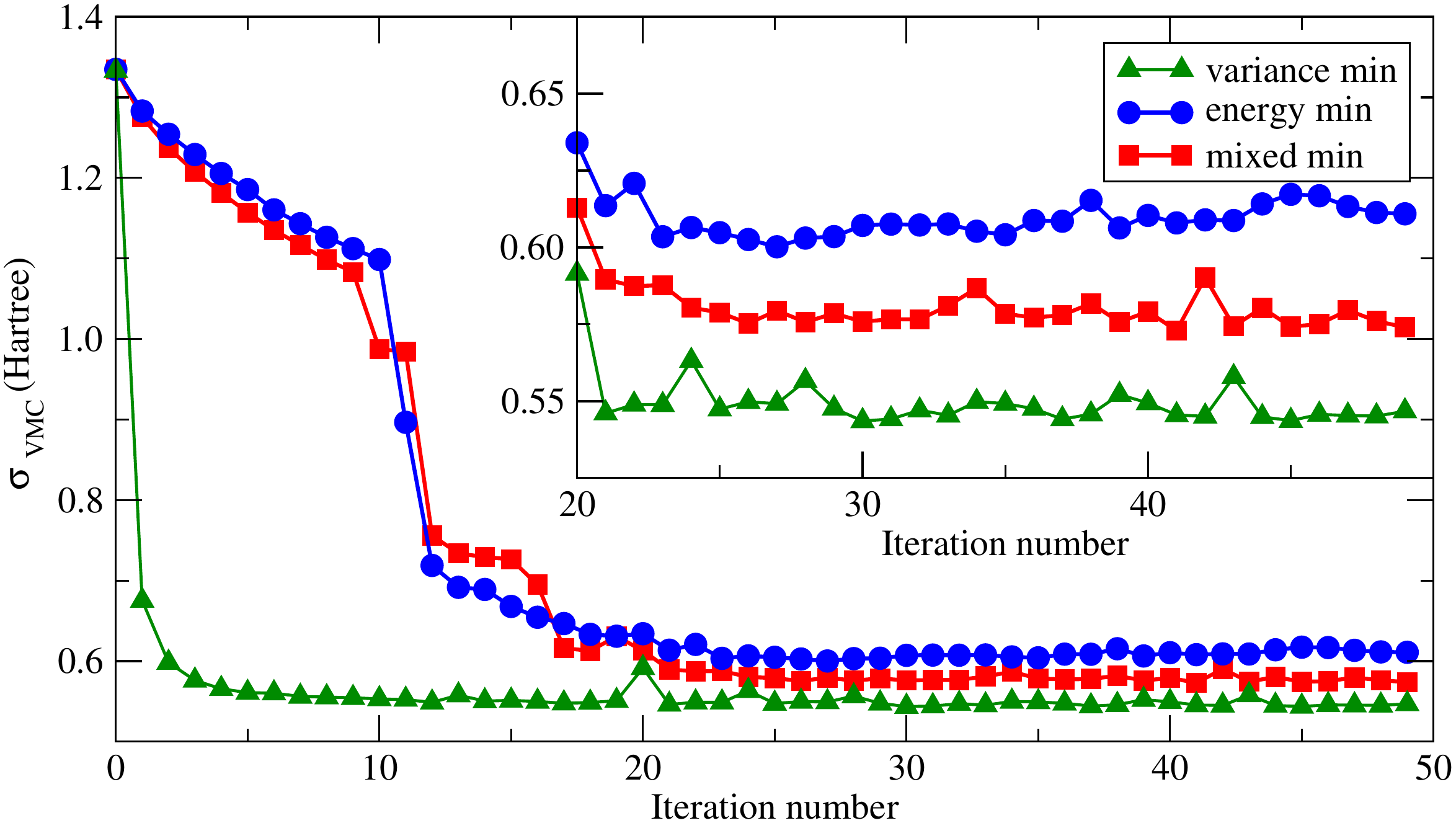}
\end{minipage}
\caption{
Energy $E$ and dispersion of the local energy $\sigma$ of the N$_2$ molecule versus the minimization step number.
Notation: minimization of variance (green triangles), energy (blue circles) and 95\% mixture of energy and 5\% of variance (red squares).
Upper figure: Energy versus the iteration number. The error bars are of the symbol size or smaller. 
Inset: the later iterations on expanded scale. 
Note the variance minimized energy being higher by almost 10 mH. 
Lower figure: Same as the upper figure but for the dispersion of local energy $\sigma$ rather than energy.
Inset shows that the mixture minimized $\sigma$ falls in between $\sigma$ from energy minimization and variance minimization.}
\label{fig:min_n2}
\end{figure}

We illustrate this for the ground state wave function of the N$_2$ molecule (see Fig~\ref{fig:min_n2}).
The employed NEWTON\_OPT method minimized 23 Jastrow parameters 
(the curvatures of correlation basis functions and all linear coefficients)
of single determinant Slater-Jastrow wave function on set of 56000 walkers.
The minimized energy from mixed objective function is almost as good as from the energy optimization, 
while the corresponding dispersion $\sigma$ falls in between $\sigma$ from energy minimization and variance minimization.
The wave functions obtained from mixed objective functions therefore appear as the 
the most efficient compromise for further DMC calculations.

\subsection{Conclusion}
QWalk represents a state of the art, usable, and extensible program
for performing quantum Monte Carlo calculations on several types of quantum
systems. 
It is able to handle medium to large systems of electrons; the 
maximum size is mostly limited by the available computer time.
It works in parallel very efficiently,
so it can take advantage of large clusters, multi-core computers, etc.
Since QWalk is available without charge and under the GNU Public license,
it is hoped that it will help bring both development and use of quantum 
Monte Carlo methods to a wide audience. Due to its modular form it is 
straightforward to expand the QWalk's applicability to quantum systems
beyond the electron-ion Hamiltonians in continuous space. 
With some knowledge of C++ it is easy to modify the system module 
to incorporate other types of interactions and to expand the one-particle and pair
orbitals using the coded basis functions.   


%% file: Conclusions/Conclusions.tex
\section{Summary}\label{ch:last}
We have presented an overview of QMC methods and some of the recent developments
in the analysis of the fermion nodes, constructions of pairing wave functions
and explorations of backflow correlations. Most of the applications were using small
systems both because of educational purposes and in some cases because these were the first
benchmarks of the new developments. 
QMC methodology has proved to be a very powerful technique for studies of quantum
systems and also real materials as illustrated on the FeO solid calculations.
In essence, QMC has a number of advantages when compared with 
other approaches:

\begin{itemize}

\item[-] direct and explicit many-body wave function framework for solving the stationary 
Schr\"odinger equation;

\item[-] favorable scaling with the systems size;

\item[-] wide range of applicability;

\item[-] the fixed-node approximation  which enables to obtain 90-95\% of the correlation effects;
 
\item[-] scalability on parallel machines;

\item[-]  new insights into the many-body phenomena.

\end{itemize}

The remaining fixed-node errors account for small, but still important 
fraction of correlations. As we have shown, further
improvement calls for better understanding of the nodes and their 
efficient descriptions. It is encouraging that studies of the nodal properties 
brought new understanding such as the two nodal cell property for generic
fermionic states and revealed new research directions which are yet to be
explored. 

We have shown that the pairing orbitals and Pfaffian wave functions 
show significant improvements in providing more systematic results and higher accuracy.
The key virtue of the pairing functions is that they exhibit the correct nodal topologies 
and as such are capable to describe quantum condensates and pairing effects.  
As we explained, the backflow has perhaps more limited scope of usefulness
and appears to be effective especially for homogeneous or  marginally inhomogeneous
systems. Many generalizations of QMC exist. For example, for systems with magnetic fields
and stationary currents the fixed-node methods can be generalized to the fixed-phase
QMC \cite{ortiz,vagner}. For finite temperature equilibrium properties Path Integral Monte Carlo 
has been used very successfuly especially for bosonic quantum liquids.

The overall impact of the electronic structure QMC methods has been so far mainly
in providing high accuracy and benchmark results which were used by other methods as a reference.
We believe that the impact of QMC methods is on the rise and 
their use will gradually become more routine.
Already at present, there are cases for which developing elaborate mean-fields take more human time
than direct solutions by many-body stochastic approaches.  However, as we  stated in the introduction,
much needs to be done to make the methods more insightful, efficient and easy to use.

Acknowledgements. Support by NSF EAR-05301110 and DMR-0804549 grants and by DOE DE-FG05-08OR23336
Endstation grant is gratefully acknowledged. We acknowledge also  allocations at ORNL through 
INCITE and CNMS initiatives
as well as allocations at NSF NCSA and TACC centers. We would also like to thank 
L.K. Wagner and J. Koloren\v c for discussions.